\documentclass[11pt]{article}
\usepackage{arxiv}
\newcommand{\vect}[1]{\bm{#1}}
\newcommand{\degree}{^{\circ}}

\title{Wave-Propagation Geometry Emerges from Scalar Travel-Time Relations}
\author[1]{Ziye Yu\thanks{Corresponding author: \href{mailto:yuziye@cea-igp.ac.cn}{yuziye@cea-igp.ac.cn}. ORCID: \href{https://orcid.org/0000-0002-1720-3811}{0000-0002-1720-3811}.}}
\affil[1]{Institute of Geophysics, China Earthquake Administration, Beijing 100081, China}
\date{}
\hypersetup{pdftitle={Wave-Propagation Geometry Emerges from Scalar Travel-Time Relations},pdfauthor={Ziye Yu}}

\begin{document}

\maketitle

\begin{abstract}
Scalar travel-time relations can identify propagation geometry without velocity inputs, ray labels or equation supervision during learning. Physical interpretation is applied after learning through the high-frequency isotropic eikonal relation~\citep{AkiRichards2002}. Here we show that local value differences constrain derivatives through sampling geometry and smoothness. Controlled neural fields recover propagation directions with $5.34\pm0.07\degree$ median angular error, and local P- and S-wave velocities in a prospectively locked, test-sealed realization. Classical representations also recover geometry. Scalar-matched perturbations separate value accuracy from derivative reliability; regularized inversion can recover structure when direct gradients fail. A field trained on 2.90 million Chinese arrivals retains continental crust--mantle boundary structure: correlation with a withheld reference is $\rho=0.910$ raw and 0.592 after quadratic detrending (spatial-shift $p=0.020$). Direct catalogue values yield $\rho=0.909$, and thickness contrasts are compressed. Manual-only California arrivals support broad velocity organization with limited lateral fidelity. Synthetic quantum fields containing phase information support probability-flow and trajectory readouts without supervising those quantities. These results establish conditional identification across representations and physical settings: informative scalar relations can constrain geometry, while known physical relations supply its interpretation. They do not imply that accurate scalar fitting alone guarantees reliable physical derivatives.
\end{abstract}

Seismic imaging turns earthquake arrival times into models of Earth's interior. Conventional travel-time tomography represents the unknown medium through a velocity field and predicts arrivals through a propagation model~\citep{Rawlinson2010}. The same model supplies the ray directions needed to interpret earthquake radiation. The question is whether an arrival-time archive can itself supply the propagation geometry needed to interpret Earth structure and earthquake radiation, before a velocity field is prescribed to the learner.

Differentiable travel-time fields already support physical computation. EikoNet and GlobeNN learn travel-time solutions using prescribed velocity models and eikonal constraints~\citep{Smith2021,Taufik2023}; PINNeik also uses equation residuals~\citep{Waheed2021}, within the broader practice of equation- or derivative-supervised learning~\citep{Raissi2019,Czarnecki2017}. Shi et al.'s HARPA couples travel-time neural fields to a learned low-dimensional wave-speed prior for joint phase association, location and medium inference~\citep{Shi2026}. These advances establish continuous travel-time parameterization and its applications. Our question concerns the information needed to identify its derivatives when velocity-conditioned inputs and equation or derivative supervision are absent. We connect a local derivative-error bound to coverage interventions, scalar-matched negative controls and a prospectively locked metric test. The distinction is the conditional identification of unsupervised physical readouts, rather than the use of a neural travel-time field itself.

The field $T(\vect{x}_r,\vect{x}_s)$ connects overlapping event--station observations. On a differentiable branch in the high-frequency isotropic limit, the negative source-gradient direction gives the initial propagation direction, while the gradient norm gives local slowness~\citep{AkiRichards2002}. For a fixed receiver, nearby source coordinates constrain this vector through

\begin{equation}
T(\vect{x}_i)-T(\vect{x}_0)\simeq
\nabla T(\vect{x}_0)^{\mathsf T}(\vect{x}_i-\vect{x}_0).
\label{eq:local_taylor}
\end{equation}
Each difference constrains one gradient projection. Directional diversity, proximity, curvature and noise determine how accurately these projections identify the vector.

We test this route with raw-coordinate neural fields, classical scalar estimators, controlled media and earthquake catalogues. Independently solved reciprocal fast-marching fields supply gradient references; matched controls and coverage interventions identify what makes derivatives physical. Fields trained separately for each medium or catalogue are then queried for velocities, rays, mechanisms and crustal thickness without task-specific retraining. The geophysical advance is to make the observed travel-time relation itself a reusable object of inference, rather than requiring arrivals to first be converted into a velocity model. Its values retain continental structure and support regularized velocity extraction; its derivatives supply local propagation geometry where their metric is reliable. Each application reuses its own trained field; no common weights are assumed across regions or media. A companion quantum test asks whether phase-sensitive field values support analogous geometric queries under a different evolution equation.

\begin{figure}[t]
  \centering
  \includegraphics[width=0.99\textwidth]{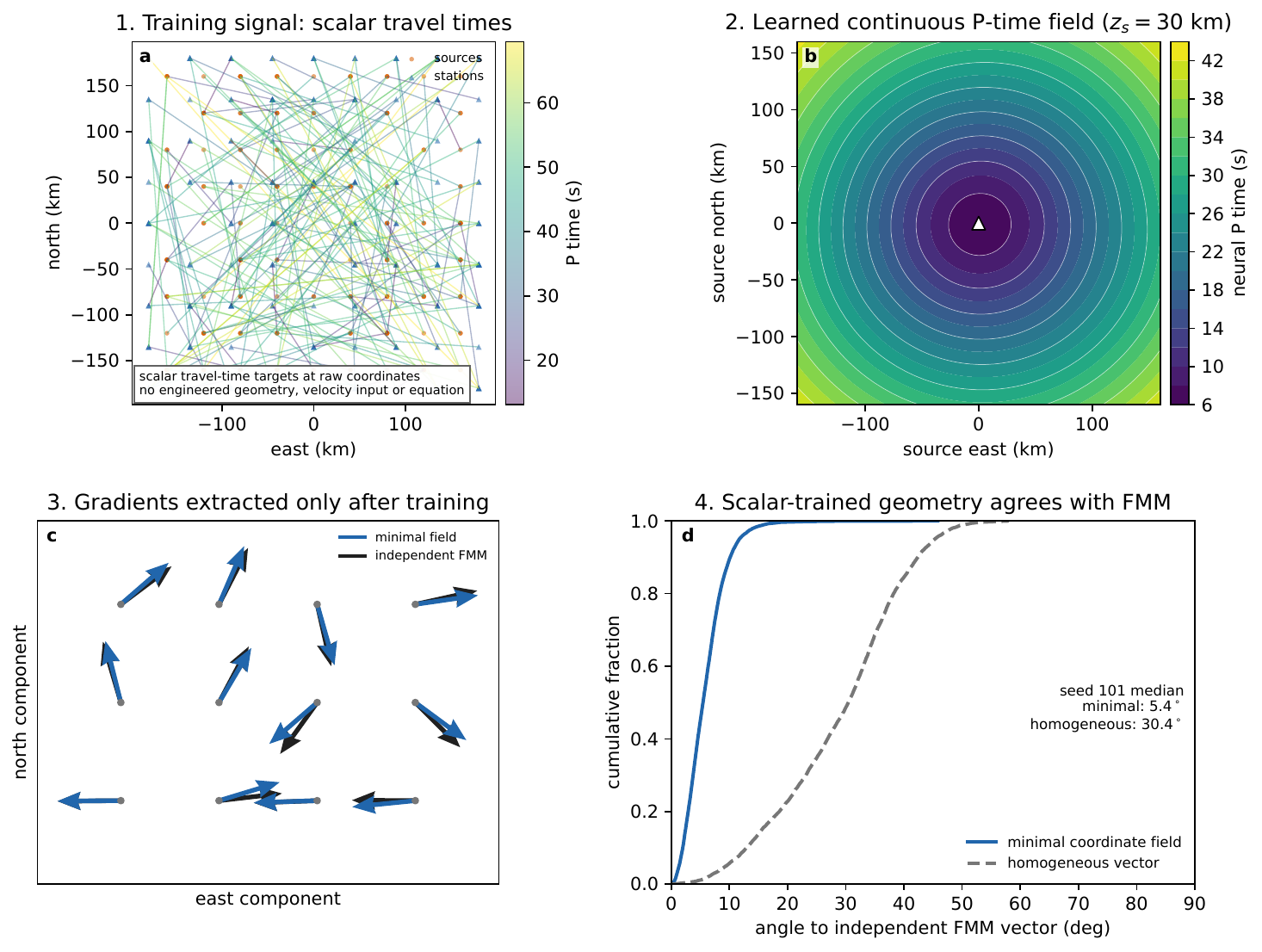}
  \caption{\textbf{Scalar travel-time relations become a queryable geometric field.} (a) Event--station values and the coordinate relations that connect them. (b) The continuous P-time field at 30-km source depth for one fixed receiver. (c) Unit horizontal projections of 12 illustrative source takeoff vectors from the learned field (blue) and independent reciprocal fast-marching-method (FMM) fields (black). (d) Cumulative distributions of three-dimensional angular error over all 6,885 held-out paths for the seed-101 learned field and homogeneous-reference geometry.}
  \label{fig:emergence}
\end{figure}

\begin{samepage}
\section*{Results}

The contribution is conditional identifiability from scalar relations, demonstrated across representations. Neural fields provide a catalogue-scale continuous representation: the China model fits 2.90 million arrivals across four phases and supports post-training value and derivative queries within its coordinate domain. This demonstrated scale and reuse do not establish accuracy or computational superiority over classical alternatives. Controlled experiments test directional and metric recovery against known media; catalogue experiments test physical queries against withheld observations, and the quantum companion probes the scope of the readout principle.
\par
\end{samepage}

\subsection*{Scalar relations identify physical geometry}

Trained on scalar P- and S-wave times at raw source--receiver coordinates, the minimal field predicts held-out arrivals with a mean absolute error (MAE) of $0.717\pm0.008$~s across five seeds. Its source gradients agree with reciprocal fast-marching-method (FMM) P-wave gradients to a median angular error of $5.34\pm0.07\degree$ and a relative-vector error of $0.108\pm0.001$ (Fig.~\ref{fig:emergence}). The corresponding homogeneous-reference errors are $30.45\degree$ and 0.523. The median source eikonal residual is $0.0264\pm0.0011$ (mean $0.0390\pm0.0005$), despite the absence of velocity or equation targets. In these multi-seed summaries, central values are seed means and $\pm$ denotes seed standard deviation.

Reciprocal station-centred FMM vectors enter only evaluation. The learned source takeoff direction is

\begin{equation}
\widehat{\vect{q}}_s=-\frac{\nabla_{\vect{x}_s}T_{P,\theta}}
{\|\nabla_{\vect{x}_s}T_{P,\theta}\|}.
\label{eq:ray_direction}
\end{equation}

A parameter-matched tanh field recovers the same structure, with $0.917$~s MAE, $6.18\degree$ angular error and 0.131 relative-vector error. Engineered distance and reference-time inputs are unnecessary. Representation nevertheless matters: adding a fixed, smooth 0.05-s oscillation changes scalar MAE by $0.0003\pm0.0005$~s but raises angular error to $38.94\degree$ and relative-vector error to 0.987. This constructive control shows that scalar accuracy alone cannot certify a derivative; the unperturbed fields recover both values and geometry.

Geometric recovery extends beyond neural parameterizations. With identical source partitions, validation-selected cubic moving least squares (MLS) gives 0.055~s MAE and $2.73\degree$ median angular error in the full-coverage heterogeneous medium, compared with 0.717~s and $5.34\degree$ for the neural field. At 25\% coverage, MLS retains lower scalar error (0.497 versus 0.725~s) but has higher full-derivative angular error ($6.56\degree$ versus $5.64\degree$). A layered-medium comparison also recovers geometry with both representations. The same scalar relations therefore support geometric recovery through distinct representations, whose accuracy depends on coverage and query type (Supplementary Information).

\subsection*{Relational coverage governs identifiability}

Geometry appears where observations contain enough independent relations to determine it. Equation~\ref{eq:local_taylor} leads to the weighted local design tensor
\begin{equation}
\vect{G}(\vect{x}_0)=\sum_{i\in\mathcal{N}(\vect{x}_0)}
w_i\,\Delta\vect{x}_i\Delta\vect{x}_i^{\mathsf T},\qquad
w_i\propto\exp\left[-\frac{\|\Delta\vect{x}_i\|^2}{2h^2}\right].
\label{eq:design_tensor}
\end{equation}
The tensor records the directions sampled by observations. Its rank determines which gradient components are constrained at first order, while its eigenvalues measure the strength of those constraints.

Writing
$\Delta\vect{T}=\vect{X}\vect{g}+\vect{r}+\vect{\epsilon}$, with rows $\vect{X}_{i:}=\Delta\vect{x}_i^{\mathsf T}$, gradient $\vect{g}=\nabla T(\vect{x}_0)$, curvature remainder $\vect{r}$ and observation noise $\vect{\epsilon}$, weighted local linear estimation gives
\begin{equation}
\widehat{\vect{g}}=(\vect{X}^{\mathsf T}\vect{W}\vect{X})^{-1}
\vect{X}^{\mathsf T}\vect{W}\Delta\vect{T}.
\label{eq:local_gradient_estimator}
\end{equation}
Here $\vect{W}=\operatorname{diag}(w_i)$. The same design tensor directly controls a learned derivative. Define the local
relation mismatch
$\delta_i=[T_\theta(\vect{x}_i)-T_\theta(\vect{x}_0)]-[T(\vect{x}_i)-T(\vect{x}_0)]$.
If $T$ and $T_\theta$ have Hessian norms bounded by $M$ and $M_\theta$ in the
neighbourhood and $\vect{G}=\vect{X}^{\mathsf T}\vect{W}\vect{X}$ is full rank, then
\begin{equation}
\|\nabla T_\theta(\vect{x}_0)-\nabla T(\vect{x}_0)\|_2\leq
\frac{\|\vect{W}^{1/2}\vect{\delta}\|_2+
\frac{M_\theta+M}{2}\left(\sum_i w_i\|\Delta\vect{x}_i\|_2^4\right)^{1/2}}
{\sqrt{\lambda_{\min}(\vect{G})}}.
\label{eq:gradient_error_bound}
\end{equation}
The bound connects relational coverage to learned-gradient accuracy: directional diversity limits error amplification, proximity limits curvature bias, and scalar mismatch enters through $\vect{\delta}$. Observational noise adds a weighted relation-error term. It extends the familiar local-polynomial and moving-least-squares argument to a smooth learned field~\citep{LancasterSalkauskas1981,Levin1998,DeBrabanter2013}. Across 30 locked minimal-coordinate fits, the run-median weakest design fraction $\widetilde{\lambda}_{\min}=\lambda_{\min}/\operatorname{tr}(G)$ orders angular and relative-vector error with Spearman $\rho=-0.911$ and $-0.928$. At fixed source count, angular/vector correlations remain $-0.636/-0.661$ at 10\% coverage and $-0.648/-0.733$ at 5\%; at 25\%, errors have saturated at low levels. With architecture and update budget matched, variation in relational design predicts variation in the learned derivative.

\begin{figure}[t]
  \centering
  \includegraphics[width=0.99\textwidth]{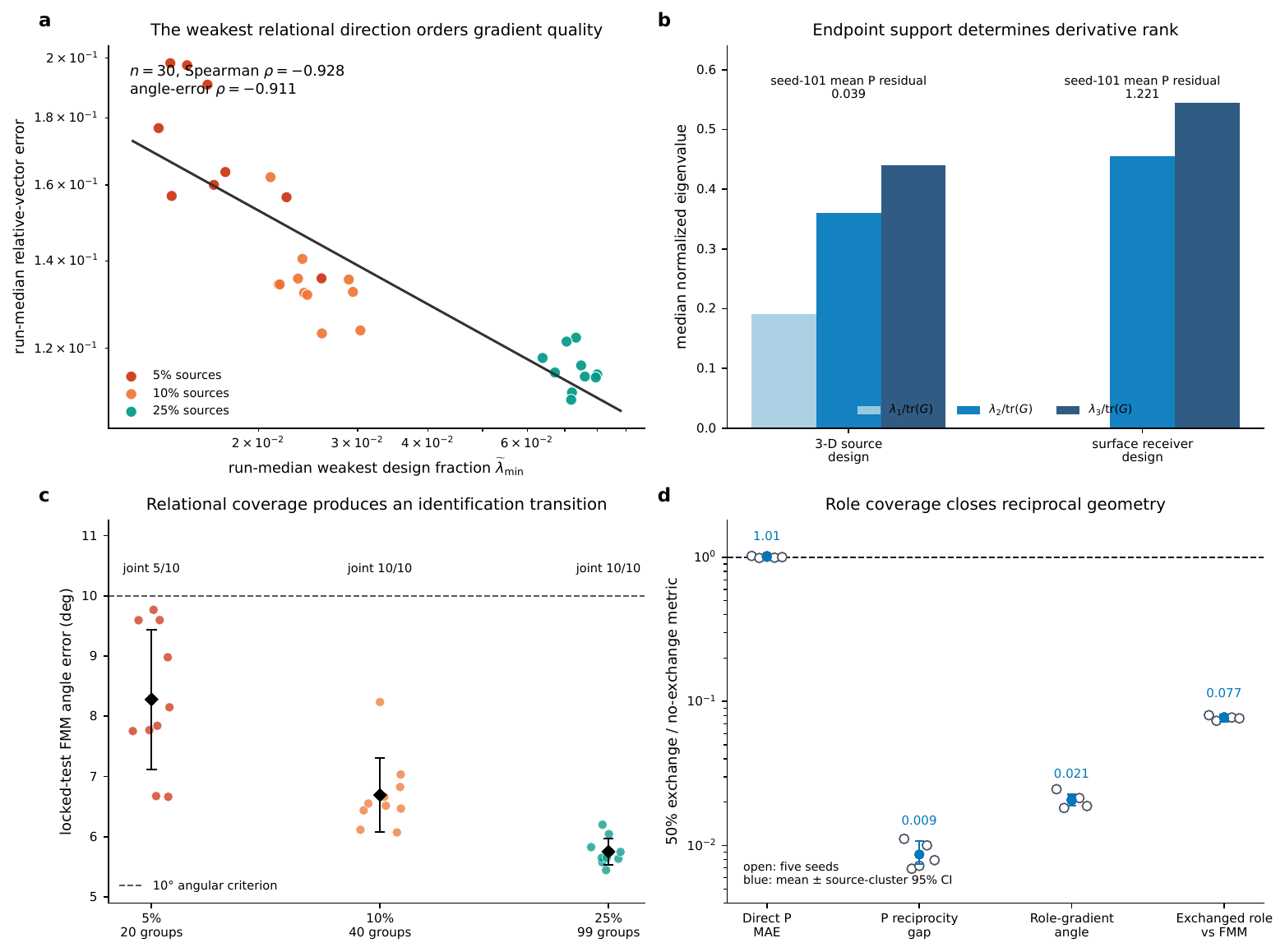}
  \caption{\textbf{Relational support controls geometric identifiability in the minimal-coordinate field.} (a) Across 30 locked minimal-field fits, the run-median weakest design fraction orders relative-vector and angular error; the relation also persists within the equal-count 10\% and 5\% ensembles. (b) Surface confinement removes one receiver-design eigenvalue and leaves the endpoint derivative metrically uncalibrated, whereas the three-dimensional source design remains full rank. (c) Ten independent minimal-field fits per coverage level reveal the transition from data-identified to underdetermined geometry. (d) Reversing 50\% of the identical scalar-pair presentations closes the unsupported endpoint role; open points are five seeds and blue points show means with source-cluster 95\% intervals. Panels a--c establish identifiability in the unaugmented minimal field; panel d shows that scalar role completion closes reciprocal geometry without derivative labels or equations.}
  \label{fig:identifiability}
\end{figure}

Source--receiver asymmetry exposes the missing degree of freedom (Fig.~\ref{fig:identifiability}b). Sources span three dimensions, whereas surface stations have no receiver-depth offsets. Mean P eikonal residuals are $0.0390\pm0.0005$ at the source and $1.123\pm0.134$ at the receiver. Differentiability alone does not identify the unsupported depth derivative.

Endpoint exchange tests whether completing the observed roles closes reciprocity. Five matched fields use the same scalar pairs, with a deterministic 50\% of each batch presented in reversed order. This augmentation explicitly supplies reciprocity symmetry. The P-time reciprocity gap falls from $13.70\pm0.65$ to $0.119\pm0.028$~s and the direct--exchanged gradient angle from $72.52\pm1.51\degree$ to $1.50\pm0.16\degree$ (Fig.~\ref{fig:identifiability}d). Exchanged-role angular error against FMM falls from $72.70\pm1.37\degree$ to $5.60\pm0.13\degree$, while direct P-time MAE changes by only 0.0035~s. Endpoint-role coverage therefore controls reciprocal closure without derivative labels or equation residuals.

All ten subsets at both 25\% and 10\% coverage recover geometry, with subset-averaged median angular errors of $5.75\degree$ and $6.69\degree$. At 5\%, five of ten meet the joint scalar--direction--magnitude criterion (Fig.~\ref{fig:identifiability}c). Relational design, beyond sample count, governs this transition.

Across 206,550 locked queries, the pooled design--error association is $\rho=-0.222$. Removing run means in log space gives a within-run slope of $-0.040$ (run-cluster 95\% interval $-0.066$ to $-0.015$; Supplementary Information). This local association complements the stronger coverage-level transition; it does not isolate the scalar-mismatch, curvature and scale terms that also enter Eq.~\ref{eq:gradient_error_bound}.

\subsection*{Direction and metric emerge without equation supervision}

Residual-SiLU and tanh fields recover physically admissible gradients from the same scalar targets. In a layered medium, value-only learning reaches $0.044$~s MAE and $2.22\degree$ angular error and remains jointly accurate at 5\% coverage. In a smooth three-dimensional heterogeneous medium, it reaches $0.270$~s and $3.58\degree$, retaining joint recovery at 25\%. Greater field complexity raises the required relational density without changing the mechanism.

\begin{figure}[t]
  \centering
  \includegraphics[width=0.99\textwidth]{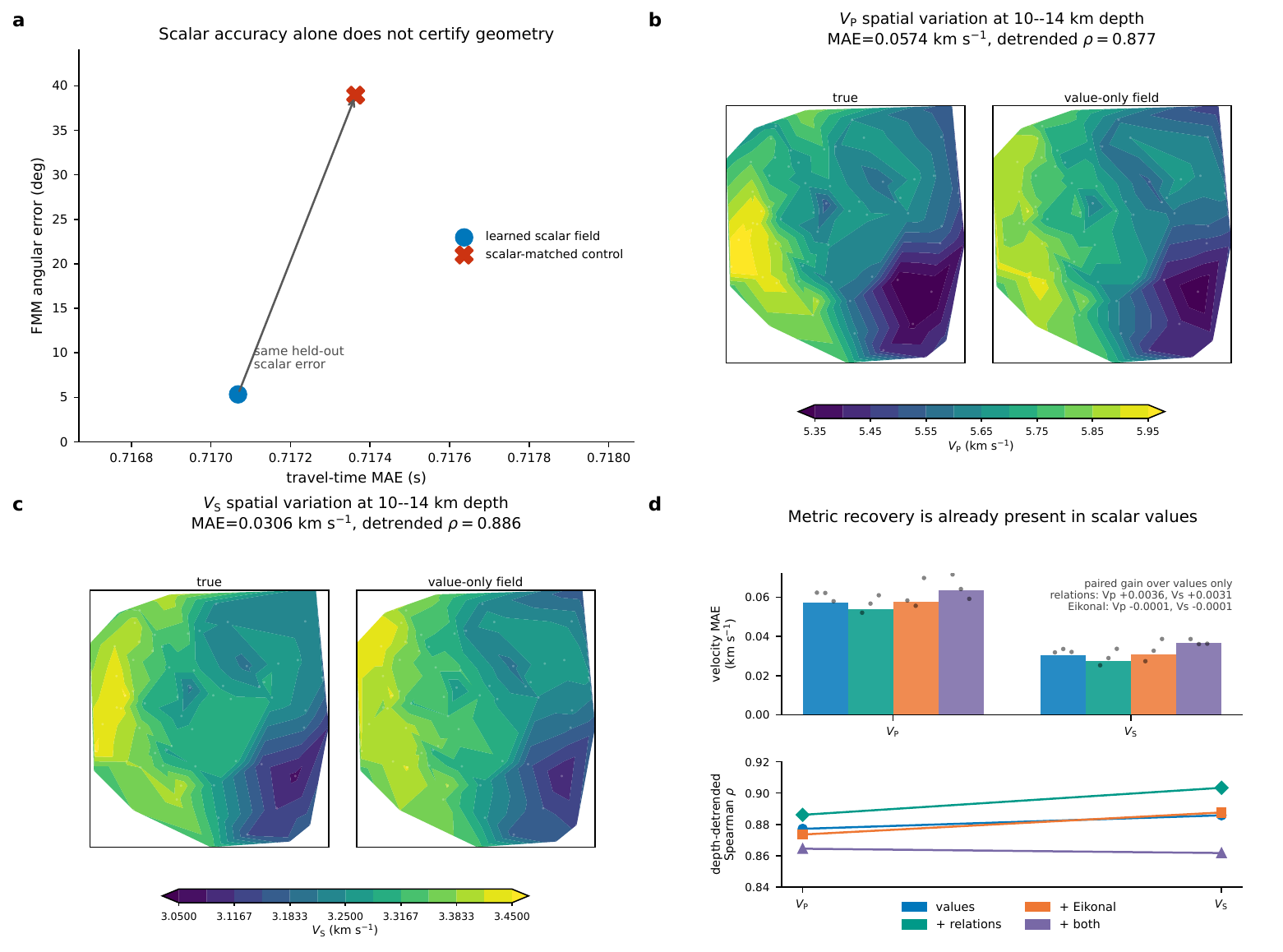}
  \caption{\textbf{Direction and metric structure emerge without equation supervision during learning.} (a) A scalar-matched oscillatory control preserves held-out travel-time error while disrupting gradient direction; points show means across five seeds. (b,c) True and value-only $V_\mathrm{P}$ and $V_\mathrm{S}$ variations in an independently generated medium; maps interpolate held-out sources at 10--14~km depth, whereas annotated metrics use all 307 test sources. (d) Four fields use identical data, architecture, optimization budget and three seeds. Bars and curves show velocity metrics after median aggregation over receivers and seeds; black points show individual-seed MAEs. In this first realization, values alone recover both velocity fields, scalar relations provide a modest refinement, and endpoint-eikonal supervision gives no statistically resolved advantage over values alone.}
  \label{fig:metric_emergence}
\end{figure}

Metric recovery goes beyond orientation: the gradient norm must reproduce local slowness, $v_\alpha=1/\|\nabla_{\vect{x}_s}T_\alpha\|$. Four fields were tested on one independently generated three-dimensional medium with identical source-disjoint data, architecture, budget and seeds (Fig.~\ref{fig:metric_emergence}). Values alone recover $V_\mathrm{P}$ with 0.0574~km~s$^{-1}$ MAE and depth-detrended $\rho=0.877$, and $V_\mathrm{S}$ with 0.0306~km~s$^{-1}$ MAE and $\rho=0.886$ across 307 held-out events. Spatial slices reproduce the true velocity anomalies. Metric structure is present before local finite-difference relations or an equation enter the objective.

Same-station secants constructed from scalar values refine $V_\mathrm{P}$ and $V_\mathrm{S}$ MAE to 0.0537 and 0.0274~km~s$^{-1}$, with paired gains of 0.00363 (95\% interval 0.00140--0.00586) and 0.00313~km~s$^{-1}$ (0.00191--0.00439). Exact endpoint-velocity supervision through the eikonal equation gives 0.0575 and 0.0306~km~s$^{-1}$, statistically unresolved from values alone. Combining both terms gives 0.0636 and 0.0366~km~s$^{-1}$. Scalar observations identify metric scale before either auxiliary term is introduced.

Coverage determines where the equation contributes. At 100\% and 25\% coverage, every value-only seed meets the joint criterion; eikonal supervision changes angular error by $-0.255\degree$ and $-0.171\degree$. At 10\%, it reduces time MAE from 0.824 to 0.770~s, angular error from $6.68\degree$ to $5.71\degree$ and relative-vector error from 0.138 to 0.112. At 5\%, qualification rises from one to four of five seeds despite unresolved scalar gain. The equation supplies additional constraints where scalar relations are sparse.

A prospectively locked confirmation froze all four training cells before generating a second velocity realization and sealed test truth until all 12 checkpoints were hashed. On 307 new held-out events, values alone recover $V_\mathrm{P}$ with 0.0606~km~s$^{-1}$ MAE and depth-detrended $\rho=0.791$, and $V_\mathrm{S}$ with 0.0258~km~s$^{-1}$ and $\rho=0.862$, passing all four predeclared metric-emergence gates. Eikonal-minus-value MAE differences are $-0.00727$ and $-0.00130$~km~s$^{-1}$; both hierarchical-bootstrap intervals include zero. Non-inferiority remains unresolved across three seeds, while scalar secants provide resolved refinements (Supplementary Information). The new realization confirms metric recovery without equation supervision during learning, using the same eikonal readout after training; it does not establish equivalence between objectives.

Event-correlated perturbations preserve joint recovery at the moderate tested level ($5.27\degree$ angular error and 0.105 relative-vector error), while the stronger level approaches the identifiability boundary ($9.14\degree$ and 0.187; Supplementary Information).

\subsection*{Post-training queries probe continental and source structure}

The same learned relation can be queried as an arrival time, a propagation direction or a local slowness. We use frozen travel-time fields to recover crustal and source structure through physical readouts applied after learning.

A four-phase field learns 2.90 million Pg, Sg, Pn and Sn values from 791,404 Chinese earthquakes recorded in 2009--2022. Event-disjoint test MAEs are 0.398, 0.578, 0.934 and 1.631~s, improving on phase-specific scalar regression by 8.7, 8.2, 22.4 and 11.2\%. No crustal thickness, velocity field or head-wave equation enters training. Training-data branch fits remove 12,003 nominal Pn records better explained by Pg before the physical query.

Crustal thickness is queried through a regional head-wave approximation after training. The gradient-based readout defines
\begin{equation}
p_c=\|\nabla_{\vect{x}_s}T_{Pg}\|,\qquad
p_m=\|\left(\nabla_{\vect{x}_s}T_{Pn}\right)_{xy}\|,\qquad
\eta=\sqrt{p_c^2-p_m^2},
\label{eq:relational_slowness}
\end{equation}
where $p_c$ and $p_m$ are crustal and refractor slowness estimates. The Pn value supplies the intercept in
\begin{equation}
H_s+H_r=\frac{T_{Pn}-p_m d+\eta z_s}{\eta},
\label{eq:relational_moho}
\end{equation}
where $d$ is epicentral distance, $z_s$ is source depth and $H_s,H_r$ are endpoint crustal thicknesses. The readout combines 62,086 qualified relations from 12,231 held-out earthquakes into a 261-cell field on a 2$\degree$ grid, without adjusting neural weights.

\begin{figure}[t]
  \centering
  \includegraphics[width=0.99\textwidth]{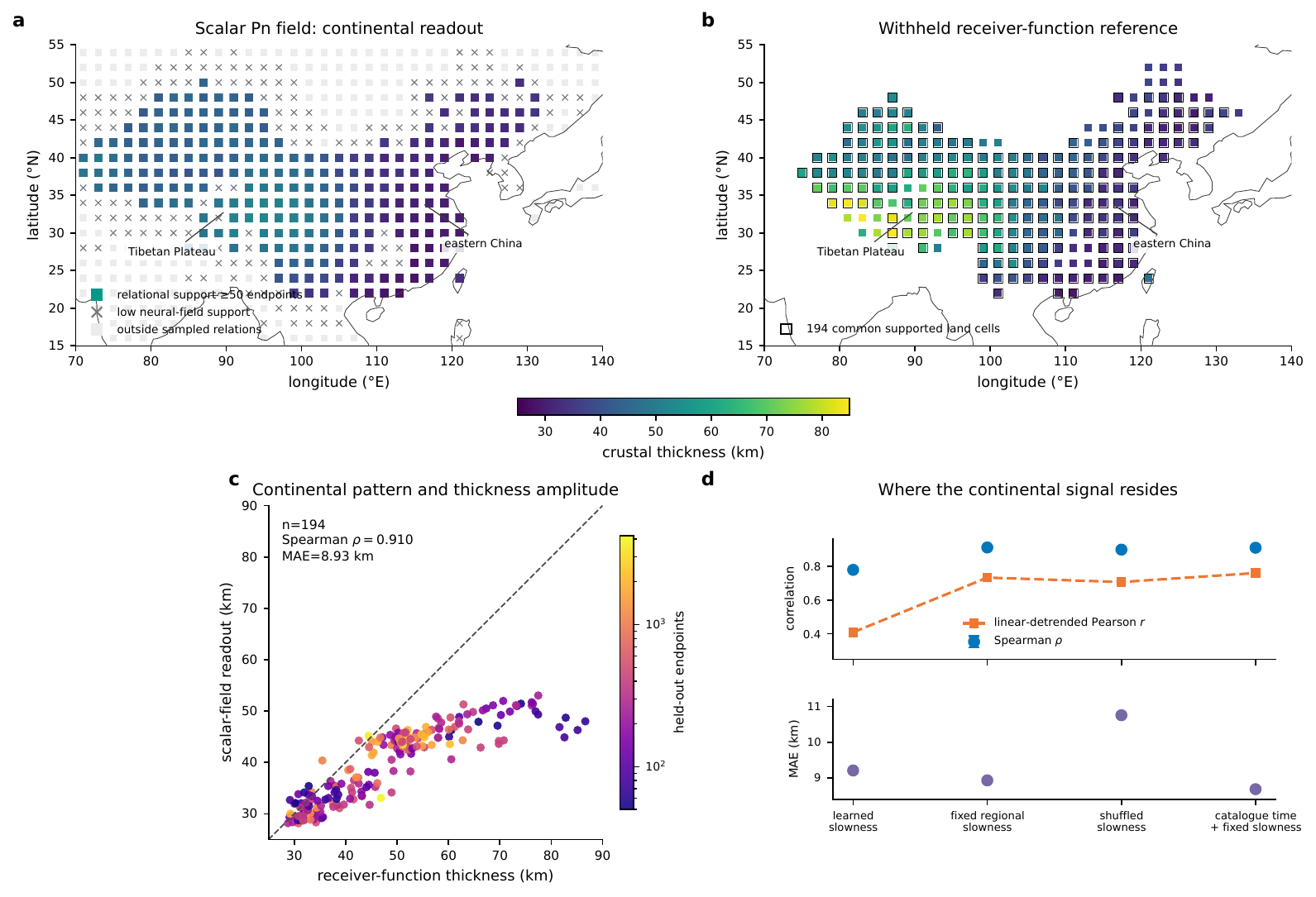}
  \caption{\textbf{Scalar Pn relations expose continental crustal thickness.} (a) The learned Pn field with training-arrival-derived regional slownesses, shown on a 2$\degree$ grid in cells with at least 50 qualified endpoints. (b) Withheld receiver-function thicknesses on their own observational support; outlines mark the 194 common cells. Both maps use the same colour scale. Geographic labels locate the broad contrast and do not delimit tectonic units. (c) Comparison on common cells, coloured by endpoint count ($\rho=0.910$, MAE 8.93~km); the dashed line is identity, exposing amplitude compression. (d) Three learned-time readouts using spatial, fixed and 100 shuffled slowness pairs, followed by the direct-catalogue-time control with fixed slownesses. Upper points show raw Spearman and linear-detrended Pearson correlations; lower points show MAE. Whiskers on shuffled results span their 5th--95th permutation percentiles. The reference enters neither neural training nor map fitting.}
  \label{fig:china_moho}
\end{figure}

\begin{samepage}
The component audit locates the continental signal primarily in scalar Pn values. Using regional slownesses fitted to training arrivals, the neural-field readout gives $\rho=0.910$ raw and 0.592 after quadratic longitude--latitude detrending (spatial-shift $p=0.020$) against 194 withheld receiver-function cells (Fig.~\ref{fig:china_moho}).
\par
\end{samepage}

Its MAE is 8.93~km, and the detrended correlation has a 10$\degree$ spatial-block 95\% interval of 0.345--0.750. Direct catalogue times with the same fixed slownesses give $\rho=0.909$ and 8.69~km MAE. The neural-field map recovers a Tibetan thickness contrast of 12.69~km, compared with 23.52~km in the reference. Thus the field retains the catalogue's continental pattern with compressed amplitude; these results do not establish superior Moho inference from the neural representation. Learned gradient magnitudes give $\rho=0.778$ and 9.21~km MAE; shuffled magnitudes give median values of $\rho=0.897$ and 10.75~km MAE. Spatially varying gradient magnitudes modify the readout but do not improve the external comparison. Event-correlated coordinate perturbations retain raw $\rho\geq0.900$ at both tested levels.

A separate source-geographic test withholds eight 4$\degree$ blocks and excludes training sources within 25~km of held-out sources. On all 25,222 nominal Pn test arrivals, the scalar ensemble reaches 0.762~s MAE versus 0.884~s for regional regression (paired spatial-block difference $-0.122$~s, 95\% interval $-0.291$ to $-0.048$~s). Arrival prediction transfers geographically; Moho rank agreement remains unresolved on the 14 supported cells inside withheld source blocks (Supplementary Information).

A matched three-seed reciprocity augmentation reverses 50\% of presentations while preserving phase and scalar time. The Pn reciprocity gap falls from $1.763\pm0.499$ to $0.094\pm0.006$~s and the reciprocal-gradient angle from $52.28\pm4.42\degree$ to $2.49\pm0.37\degree$. Original-orientation MAE changes from $0.928\pm0.007$ to $0.939\pm0.008$~s. The direct Moho map remains within seed variability, while symmetric two-end readout MAE falls from $13.40\pm1.68$ to $8.72\pm0.20$~km. Geometry emerges endpoint by endpoint from observed relations; supplied reciprocity symmetry connects the two roles.

The controlled medium allows directional queries to be tested against known propagation geometry. Identified gradients integrate into 40 support-qualified trajectories, with $5.77\degree$ median initial-direction difference and 4.70~km median separation from reciprocal-FMM paths. All 48 tested trajectories reach a 7.5-km receiver neighbourhood; eight leave the sampled source domain. Ray tracing was never a training task.

\begin{figure}[t]
  \centering
  \includegraphics[width=0.99\textwidth]{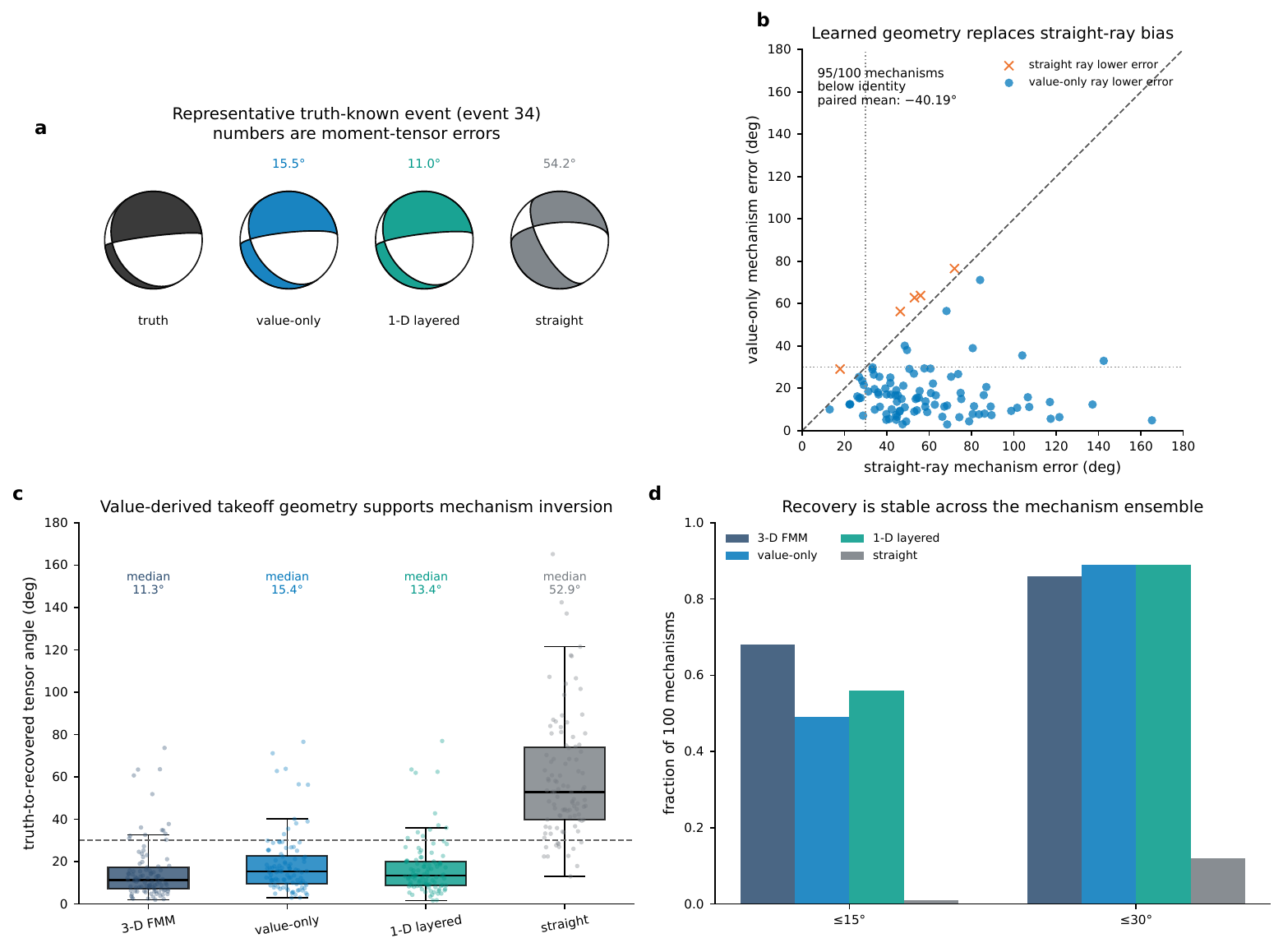}
  \caption{\textbf{Value-derived takeoff geometry recovers synthetic source mechanisms.} (a) Truth and recovered focal mechanisms for a representative held-out event; numbers give moment-tensor angular error. (b) All 100 paired value-only and straight-ray errors at 73 source locations, displayed over the full 0--180$\degree$ range. (c) Tensor-angle distributions using identical polarities and the same inversion. Boxes span the interquartile range with median lines; whiskers extend to the most extreme observations within 1.5 interquartile ranges of the box. Overlaid points show every mechanism. (d) Recovery fractions within $15\degree$ and $30\degree$. Angles use the normalized Frobenius inner product of recovered and true tensors (Methods). Mechanism parameters and polarities never enter travel-time training.}
  \label{fig:transfer}
\end{figure}

Across 100 held-out synthetic double-couple mechanisms at 73 source locations, learned takeoff geometry gives $15.38\degree$ median moment-tensor angular error, compared with $52.87\degree$ for straight rays, $13.38\degree$ for a layered 1-D model and $11.34\degree$ for the 3-D FMM oracle (Fig.~\ref{fig:transfer}). Recovery within $30\degree$ is 89\% versus 12\% for straight rays. The mean paired learned-minus-straight difference is $-40.19\degree$ (source-cluster 95\% interval $-47.82$ to $-34.15\degree$). The same scalar-trained field thus supplies paths and takeoff vectors for interpreting source radiation, without ray or mechanism labels in its training objective.

The same post-training physical queries also operate on classical scalar fields. On the same 100 mechanisms, scalar-selected MLS and local-polynomial slopes give median errors of $13.00\degree$ and $13.31\degree$; neither paired mean difference from the neural field is statistically resolved. Across all 48 paths, MLS has lower median path separation (2.10 versus 4.44~km), whereas the neural field reaches more receiver neighbourhoods (48 versus 42). Both representations supply ray and source geometry without task-specific training.

A separate historical California experiment, whose travel-time field used the mixed manual/automatic P cache, supplies an observational query using fixed waveform moment tensors and 447 manual P polarities from 16 events excluded from travel-time training. The primary event-mean mismatch is 0.158 for learned rays, 0.219 for MUSCAL 3-D, 0.145 for HASH-derived 1-D rays and 0.103 for the three-dimensional chord. Learned rays exceed the chord's mismatch by 0.055 (event-bootstrap 95\% interval 0.017--0.096), failing the prespecified non-inferiority rule. Requiring normalized radiation amplitudes of at least 0.05 under every compared ray reduces the difference to 0.024 ($-0.006$--0.061; Supplementary Information). These fixed-tensor sign predictions test observational transfer; controlled gradients and synthetic mechanisms provide the direct accuracy tests.

\subsection*{Regularized queries recover information beyond pointwise gradients}

We next test whether a frozen scalar field retains velocity information when its pointwise gradient is unreliable. This experiment trains new California fields using only explicitly manual P/S arrival picks; neither automatic picks nor PINN losses enter learning or extraction. The three-field ensemble predicts 12,914 P and 15,506 S arrivals from 940 held-out earthquakes with MAEs of 0.129 and 0.234~s. After training, we sample both 1,024 new earthquake locations and 24 new surface-station locations within training-supported geometry. The resulting 14,103 model-generated P/S pairs feed conventional ray-based tomography with a finite slowness grid and validation-selected regularization (Methods). The inverse model receives scalar values, without differentiating or updating the neural field.

\begin{figure}[p]
  \centering
  \includegraphics[width=0.96\textwidth]{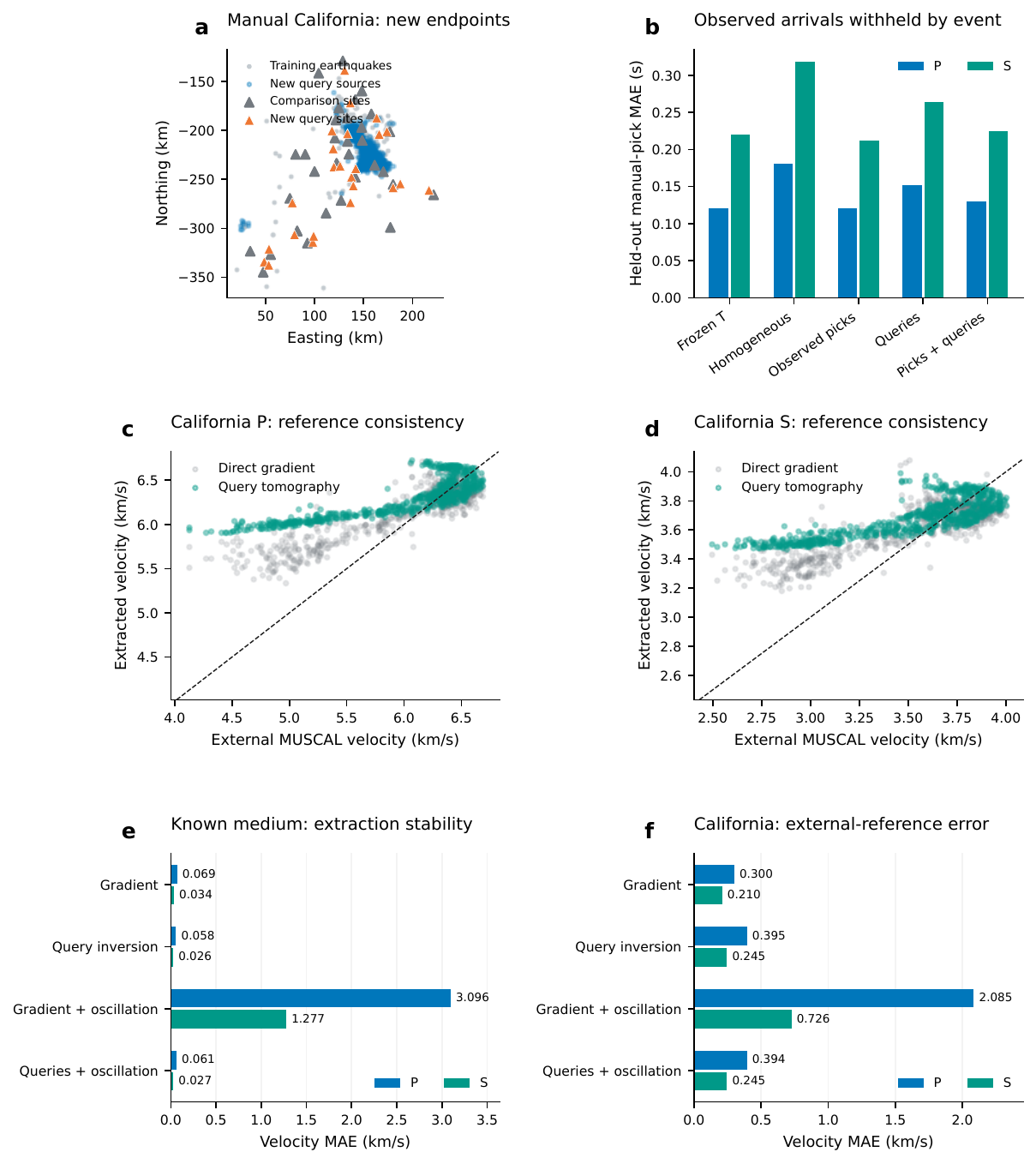}
  \caption{\textbf{Scalar information remains extractable when direct gradients fail.} (a) Manual-pick California training geometry and newly sampled source and surface-station locations. (b) Predictions of original regional manual arrivals withheld by event; the homogeneous comparator uses observed training picks. (c,d) Direct-gradient and generated-time tomography velocities against the external MUSCAL reference at held-out hypocentres. (e,f) Velocity MAEs before and after adding the same fixed 0.05-s oscillation to the frozen ensemble field. Conventional extraction uses generated scalar values, a finite grid and validation-selected regularization, without PINNs. Known-medium accuracy and California external-reference consistency are distinguished; the controlled readout recovers lateral structure, whereas the California result does not establish both lateral fields. All velocity readouts use the same phase-specific common event support within each dataset.}
  \label{fig:manual_tomography}
\end{figure}

On the regional test containing 740 earthquakes, the extracted velocity model predicts original manual picks with P/S MAEs of 0.152/0.264~s, compared with 0.181/0.320~s for its homogeneous initialization. Against the external MUSCAL reference, velocity MAEs are 0.395/0.245~km~s$^{-1}$ on 622/694 common P/S events, compared with 0.484/0.302~km~s$^{-1}$ for the homogeneous model. The paired reductions have 20-km-block 95\% intervals of 0.061--0.148 and 0.035--0.078~km~s$^{-1}$. The extracted P/S rank correlations are 0.839/0.817, and the independently fitted phase models retain $V_P>V_S$ throughout their common event support. This demonstrates a usable conventional velocity readout from a field trained exclusively on observed scalar arrivals (Fig.~\ref{fig:manual_tomography}).

The field comparison also limits the claim. Unperturbed direct gradients have lower MUSCAL discrepancies, 0.300/0.210~km~s$^{-1}$. Removing depth trends leaves query-model correlations of 0.231 for P and $-0.174$ for S, with strongly attenuated lateral amplitudes. Thus the California result establishes broad velocity organization and prediction of held-out manual arrivals, rather than independently resolving both lateral velocity fields. MUSCAL is a reference model, not true-Earth velocity. The fixed oscillation nevertheless distinguishes the extraction methods: direct-gradient discrepancies rise to 2.085/0.726~km~s$^{-1}$, while conventional extraction remains at 0.394/0.245~km~s$^{-1}$. All candidate readouts, including the observed-pick and combined inversions, are reported in the Supplementary Information.

The same readout on the previously blinded known medium tests whether this procedure retrieves metric structure. On 306 common held-out sources, generated-time tomography gives P/S velocity MAEs of 0.0579/0.0255~km~s$^{-1}$ and depth-detrended rank correlations of 0.844/0.922. The direct gradient of the same ensemble field gives 0.0688/0.0335~km~s$^{-1}$ and 0.759/0.813. The paired P/S MAE improvements are 0.0109/0.0080~km~s$^{-1}$, with 20-km-block 95\% intervals of 0.0031--0.0191 and 0.0034--0.0130. Homogeneous models fitted to the generated values give 0.1995/0.1108~km~s$^{-1}$, so the improvement includes spatial structure. Conventional inversion of the original regional observations remains more accurate, at 0.0285/0.0167~km~s$^{-1}$; the query experiment tests information retained by $T$, without treating generated pairs as additional independent evidence.

The fixed 0.05-s oscillation provides a direct test of extraction stability (Fig.~\ref{fig:manual_tomography}). It changes the ensemble's P/S scalar MAEs from 0.060/0.088 to 0.068/0.094~s, but raises direct-gradient velocity MAEs to 3.096/1.277~km~s$^{-1}$ and removes anomaly correlation. Query-only tomography of this same perturbed field still gives 0.0607/0.0268~km~s$^{-1}$ and depth-detrended correlations of 0.815/0.905. Removing the explicit spatial penalty while retaining weak damping and the finite grid gives 0.1154/0.0513~km~s$^{-1}$. Thus the restricted representation already stabilizes extraction, and validation-selected regularization improves it further. A failed pointwise derivative readout need not mean that recoverable velocity information has disappeared from the scalar field. This result concerns the specified oscillatory perturbation; it does not establish recovery under every picking or location-error process.

\subsection*{Phase-sensitive quantum fields support geometric readouts}

A synthetic quantum test asks whether the value-to-geometry route extends beyond travel times to fields governed by the dispersive Schr\"odinger equation. Separate networks learn a Gaussian packet's real phase or the real and imaginary components of two-dimensional vortex and interfering wavefunctions. Only field values supervise training; potentials, currents, velocities, trajectories and equation residuals do not. For a spinless particle without a vector potential, the known post-training readout is~\citep{Bohm1952,StruyveValentini2009}
\begin{equation}
\vect{j}=\frac{\hbar}{m}\operatorname{Im}(\psi^*\nabla\psi),
\qquad
\vect{v}_{\mathrm B}=\frac{\vect{j}}{|\psi|^2}
=\frac{\nabla S}{m},\qquad \psi=R e^{iS/\hbar}.
\label{eq:quantum_readout}
\end{equation}
Thus a complex scalar carries phase information through two real output channels; density alone does not supply this readout.

\begin{figure}[t]
  \centering
  \includegraphics[width=0.99\textwidth]{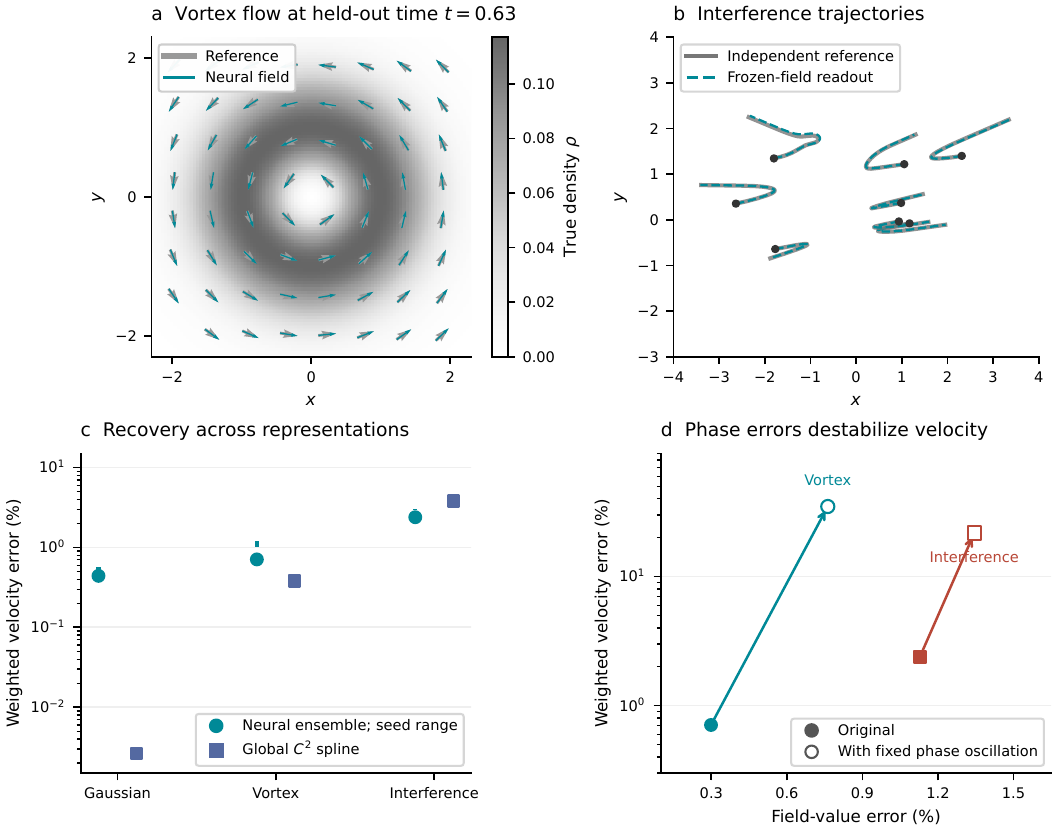}
  \caption{\textbf{Phase-bearing quantum field values support physical readouts and expose their limits.} (a) Vortex velocity directions at $t=0.63$, inside the time interval excluded from training and validation. Grey and teal arrows show unit reference and neural-ensemble directions; the background is true density. Arrows are displayed where $\rho\geq10^{-3}$. (b) The first eight of 32 prescribed interference trajectories, with initial points marked; all 32 enter scoring. (c) Probability-weighted velocity errors for the neural mean field and global $C^2$ spline. Lines span the three individual optimization seeds, not a confidence interval. (d) Fixed phase oscillations move the vortex and interference readouts from filled to open symbols while preserving their predicted densities. Panels (c,d) use the common true-density support $\rho\geq10^{-4}$; value errors use all 8,192 test points. Units are $\hbar=m=1$.}
  \label{fig:quantum_readouts}
\end{figure}

On 8,192 unseen coordinates, three-seed mean fields recover vortex and interference currents with relative $L_2$ errors of 0.585\% and 2.481\%, respectively. Probability-weighted velocity errors are 0.708\% and 2.387\% on the fixed true-density support $\rho\geq10^{-4}$ (Fig.~\ref{fig:quantum_readouts}). Independently referenced trajectories complete all 32 prescribed initial conditions in each case, with median endpoint errors of 0.00307 and 0.01888 in dimensionless units. These are within-state interpolation tests: both training and validation exclude a spatial block and a time interval. The interference velocity error rises to 6.595\% in the dedicated time holdout. A global $C^2$ spline fitted to the same values also recovers the readouts, outperforming the neural field for the Gaussian and vortex and giving comparable interference-current error (2.488\%). The comparison supports recovery across representations, with no general neural advantage.

The same distinction between scalar fidelity and derivative reliability reappears. Multiplying the frozen complex field by $e^{i\phi}$, with $\phi=0.01\sin(40x+29y)$, preserves its density exactly. Vortex field-value error increases only from 0.299\% to 0.762\%, while velocity error rises from 0.708\% to 34.908\%. This intervention probes readout sensitivity; the altered field is not asserted to satisfy the original Hamiltonian dynamics. Information content imposes a separate limit: opposite-circulation harmonic-oscillator states have identical densities at every time and opposite currents~\citep{Gale1968}. Even an exact density fit cannot distinguish them. The quantum test therefore extends the paper's central mechanism to phase-sensitive fields under another evolution equation, while retaining its two conditions: the observations must contain the queried information, and the learned representation must support stable differentiation. It does not test new-state generalization or discovery of the readout law.

\section*{Discussion}

Scalar travel-time relations can identify propagation geometry without equation supervision during learning. Physical interpretation uses the high-frequency isotropic eikonal relation after training. In controlled media, scalar-trained fields recover propagation directions and local velocities, and support ray and source-mechanism queries without task-specific supervision. Recovery through neural fields and classical moving least squares makes this observational route accessible through distinct representations. The scalar-matched control and coverage interventions distinguish accurate value fitting from reliable geometric identification. A continuous travel-time field therefore provides a reusable observational representation: one learned relation supports several physical queries without task-specific retraining.

Equation~\ref{eq:gradient_error_bound} gives the identification mechanism. A spanning set of nearby value differences constrains the gradient, while the weakest sampled direction amplifies scalar and curvature errors. The coverage experiments connect this classical differentiation result to learned propagation fields. For seismic acquisition, this makes the geometry of sampling a target alongside the number of arrival picks: observations should provide directional support around the coordinates at which physical derivatives will be queried.

Across two independently generated full-support media, including the test-sealed realization, the value-only field supports local-velocity recovery through a post-training eikonal readout, despite receiving no equation supervision. Scalar secants provide resolved refinements; the tested eikonal term has no statistically resolved advantage over values alone. Its contribution grows as relational coverage is removed. In this setting, the eikonal relation supplies the physical readout and can add training constraints when scalar coverage is insufficient.

The China component comparison shows where structural information resides within this representation. Continental Moho variation is carried primarily by scalar Pn values, with regional slownesses setting the thickness scale. The direct-catalogue control links that pattern to the observations themselves. In controlled media, local velocities and ray directions are read from spatial derivatives. Values and derivatives therefore serve different physical queries: the large-scale thickness signal need not depend on the spatially varying gradient magnitudes that encode local metric structure in the controlled tests. The contribution is to organize these relations into a field whose physical use can be specified after training.

Random endpoint queries provide a second route to velocity: a conventional regularized inverse problem can extract metric information from scalar values even when pointwise differentiation fails. A small high-frequency value perturbation has a derivative proportional to its frequency, whereas the conventional inverse map filters scalar discrepancies through path consistency and its regularizer. The known-medium intervention establishes this distinction without a PINN. Manual-only California arrivals extend the workflow to observed data, with successful broad-scale extraction but unresolved lateral fidelity, especially for S waves. Sampling additional pairs interrogates information and inductive biases already in the frozen field; it does not create independent observations or guarantee a unique medium. The classical propagation model and regularizer are explicit assumptions of the readout, rather than hidden supervision of the neural field.

Follow-up diagnostics identify which part of this failure is regularizable (Supplementary Information, Section~23.3). Averaging source gradients over local three-dimensional queries at each fixed surface receiver largely removes the imposed oscillatory error, but does not recover the missing California lateral structure. A training penalty on neighbouring gradient magnitudes, without velocity targets or equation residuals, improves known-medium P/S velocity MAEs from 0.0919/0.0432 to 0.0688/0.0275~km~s$^{-1}$ relative to an equal-budget unregularized continuation. Its observational benefit remains unresolved, and restricting epicentral distance to 10--50~km does not itself improve recovery. These post-test diagnostics distinguish derivative instability that smoothness can reduce from metric ambiguity that scalar accuracy alone does not resolve.

The quantum extension isolates the scope of this argument. Phase-bearing scalar fields governed by Schr\"odinger evolution support probability-flow and trajectory readouts without supervising those quantities. Density-only observations have an exact circulation ambiguity, and phase oscillations expose derivative sensitivity even at unchanged density. Classical splines also recover these readouts. The shared result is therefore conditional identification from informative, sufficiently regular field values; the experiment does not establish a universal learned dynamical model or a new physical law.

The observational tests separate training information from physical evaluation. Receiver-function thicknesses, manual P polarities and fixed waveform moment tensors are withheld from neural fitting and selection. Catalogue locations and origin times retain propagation-model assumptions, and waveform tensors depend on model-based Green's functions. Orthogonality here concerns the withheld observables, not complete independence of upstream models. The China comparison tests continental structure encoded in catalogue relations; the historical mixed-pick California comparison tests radiation signs for fixed source mechanisms, while the new manual-only experiment tests scalar-to-velocity extraction. MUSCAL agreement is external model consistency and can share upstream data assumptions. Together with controlled recovery and catalogue-coordinate perturbations, these queries delineate transfer to distinct physical observables.

Deeper source coverage is the direct route to identifying deeper ray segments. Extension to discontinuities and anisotropy will require branch-aware representations and the appropriate relation between slowness and ray direction.

\begin{samepage}
Physical geometry need not always be prescribed to a learner through velocity inputs or equation supervision. Informative scalar relations can constrain its derivatives under adequate sampling and smoothness. Recovery across neural and classical representations links the queried geometry to those relations, with sampling and smoothness constraining the identifiable degrees of freedom. Equations can then validate, read and refine that geometry. The quantum companion test supports the same conditional route for phase-bearing fields: value observations can constrain unsupervised geometric readouts across these distinct physical settings, provided the required information and derivative regularity are present.

\par
\end{samepage}

\section*{Methods}

\subsection*{Coordinates, data and locked partitions}

WGS84 coordinates are projected with local azimuthal-equidistant systems using PROJ \citep{PROJ}. The California centre is $(-119.242055\degree,37.722624\degree)$ and the China centre is $(105\degree,35\degree)$. Coordinates use kilometres, with $x$ east, $y$ north and $z$ positive down. Travel times use seconds and derivatives have units s~km$^{-1}$. Affine normalization is internal to the networks. For normalized input $\widetilde{x}_j=(x_j-\mu_j)/a_j$ and physical output $T=b f_\theta$, the chain rule gives $\partial T/\partial x_j=(b/a_j)\partial f_\theta/\partial\widetilde{x}_j$. Physical gradients therefore include both coordinate and output scales.

Observed targets are from SeismicX-Cont v1.0.0~\citep{SeismicXCont}, derived from CEED and SCEDC/NCEDC archives~\citep{ZhuCEED2025,SCEDC,NCEDC,SCSN,NCSN,BDSN}. Travel time is arrival time minus catalogue origin time. Training therefore uses no velocity field, traced ray or velocity-model-computed travel-time target. The released hypocentres and origin times remain upstream products of catalogue location workflows that use propagation models. The historical mixed-pick cache used for the fixed-mechanism and earlier annotation diagnostics contains 7,519 events and 172,534 event--station records. Its P targets include 102,204 manual and 50,904 automatic picks; all 121,630 selected S targets are manual. These arrival-time targets are distinct from the manual P first-motion labels used only for mechanism evaluation. For the external-mechanism experiment, all 26 exact moment-tensor associations are assigned to test. The remaining events are deterministically divided into 6,369 training and 1,124 validation events, containing 143,657 and 26,467 records. The final test contains 2,410 records. All overlaps are zero. Checkpoints, physics weights, quality rules and the identifiability descriptor are fixed without test-performance outcomes.

The China experiment uses the unified China Earthquake Network catalogue and phase archive for 2009--2022 \citep{CENCUnifiedCatalogue}. Sources are restricted to 70--140$\degree$E, 15--55$\degree$N and 0--80~km depth. Receiver coordinates are reconstructed with WGS84 forward geodesics from each reported distance and azimuth. Records require $0<T\leq300$~s, 1--1,500~km distance, finite azimuth, and broad apparent-velocity ranges of 2--12~km~s$^{-1}$ for P or 1--8~km~s$^{-1}$ for S. All Pn and Sn values and a deterministic 15\% hash sample of Pg and Sg values are retained, giving 1,001,527 Pg, 990,955 Sg, 673,362 Pn and 237,718 Sn records. Event hashes define 80/10/10 train/validation/test partitions with zero event overlap. The raw archive contains 1,361,054 event headers. The in-domain cache contains 791,404 unique events.

\subsection*{Value-only fields and training}

Each controlled medium, the China catalogue and the California catalogue has its own trained field. The physical queries reuse the relevant field's selected weights; there is no joint training or weight transfer between these datasets. Coordinate normalization, phase identity and a smooth function class are inductive biases shared by the value-only approach. Reciprocity-augmented fields additionally receive an explicit endpoint-exchange symmetry and are distinguished from the unaugmented primary field.

The primary evidence field is a 256-unit, four-block residual multilayer perceptron (MLP) with sigmoid linear unit (SiLU) activations. Its input is the concatenation of normalized receiver and source coordinates and its two outputs are unbounded P and S times. It contains no coordinate differences, midpoint, separation, reference travel time, P/S ordering constraint, velocity input or equation term. A fixed 80-s output multiplier is an algebraically absorbable numerical parameterization of the final linear layer, not a bound or reference solution. The phase-balanced Huber objective contains only travel-time values. AdamW \citep{LoshchilovHutter2019} uses learning rate $5\times10^{-4}$, weight decay $10^{-5}$, cosine decay and gradient clipping at 5. Full dense-medium runs use 60 epochs. Coverage experiments scale epoch count inversely with retained fraction to keep the number of optimizer updates approximately fixed. PyTorch automatic differentiation provides all post-training gradients \citep{Paszke2019}.

The synthetic reciprocal-role intervention uses the same primary medium, source-disjoint split, minimal-coordinate architecture, scalar objective, 60-epoch budget and seeds 101--105. The control retains the observed station--event ordering. In the matched arm, a deterministic 50\% of presentations are reversed while retaining the identical P/S scalar value; the batch size and number of updates do not change. Evaluation is paired by held-out row and compares original and exchanged predictions, endpoint gradients and independent reciprocal-FMM gradients. Twenty thousand paired bootstrap replicates resample all 81 paths of each of 85 held-out source groups together. The intervention was added after the seed-101 endpoint asymmetry was observed and is identified as a fixed post-test causal confirmation, not as prospective evidence or a minimal-prior model.

The China field has five 256-unit residual-SiLU blocks and four unconstrained outputs ordered Pg, Sg, Pn and Sn. It receives only the six AEQD source and receiver coordinates. The $x,y$ coordinates are internally scaled by 1,000~km and $z$ by 50~km. Thirty AdamW epochs use learning rate $2\times10^{-3}$ with cosine decay, weight decay $10^{-6}$ and phase-balanced smooth-Huber batches. Source-depth bins are balanced inside each phase so that the depth coordinate is not numerically erased by the abundance of shallow earthquakes. This changes sampling frequency, not target content: every loss term remains a scalar catalogue travel time. Checkpoint selection uses a deterministic depth-balanced validation set. The test partition is evaluated once at its natural depth distribution.

The reciprocal-role ablation uses the same cleaned catalogue, event split, phase/depth-balanced sampler, architecture, 30-epoch budget and seeds 20260807--20260809. Both matched groups use one shared endpoint centre and scale for the two coordinate slots. The control preserves the observed $(\vect{x}_r,\vect{x}_s)$ ordering. In the exchange group, a deterministic 50\% of every training batch is presented as $(\vect{x}_s,\vect{x}_r)$ with its original phase and scalar time; exchange replaces orientation and does not double the batch. Validation and test scalar errors are evaluated only in the original catalogue ordering. Reciprocal diagnostics compare $T_\theta(\vect{x}_r,\vect{x}_s)$ with $T_\theta(\vect{x}_s,\vect{x}_r)$ and compare $\nabla_{\vect{x}_s}T_\theta(\vect{x}_r,\vect{x}_s)$ with $\nabla_{\vect{x}_r}T_\theta(\vect{x}_s,\vect{x}_r)$. Because exchange explicitly supplies travel-time symmetry, it is reported as reciprocity-informed augmentation rather than as the minimal-prior field.

A secondary structured field contains source--receiver differences, midpoint, separation, homogeneous P and S--P references and a smooth causal head. It is retained only as an architectural reference and is not the source of the primary identifiability or external-mechanism evidence. Equal-parameter ablations remove each component. A coordinate-only tanh MLP and a fixed-Fourier-feature tanh MLP are parameter matched at approximately 0.53 million parameters and receive identical 40-epoch budgets. The Fourier matrix is fixed from the prespecified seed with 32 frequencies per six-dimensional coordinate pair.

To test whether matched scalar accuracy alone certifies a derivative, we construct a fixed differentiable control from each selected primary field,
\begin{equation}
T_{\mathrm{control}}=T_{\theta}+0.05\sin\!\left(64\pi\vect{d}^{\mathsf T}\widetilde{\vect{x}}_s\right),\qquad
\vect{d}=\frac{(1,\sqrt{2},\sqrt{3})}{\|(1,\sqrt{2},\sqrt{3})\|},
\end{equation}
where $\widetilde{\vect{x}}_s$ is the normalized source coordinate. The value perturbation is bounded by 0.05~s, while its analytically known derivative is not small. Its amplitude, frequency and direction are fixed by construction and are not selected from test gradients.

\subsection*{Classical scalar-only controls}

Gaussian local-polynomial fits use the identical training sources, validation sources and test rows as the primary neural field~\citep{LancasterSalkauskas1981,Levin1998,Mirzaei2012}. Fits are independent for each observed receiver and phase, with a three-dimensional source-coordinate basis. Degrees 1--3, horizontal bandwidths 20/40/80~km, vertical bandwidths 10/20/40~km and non-intercept ridge penalties 0/$10^{-6}$ define 54 candidates. All training sources enter each Gaussian sum. Candidates with any validation Gram condition above $10^{12}$ are rejected. Validation P/S MAE selects parameters within each degree and the overall order before classical test evaluation. Both the fitted polynomial slope and the full derivative of the moving scalar estimate are reported, including weight and basis derivatives in the latter. Finite differences verify the implementation. Three new layered-medium neural seeds use the unchanged minimal-coordinate architecture and 60-epoch budget; heterogeneous runs are reused. Tests retain all 6,885 rows and the same 48 paths and 100 mechanisms. These are new controls on existing benchmarks, not new blinded realizations. Full settings, paired source-bootstrap intervals and all polynomial orders are in the Supplementary Information.

\subsection*{Local identifiability analysis}

For each evaluation path, source tensors hold receiver identity fixed and use training sources observed at that station. Receiver tensors use retained station coordinates. Gaussian weights have bandwidth half the $k$th-neighbour distance for $k\in\{8,16,32,64\}$. Candidate descriptors include fill and nearest distances, condition number, effective rank, directional isotropy, normalized determinant, depth support and boundary distance. Candidates are compared across 17 validation-only random, structured and station-thinning conditions using within-condition rank association with independent validation-FMM angle and vector errors. Descriptor identity is frozen before test evaluation.

Ten independent source subsets are evaluated at 25, 10 and 5\% coverage using one optimization seed. The original subset at each level is also run with five optimization seeds. Structured screens include shallow depth, restricted extent, azimuth wedge, spatial cluster, depth gap and boundary-biased selection. Subset identifiers, selected groups, predictions and local descriptors are retained in machine-readable files.

\subsection*{Controlled media and FMM evaluation}

The original MUSCAL comparison dataset contains 559 source groups and 801,459 P/S records on a 2-km grid \citep{Doody2023,YehBenZion2026}. Its locked split contains 397/92/70 train/validation/test sources. Independent reciprocal P fields are solved from stations by fast marching \citep{Sethian1996}, differentiated on the grid and sampled at held-out sources.

The primary controlled medium occupies $[-200,200]^2\times[0,80]$~km on a 5-km grid. It contains 567 sources on a $40\times40\times10$~km lattice spanning depths 0--60~km and 81 surface stations. The source-disjoint split contains 397/85/85 train/validation/test sources. P velocity is $4.6+0.03z$~km~s$^{-1}$ plus smooth sinusoidal and Gaussian anomalies with an approximately 50-km characteristic scale. Source-centred FMM supplies scalar P times. Reciprocal station-centred FMM independently supplies test gradients. S times are 1.75 times P times. A source-aligned layered medium and the original 147-source benchmarks provide transfer controls.

For learned gradient $\vect{g}$ and FMM gradient $\vect{g}^{*}$,
\begin{align}
\theta&=\cos^{-1}\frac{\vect{g}\cdot\vect{g}^{*}}
{\|\vect{g}\|\|\vect{g}^{*}\|},\\
e_{\mathrm{vec}}&=\frac{\|\vect{g}-\vect{g}^{*}\|}{\|\vect{g}^{*}\|},\qquad
e_{\mathrm{eik}}=|v_P\|\vect{g}\|-1|.
\end{align}
Joint adequacy requires combined MAE $\leq1.0$~s, median $\theta\leq10\degree$ and median $e_{\mathrm{vec}}\leq0.25$.

\subsection*{Manual-arrival fields and conventional scalar tomography}

The California velocity readout uses newly trained fields and an exclusively manual-pick cache. A pick must have the explicit annotation status \texttt{manual} before duplicate selection; automatic picks are never used as a fallback. Highest score and then earliest valid arrival resolve duplicate manual records, with $0<T\leq120$~s. The cache contains 121,630 event--station records, with 102,204 P and 121,630 S targets. The existing event-identifier partition gives 5,423/1,156/940 training/validation/test earthquakes, 87,697/18,427/15,506 records, and 73,777/15,513/12,914 P targets. All S targets are present. Three minimal-coordinate residual-SiLU fields use seeds 101--103, an initial 100 epochs and batches of 8,192, with the scalar objective and optimizer specified above. Because the manual-only validation curves had not stabilized, a further 200 epochs start from each selected checkpoint at learning rate $5\times10^{-5}$, with cosine decay to $2.5\times10^{-6}$. This refinement was fixed from validation behaviour before final-test access; it adds no targets or equation terms. Checkpoints minimize original validation-time MAE. No velocity label, derivative target, equation loss or PINN enters this new experiment. Input longitude, latitude, earthquake depth and station elevation are converted to the six physical-kilometre coordinates; the network is then frozen.

Random endpoint queries are restricted to support defined by training geometry. Sources lie at 2--20~km depth and pairs at 20--70~km epicentral distance inside $[20,280]\times[-380,-120]$~km. A deterministic, count-weighted farthest-site procedure selects up to 32 actual stations for the regional observational comparison. Twenty-four new surface locations are sampled inside the training-station convex hull, within 12~km of an observed site and at least 5~km apart; their elevations use inverse-distance interpolation from the three nearest stations. The 1,024 new source locations perturb training hypocentres uniformly by up to 5~km horizontally and 1.5~km vertically. The fourth-nearest training source must lie within scaled distance 1.5 using 10/10/3-km coordinate scales. The mean of the three frozen fields supplies 14,103 P/S query pairs in California. These are model-generated values, not additional observations. An identically specified readout of the previously blinded, value-only controlled-medium fields supplies 4,768 query pairs. No PINN-trained checkpoint is used in either readout.

Conventional isotropic first-arrival tomography fits a positive slowness field $m=m_0(1+u)$, represented by trilinear nodes at $20\times20\times5$~km spacing. Dynamic shortest-path ray tracing~\citep{Giroux2021} uses 5-km cells, two secondary and two tertiary nodes per edge, and path-integrated travel times. The grid extends to $[0,300]\times[-400,-100]\times[-10,50]$~km. For each phase, the objective is
\begin{equation}
\sum_i w_i\rho_{0.2}\!\left(F_i[m]+b-t_i\right)
+\frac{\lambda}{2N_D}\|Du\|_2^2
+\frac{0.001+0.01\lambda}{2N_m}\|u\|_2^2
+\frac{10^{-4}}{2}b^2,
\label{eq:scalar_tomography}
\end{equation}
where $F_i$ is the classical forward time at fixed source and receiver positions, $D$ takes equally weighted differences between adjacent grid nodes, $\rho_{0.2}$ is the Huber loss with a 0.2-s transition, and $b$ is a phase-wide intercept. Four arms fit original training picks, generated queries, an equal-weight combination, or queries with the fixed 0.05-s source oscillation. Each source receives equal total weight within an arm; the combined arm assigns half its weight to each source of values. Each arm estimates its own initial $m_0,b$ by robust distance--time regression on its input values. Query-only initialization therefore receives no original travel-time targets. A maximum of five robust Gauss--Newton updates uses preconditioned least squares, a shared step limited to 0.2 in relative slowness, and backtracking on the training objective; $u\in[-0.6,1.2]$ and $b\in[-3,3]$~s. Grouping ray origins explicitly aligns the solver's travel times with sensitivity-matrix rows; homogeneous and heterogeneous path-integral closure checks pass at floating-point precision.

The iteration (including initialization) and $\lambda\in\{0,0.1,1,10\}$ minimize original validation-arrival MAE, with all readouts for each dataset frozen before its final test. The manual experiment retains its locked settings and automatic validation-only selection while the completed controlled experiment is evaluated. The zero-$\lambda$ control removes explicit spatial roughness but retains weak amplitude damping, the finite grid and bounds. Homogeneous initial models are also reported. The oscillatory intervention perturbs the already trained scalar field; it is distinct from retraining on random measurement noise or perturbed hypocentres. Neither inversion nor selection uses neural gradients or an external velocity model. Classical propagation, smoothness and bounded slowness enter only the post-training extraction step.

Final velocity comparisons use held-out events with at least three finite phase paths, taking the inverse norm of the three-seed mean gradient and then its median over receivers for the direct readout. Thus values and gradients belong to exactly the same ensemble field. Conventional models are interpolated at the same hypocentres. The controlled reference is known synthetic truth; California uses MUSCAL as an external consistency reference, not a measurement of true Earth velocity. Cubic depth trends are removed separately before anomaly rank correlation. Paired 95\% intervals use 2,000 event or 20-km horizontal-block bootstrap replicates, conditional on the fitted teachers, geometry and reference. This is a new readout on previously studied event partitions, not a new independently blinded benchmark. Synthetic query counts do not enter the independent sample count.

\subsection*{Relational Pg--Pn crustal-thickness readout}

The readout uses a regional isotropic head-wave approximation with crustal and refractor slownesses and source- and receiver-side thickness terms. It is a physical interpretation applied after scalar training, not an equation-free thickness inversion. Pn branch mixing, interface dip, lateral velocity variation and catalogue origin-time errors can affect its absolute thickness scale. The 2$\degree$ grid specifies the reported map spacing rather than an independently measured resolving length. Gradient magnitudes in the China comparison are computed in the AEQD chart; the direct local-velocity accuracy test uses Cartesian controlled media.

Competing regional branches are fitted on training data only. Pg uses robust direct-path slowness over 10--160~km. Pn uses a robust distance intercept with the source-depth coefficient updated from the fitted Pg and Pn slownesses. A nominal Pn record is removed when the Pg likelihood exceeds the Pn likelihood by a factor of ten or when its Pn residual exceeds four robust scales. Neither rule has access to a crustal-thickness reference.

For every retained held-out Pn coordinate, automatic differentiation supplies Pg and Pn source gradients. We compute Eq.~\ref{eq:relational_slowness} and retain finite relations with $p_c\in[0.14,0.22]$, $p_m\in[0.08,0.18]$ and $\eta\in[0.04,0.18]$~s~km$^{-1}$ and pair thickness 30--160~km. Autodiff gradient directions agree with 0.5-km centred differences to $0.028\degree$ at the 95th percentile. Each accepted relation contributes one linear equation for the source and receiver cells in Eq.~\ref{eq:relational_moho}. Robust sparse least squares on a 2$\degree$ grid selects a nearest-neighbour smoothing weight from $\{0,0.01,0.1,1,10\}$ by validation travel-time MAE; a cell requires at least 50 held-out ray endpoints. No external thickness enters fitting or support selection.

A fixed component audit uses the same 62,086 observations, derivative-qualified support mask, grid and smoothing weight for every readout. It compares learned Pn time with learned Pg/Pn magnitudes, fixed regional slownesses, and 100 deterministic joint permutations of learned magnitude pairs. The fixed values, $p_c=0.16909597$ and $p_m=0.12442703$~s~km$^{-1}$, are the reciprocals of the crustal and mantle velocities estimated by robust Pg/Pn moveout fits on training arrivals. They are not medians of neural gradients and are not obtained from an external velocity model or thickness reference. Additional controls use catalogue Pn time with the same fixed slownesses and a global moveout null. The fixed-slowness readout is highlighted following this component comparison; no neural model is retrained. Spatial inference is applied to the frozen maps without selecting parameters. A quadratic surface in longitude and latitude is fitted separately to prediction and reference and removed before rank correlation. Spatial-block bootstraps resample all cells inside 6$\degree$ or 10$\degree$ blocks for 10,000 replicates. A map-preserving null evaluates every non-zero toroidal translation for which at least 100 prediction--reference cells overlap; the prediction and its support mask are shifted together. These resampling results are conditional on the observed domain and support. The Tibetan contrast is the median thickness in 78--102$\degree$E, 27--38$\degree$N minus the median outside that window, evaluated on the same 194 common cells.

The final withheld reference is the 0.5$\degree$ receiver-function grid of He et al.~\citep{HeEtAl2014}, which combines a uniform analysis of 798 stations with earlier receiver-function estimates. Its reported sea-level Moho depth is converted to thickness below the local surface using CRUST1.0 topography; CRUST1.0 contributes only this datum conversion in the final test. The He et al. supplement was downloaded after the neural checkpoint, gradient qualification and original gradient-based map were frozen. CRUST1.0 itself was used earlier as a development comparison and is therefore reported as secondary consistency evidence rather than the final independent endpoint~\citep{LaskeEtAl2013}. Land cells are defined from Natural Earth physical polygons~\citep{NaturalEarth}.

\subsection*{Source-geographic holdout}

A new China experiment replaces the random event split with 4$\degree$ source blocks anchored at 70$\degree$E, 15$\degree$N. Deterministic block hashing assigns test, validation and training before fitting. All rows of four identifiers with inconsistent source coordinates (13 rows) are excluded; training sources within 25~km horizontally of any validation or test source are buffered out. The retained partitions contain 517,388/235,383/14,109 training/validation/test events. Three fields use the original architecture and 30-epoch budget, seeds 101--103, and no endpoint exchange or equation loss. Branch fits, normalization and the regional polynomial baseline use new training rows only. All three validation-selected checkpoints are hashed before test scoring. Primary MAEs include every existing-cache test arrival, irrespective of its held-out branch residual; branch-screened results are secondary. Paired intervals resample whole source blocks 2,000 times, conditional on the three trained seeds.

The separate thickness query combines 24,789 branch-qualified Pn test pairs at distances $\geq100$~km using fixed training-derived slownesses and smoothing weight 0.1. Its mask requires 200 incident endpoints per cell and uses no learned-gradient qualification. All neural, regional-polynomial and direct-catalogue maps are locked before the existing receiver-function files are read. Results distinguish all 31 common endpoint cells from the 14 within withheld source blocks; 2,000 bootstrap replicates resample 10$\degree$ cell blocks. This is a new source-geographic partition of previously studied observations, not a new catalogue or a claim that receivers and propagation paths avoid the same subsurface regions. Full settings, per-seed results and both geographic masks are provided in the Supplementary Information.

\subsection*{Equation-necessity test}

For phase $\alpha\in\{P,S\}$ and endpoint $e\in\{r,s\}$,
\begin{equation}
\mathcal{L}_{\mathrm{eik}}=\frac14\sum_{\alpha,e}
\operatorname{mean}\left[(v_\alpha(\vect{x}_e)
\|\nabla_{\vect{x}_e}T_\alpha\|-1)^2\right].
\label{eq:eikonal_loss}
\end{equation}
The total loss is $\mathcal{L}_{\mathrm{data}}+\lambda\mathcal{L}_{\mathrm{eik}}$, with the physics term evaluated on 25\% of each batch. At every coverage, $\lambda\in\{0.03,0.1,0.3,1,3,10,30,100\}$ is selected by validation MAE before five-seed final evaluation. Reciprocal-FMM gradients enter neither training nor selection.

\subsection*{Independent same-data metric confirmations}

The confirmation protocol, random seeds and success rule were hashed before the new medium was generated. The medium spans $[0,300]\times[-400,-100]\times[0,60]$~km on a 2-km grid. Its $V_\mathrm{P}$ field combines the depth trend $5.15+0.038z$~km~s$^{-1}$ with 12 seeded three-dimensional Gaussian anomalies and two long-wavelength components; the spatially varying $V_\mathrm{P}/V_\mathrm{S}$ ratio lies between 1.692 and 1.780. A separately seeded scrambled-Sobol design supplies 2,048 underground sources, and a jittered $15\times15$ surface array supplies 225 receivers. Reciprocal station-centred FMM produces 184,469 P/S source--station records at 20--100~km epicentral distance. The source-disjoint split contains 1,434/307/307 train/validation/test events.

Four comparators use the same minimal-coordinate residual-SiLU field, 100 epochs, batch size 8,192 and optimization seeds 101--103; checkpoints are selected only by scalar validation MAE. The value-only field minimizes only the scalar P/S loss. The relational field adds same-station secants between each source and its three nearest neighbours, with horizontal and vertical coordinate scales of 15 and 5~km, scaled radius 1.5, minimum separation 0.25~km and dimensionless weight 10. These are finite differences constructed from measured scalar values, not derivative labels or governing-equation residuals. The endpoint-eikonal field instead adds Eq.~\ref{eq:eikonal_loss} at dimensionless weight 1 on 10\% of each batch and receives the exact endpoint $V_\mathrm{P}$ and $V_\mathrm{S}$; the fourth field includes both auxiliary terms. Automatic-differentiation source-gradient norms are converted to speed and aggregated by the median over receivers at 20--70~km and over all three seeds; an event requires at least five receivers.

The relational-versus-eikonal comparison and its $V_\mathrm{P}$ success rule were prospectively locked before the new medium was generated, and all six checkpoints were complete before its test partition was first opened. The values-only and combined cells were subsequently added without changing data, architecture, budget, seeds or loss weights; these two cells are therefore a fixed post-test factorial confirmation rather than part of the prospective gate. Their six checkpoints were completed before the four-group evaluation. All comparisons use the strict common finite support of 307 events and 20,000-replicate paired event bootstraps. Input and checkpoint SHA-256 hashes and independently recomputed metrics are retained with the result.

A second realization prospectively freezes the complete four-cell design. Before the realization seed 2026090417 generated any velocity or travel time, a hashed protocol fixed the generator family, acquisition, source-disjoint split, all loss weights, seeds, readout, bootstrap and success gates. The four cells use one identical scalar cache and the settings above. All 12 scalar-validation-selected checkpoints were completed and hashed before the test velocity was accessed once. The values-only emergence gate required $V_\mathrm{P}/V_\mathrm{S}$ MAE no greater than 0.080/0.050~km~s$^{-1}$ and depth-detrended $\rho\geq0.75$ for both phases. A separate non-inferiority gate used margins of 0.010/0.006~km~s$^{-1}$. Twenty thousand paired hierarchical replicates resample optimization seeds and the 307 common test events. The full protocol, one-time access record, outcomes and hashes are archived with the paper. Test sealing was implemented within the author project using the same generator family as the preceding confirmation, with a new realization and source split. These records document prospective protocol locking and internal replication; no public preregistration or external-team replication is claimed.

\subsection*{Direct rays and focal mechanisms}

For fixed receiver, learned and FMM paths integrate $d\vect{x}/ds=-\nabla T/\|\nabla T\|$ with 1-km midpoint steps and stop within 7.5~km of the receiver. Path separation is measured after resampling both trajectories to 101 equal normalized arc-length positions. A path is support-qualified only if both trajectories remain inside the sampled source-coordinate box $[-160,160]^2\times[0,60]$~km; contact with the 80-km FMM boundary is recorded separately.

Synthetic focal truth uses 100 deterministic strike--dip--rake tensors at 73 held-out source locations, 36 stations per event and two flipped polarities per event (the nominal 5\% error rate rounded to an integer count). Polarities are generated from independent heterogeneous 3-D FMM rays. Learned, oracle FMM, layered 1-D FMM and straight rays use the same observations and double-couple search. For unit ray $\widehat{\vect{q}}$ and tensor $\vect{M}$, P radiation is $A_P=\widehat{\vect{q}}^{\mathsf T}\vect{M}\widehat{\vect{q}}$ \citep{JostHerrmann1989}. A $5\degree$ global strike--dip--rake grid is refined to $1\degree$. Mechanism error is the tensor angle $\theta_M=\cos^{-1}[\langle\vect{M}_{\rm pred},\vect{M}_{\rm true}\rangle_F/(\|\vect{M}_{\rm pred}\|_F\|\vect{M}_{\rm true}\|_F)]$, where $\langle\cdot,\cdot\rangle_F$ is the Frobenius inner product. The signed inner product retains polarity reversal, giving angles from $0\degree$ to $180\degree$; this is distinct from the minimum double-couple rotation angle.

\subsection*{Independent observational mechanism test}

We queried the USGS ANSS Comprehensive Earthquake Catalog for moment-tensor products from 1 July 2019 through 15 November 2021 in $32$--$42\degree$N and $127$--$115\degree$W \citep{USGSComCat}. Exact associated CI or NC identifiers, rather than space--time proximity, define the primary match. All 26 matches were removed from neural training before product or polarity performance was inspected. Accepted products are reviewed USGS or Southern California solutions with magnitude type Mww, Mwr or Mw, at least three waveform stations, at least three positive-weight channels and either median reported waveform fit $\geq0.5$ or variance reduction $\geq50\%$. Event time, horizontal position and magnitude differences are audited against fixed tolerances of 2~s, 15~km and 0.75 magnitude units. These metadata rules retain 18 mechanisms; 16 contain at least 15 common valid rays and manual P polarities, including at least two U and two D observations. All exclusions are written before event scoring.

QuakeML tensors use the spherical $(r,\theta,\phi)$ basis, with $r$ up, $\theta$ south and $\phi$ east. We transform them to local east--north--up coordinates, symmetrize and Frobenius-normalize them. Reported P and T axes provide an automated coordinate-convention check. Manual U and D are mapped to $+1$ and $-1$. For a fixed external tensor $\vect{M}$ and source-to-receiver unit ray $\widehat{\vect{q}}$, the predicted sign is $\operatorname{sign}(\widehat{\vect{q}}^{\mathsf T}\vect{M}\widehat{\vect{q}})$. The tensor is never refitted. The four rays are the minimal coordinate field, reciprocal MUSCAL 3-D FMM, the median of five HASH 1-D tables \citep{HardebeckShearer2002}, and the normalized full three-dimensional source--receiver chord including station elevation. Reconstructing that chord from the archived source and station coordinates agrees with the stored vectors to $1.1\times10^{-16}$ in maximum absolute component difference. A path enters the common set only when all four rays are finite and every HASH status flag is zero.

The primary endpoint is event-level mismatch using all 447 common manual polarities. Paired event resampling uses 20,000 replicates and a prespecified five-percentage-point non-inferiority margin. Prespecified learned--chord angular-separation bins are $<5\degree$, 5--$10\degree$, 10--$15\degree$ and $\geq15\degree$. Radiation sensitivity requires the absolute normalized amplitude to exceed 0, 0.02, 0.05 or 0.10 under every compared ray, retains an event only with at least eight observations and both signs, and never refits the tensor. Event-level Spearman analyses use source depth, median epicentral distance, azimuthal gap, magnitude, polarity count and waveform-fit quality; event resampling supplies 95\% intervals. The external mechanisms are waveform-derived and independent of the manual-polarity scoring step, but their Green's functions use propagation models; they are not treated as model-free ray truth \citep{USGSComCat}.

\subsection*{Event-correlated catalogue-error stress tests}

The annotation and China phase files contain no formal per-event horizontal, depth or origin-time uncertainties. We therefore label the perturbations as stress tests rather than measured catalogue-error distributions. Each realization draws one isotropic horizontal offset, one depth offset and, where a defensible scale exists, one origin-time offset for an event; that same offset is applied to every station record of the event. Horizontal perturbations have radial RMS 2 or 5~km. Depth perturbations have standard deviation 5 or 10~km and are rejection-sampled to remain inside the controlled or China domain. In 18 exact California catalogue--ComCat associations passing the fixed 2-s/15-km audit, absolute origin-time disagreement has median 0.040~s and 90th percentile 0.099~s. The controlled test therefore uses zero-mean Gaussian origin-time stresses of 0.04 and 0.10~s, explicitly as empirical association scales rather than formal standard errors; an origin-time shift $\delta t_0$ changes every event travel-time label by $-\delta t_0$.

The truth-known test uses the independent 2,048-source medium of the same-data metric confirmation. Stations, 20--100-km record selection, source-disjoint split, relational scalar objective, minimal-coordinate residual-SiLU architecture, 100-epoch budget, checkpoint rule and evaluation thresholds remain fixed. Three whole-catalogue realizations at each stress level are crossed with optimization seeds 101--103. All 18 perturbed checkpoints are completed before the fixed 96-receiver reciprocal-FMM gradient evaluation for this robustness comparison is generated. Independent reciprocal-FMM gradients then measure physical travel-time MAE, P-gradient angular and relative-vector errors, path- and event-level joint qualification and event-aggregated $V_\mathrm{P}$ error. Paired changes from the same-seed unperturbed fields use 20,000 hierarchical bootstrap replicates over events, perturbation realizations and optimization seeds.

Observational propagation is post-training and does not refit a field or downstream parameter. The California test applies the same spatial stresses to each held-out source while stations and external tensors remain fixed; origin-time changes cannot alter a spatial gradient or polarity sign and are recorded only. The China test applies the spatial stresses to the frozen four-phase field while keeping the original supported observations, cells, fixed-regional-slowness readout and smoothing weight. No China origin-time stress is applied because the archive provides neither a formal uncertainty nor an independently grounded time-error scale. Three spatial realizations and 200 event-cluster bootstrap replicates quantify the China correlation, MAE and retained derivative support.

\subsection*{Statistical analysis}

Principal and endpoint-eikonal results use five optimization seeds and two-sided 95\% Student's $t$ intervals for paired differences. Independent-subset summaries report sampling standard deviation separately from five-seed optimization standard deviation. The primary identifiability ensemble contains 30 minimal-coordinate fields: ten independently sampled source subsets at each of 25\%, 10\% and 5\% coverage. Run-level and fixed-coverage associations use Spearman correlation. The query-level Eq.~\ref{eq:gradient_error_bound} audit contains 206,550 finite FMM comparisons; its within-run exponent is estimated after demeaning log design and log error inside every training run, and its 95\% interval resamples complete runs 10,000 times. The synthetic role-completion experiment reports seed SD separately from 20,000 paired source-group bootstrap replicates, conditional on the five trained fields and fixed 81-station network. The independent metric confirmation uses 20,000 event-level paired bootstrap replicates on the strict common test support. Synthetic focal-mechanism comparisons use 20,000-replica paired source-cluster bootstrap intervals; the fixed-mechanism observational test resamples events. The China map uses 200 event-cluster bootstrap replicates over 12,231 held-out events while keeping the supported-cell mask fixed; 10,000 spatial-block replicates and the exhaustive admissible toroidal shifts test geographical dependence. China reciprocity-ablation summaries report mean and sample standard deviation across three matched optimization seeds. The dedicated event-correlated tests propagate the specified catalogue-coordinate stresses; other intervals do not propagate external-tensor, polarity-coding, neural-retraining, catalogue-location or reference-map uncertainty. No final receiver-function outcome selects a descriptor, architecture, quality threshold or checkpoint.

\subsection*{Quantum field-value test}

The synthetic companion test uses exact two-dimensional Gaussian, harmonic-oscillator vortex and free-packet interference fields in units $\hbar=m=1$. Each task has its own raw-coordinate $(x,y,t)$ network with four 128-unit SiLU hidden layers, trained on 8,192 field-value samples for 600 epochs with three optimization seeds. An additional density-only task tests the distinction between fitting density and identifying current. All labels are generated field values; no derivative, trajectory, potential or equation-residual targets enter optimization. Both training and scalar validation exclude a fixed spatial block and time slab. Checkpoints minimize validation value error, and the frozen mean of three fields supplies the ensemble readout. Separate domain, spatial-block and time-slab tests evaluate the same underlying state at unseen coordinates. A global cubic tensor spline, added after initial neural/RBF scoring, uses the same training values and scalar-only validation. Analytic or independently integrated trajectories, density-threshold scans, all individual seeds and the baseline amendment are reported in Supplementary Information, Section~24. These synthetic tests do not constitute quantum measurements or transfer to unseen Hamiltonians.

\section*{Data availability}
SeismicX-Cont is available through its versioned DOI and public repository~\citep{SeismicXCont}; MUSCAL is archived separately~\citep{YehBenZion2026}. The unified China catalogue and phase products are distributed by the National Earthquake Science Data Center~\citep{CENCUnifiedCatalogue}. The receiver-function grid is the publisher-hosted supplement to He et al.~\citep{HeEtAl2014}, and CRUST1.0 is publicly downloadable~\citep{LaskeEtAl2013}.

Public reproduction materials are available in the \path{scalar-wave-geometry} directory at \url{https://huggingface.co/cangyeone/relational-geophysics/tree/main/scalar-wave-geometry}. They include synthetic seismic and quantum data, derivatives of openly licensed California observations, selected checkpoints, fixed splits, comparison outputs, and file-level SHA-256 manifests. Domestic catalogue rows, event/station coordinates, geographic caches and reference maps are outside the public release. Regional model kernels omit geographic normalization and projection metadata and support normalized-input inference only; the public package does not independently reproduce the domestic catalogue or Moho scores.

\section*{Code availability}
Training, evaluation, figure-generation and verification code is distributed in the same public project. Its README and \path{VALIDATION.md} describe commands, dependencies and the verified scope for controlled seismology, California manual tomography, gradient regularization, quantum fields and California focal validation. Checkpoint inference and numerical verification are distinct from rerunning the complete historical optimization matrix. Paths to earlier experiment archives in the Supplementary Information identify historical evidence and are not assertions that all such files are in the public release.

\section*{Author contributions}
Z.Y.: Conceptualization, Methodology, Software, Formal analysis, Investigation, Data curation, Validation, Visualization, Writing--original draft, and Writing--review and editing.

\section*{Funding}
This research received no external funding.

\section*{Ethics declaration}
The study uses previously released earthquake catalogues and seismic annotations and involves no human participants, personal data or animal subjects.

\section*{Competing interests}
The author declares no competing interests.

\section*{AI-assisted work disclosure}
Generative AI tools assisted manuscript drafting and editing, code development, experimental-workflow preparation and analysis. Scientific interpretation, verification of the results and responsibility for the submitted work remain with the author.

\FloatBarrier
\clearpage
\bibliographystyle{unsrtnat}
\bibliography{references}

\begin{thebibliography}{36}
\providecommand{\natexlab}[1]{#1}
\providecommand{\url}[1]{\texttt{#1}}
\expandafter\ifx\csname urlstyle\endcsname\relax
  \providecommand{\doi}[1]{doi: #1}\else
  \providecommand{\doi}{doi: \begingroup \urlstyle{rm}\Url}\fi

\bibitem[Aki and Richards(2002)]{AkiRichards2002}
Keiiti Aki and Paul~G. Richards.
\newblock \emph{Quantitative Seismology}.
\newblock University Science Books, Sausalito, California, 2 edition, 2002.

\bibitem[Rawlinson et~al.(2010)Rawlinson, Pozgay, and Fishwick]{Rawlinson2010}
N.~Rawlinson, S.~Pozgay, and S.~Fishwick.
\newblock Seismic tomography: A window into deep {Earth}.
\newblock \emph{Physics of the Earth and Planetary Interiors}, 178\penalty0
  (3--4):\penalty0 101--135, 2010.
\newblock \doi{10.1016/j.pepi.2009.10.002}.

\bibitem[Smith et~al.(2021)Smith, Azizzadenesheli, and Ross]{Smith2021}
Jonathan~D. Smith, Kamyar Azizzadenesheli, and Zachary~E. Ross.
\newblock {EikoNet}: Solving the eikonal equation with deep neural networks.
\newblock \emph{IEEE Transactions on Geoscience and Remote Sensing},
  59\penalty0 (12):\penalty0 10685--10696, 2021.
\newblock \doi{10.1109/TGRS.2020.3039165}.

\bibitem[Taufik et~al.(2023)Taufik, Waheed, and Alkhalifah]{Taufik2023}
Mohammad~H. Taufik, Umair~bin Waheed, and Tariq~A. Alkhalifah.
\newblock A neural network based global traveltime function ({GlobeNN}).
\newblock \emph{Scientific Reports}, 13:\penalty0 7179, 2023.
\newblock \doi{10.1038/s41598-023-33203-1}.

\bibitem[Waheed et~al.(2021)Waheed, Haghighat, Alkhalifah, Song, and
  Hao]{Waheed2021}
Umair~bin Waheed, Ehsan Haghighat, Tariq Alkhalifah, Chao Song, and Qi~Hao.
\newblock {PINNeik}: Eikonal solution using physics-informed neural networks.
\newblock \emph{Computers \& Geosciences}, 155:\penalty0 104833, 2021.
\newblock \doi{10.1016/j.cageo.2021.104833}.

\bibitem[Raissi et~al.(2019)Raissi, Perdikaris, and Karniadakis]{Raissi2019}
Maziar Raissi, Paris Perdikaris, and George~Em Karniadakis.
\newblock Physics-informed neural networks: A deep learning framework for
  solving forward and inverse problems involving nonlinear partial differential
  equations.
\newblock \emph{Journal of Computational Physics}, 378:\penalty0 686--707,
  2019.
\newblock \doi{10.1016/j.jcp.2018.10.045}.

\bibitem[Czarnecki et~al.(2017)Czarnecki, Osindero, Jaderberg, {\'{S}}wirszcz,
  and Pascanu]{Czarnecki2017}
Wojciech~M. Czarnecki, Simon Osindero, Max Jaderberg, Grzegorz {\'{S}}wirszcz,
  and Razvan Pascanu.
\newblock Sobolev training for neural networks.
\newblock In \emph{Advances in Neural Information Processing Systems 30}, 2017.

\bibitem[Shi et~al.(2026)Shi, Poggiali, Marone, de~Hoop, and
  Dokmani{\'c}]{Shi2026}
Cheng Shi, Giulio Poggiali, Chris Marone, Maarten~V. de~Hoop, and Ivan
  Dokmani{\'c}.
\newblock High-rate phase association with travel time neural fields.
\newblock \emph{Nature Communications}, 17:\penalty0 8140, 2026.
\newblock \doi{10.1038/s41467-026-74092-y}.

\bibitem[Lancaster and Salkauskas(1981)]{LancasterSalkauskas1981}
Peter Lancaster and Kestutis Salkauskas.
\newblock Surfaces generated by moving least squares methods.
\newblock \emph{Mathematics of Computation}, 37\penalty0 (155):\penalty0
  141--158, 1981.
\newblock \doi{10.1090/S0025-5718-1981-0616367-1}.

\bibitem[Levin(1998)]{Levin1998}
David Levin.
\newblock The approximation power of moving least-squares.
\newblock \emph{Mathematics of Computation}, 67\penalty0 (224):\penalty0
  1517--1531, 1998.
\newblock \doi{10.1090/S0025-5718-98-00974-0}.

\bibitem[De~Brabanter et~al.(2013)De~Brabanter, De~Brabanter, De~Moor, and
  Gijbels]{DeBrabanter2013}
Kris De~Brabanter, Jos De~Brabanter, Bart De~Moor, and Ir{\`e}ne Gijbels.
\newblock Derivative estimation with local polynomial fitting.
\newblock \emph{Journal of Machine Learning Research}, 14:\penalty0 281--301,
  2013.
\newblock URL
  \url{https://jmlr.csail.mit.edu/papers/volume14/debrabanter13a-deleted/debrabanter13a.pdf}.

\bibitem[Bohm(1952)]{Bohm1952}
David Bohm.
\newblock A suggested interpretation of the quantum theory in terms of
  ``hidden'' variables. {I}.
\newblock \emph{Physical Review}, 85:\penalty0 166--179, 1952.
\newblock \doi{10.1103/PhysRev.85.166}.

\bibitem[Struyve and Valentini(2009)]{StruyveValentini2009}
Ward Struyve and Antony Valentini.
\newblock De {Broglie--Bohm} guidance equations for arbitrary {Hamiltonians}.
\newblock \emph{Journal of Physics A: Mathematical and Theoretical},
  42:\penalty0 035301, 2009.
\newblock \doi{10.1088/1751-8113/42/3/035301}.

\bibitem[Gale et~al.(1968)Gale, Guth, and Trammell]{Gale1968}
W.~Gale, E.~Guth, and G.~T. Trammell.
\newblock Determination of the quantum state by measurements.
\newblock \emph{Physical Review}, 165:\penalty0 1434--1436, 1968.
\newblock \doi{10.1103/PhysRev.165.1434}.

\bibitem[{PROJ contributors}(2026)]{PROJ}
{PROJ contributors}.
\newblock {PROJ} coordinate transformation software library.
\newblock Open Source Geospatial Foundation ({OSGeo}), 2026.
\newblock URL \url{https://doi.org/10.5281/zenodo.5884394}.

\bibitem[Yu et~al.(2026)Yu, Cai, Liu, Miu, Dong, Li, and Huang]{SeismicXCont}
Ziye Yu, Yuqi Cai, Xin Liu, Fajun Miu, Tengchao Dong, Tingting Li, and Xinghui
  Huang.
\newblock {SeismicX-Cont}: Continuous seismic waveforms, station-time coverage,
  and phase annotations for the 2019 {Ridgecrest} sequence and a 2021
  {California} background period.
\newblock Hugging Face dataset, 2026.
\newblock URL \url{https://doi.org/10.57967/hf/9598}.
\newblock Version 1.0.0; release revision 0f56956.

\bibitem[Zhu et~al.(2025)Zhu, Wang, Rong, Yu, Zuzlewski, Tepp, Taira, Marty,
  Husker, and Allen]{ZhuCEED2025}
Weiqiang Zhu, Haoyu Wang, Bo~Rong, Ellen Yu, Stephane Zuzlewski, Gabrielle
  Tepp, Taka'aki Taira, Julien Marty, Allen Husker, and Richard~M. Allen.
\newblock {California} earthquake dataset for machine learning and cloud
  computing.
\newblock arXiv preprint arXiv:2502.11500, 2025.
\newblock URL \url{https://arxiv.org/abs/2502.11500}.

\bibitem[{Southern California Earthquake Data Center}(2013)]{SCEDC}
{Southern California Earthquake Data Center}.
\newblock {Southern California Earthquake Data Center}.
\newblock California Institute of Technology, 2013.
\newblock URL \url{https://doi.org/10.7909/C3WD3XH1}.

\bibitem[{Northern California Earthquake Data
  Center}(2014{\natexlab{a}})]{NCEDC}
{Northern California Earthquake Data Center}.
\newblock {Northern California Earthquake Data Center}.
\newblock Northern California Earthquake Data Center, 2014{\natexlab{a}}.
\newblock URL \url{https://doi.org/10.7932/NCEDC}.

\bibitem[{California Institute of Technology and United States Geological
  Survey Pasadena}(1926)]{SCSN}
{California Institute of Technology and United States Geological Survey
  Pasadena}.
\newblock {Southern California Seismic Network}.
\newblock International Federation of Digital Seismograph Networks, 1926.
\newblock URL \url{https://doi.org/10.7914/SN/CI}.

\bibitem[{USGS Menlo Park}(1966)]{NCSN}
{USGS Menlo Park}.
\newblock {USGS Northern California Seismic Network}.
\newblock International Federation of Digital Seismograph Networks, 1966.
\newblock URL \url{https://doi.org/10.7914/SN/NC}.

\bibitem[{Northern California Earthquake Data
  Center}(2014{\natexlab{b}})]{BDSN}
{Northern California Earthquake Data Center}.
\newblock {Berkeley Digital Seismic Network} ({BDSN}).
\newblock Northern California Earthquake Data Center, 2014{\natexlab{b}}.
\newblock URL \url{https://doi.org/10.7932/BDSN}.

\bibitem[{China Earthquake Networks Center}(2026)]{CENCUnifiedCatalogue}
{China Earthquake Networks Center}.
\newblock Unified earthquake catalogue and phase data of the {China Earthquake
  Network}, 2026.
\newblock URL
  \url{https://data.earthquake.cn/datashare/report.shtml?PAGEID=earthquake_zgtwzx}.
\newblock Catalogue production from 2009 onward; accessed 7 August 2026.

\bibitem[Loshchilov and Hutter(2019)]{LoshchilovHutter2019}
Ilya Loshchilov and Frank Hutter.
\newblock Decoupled weight decay regularization.
\newblock In \emph{International Conference on Learning Representations}, 2019.
\newblock URL \url{https://openreview.net/forum?id=Bkg6RiCqY7}.

\bibitem[Paszke et~al.(2019)Paszke, Gross, Massa, Lerer, Bradbury, Chanan,
  Killeen, Lin, Gimelshein, Antiga, Desmaison, K{\"o}pf, Yang, DeVito, Raison,
  Tejani, Chilamkurthy, Steiner, Fang, Bai, and Chintala]{Paszke2019}
Adam Paszke, Sam Gross, Francisco Massa, Adam Lerer, James Bradbury, Gregory
  Chanan, Trevor Killeen, Zeming Lin, Natalia Gimelshein, Luca Antiga, Alban
  Desmaison, Andreas K{\"o}pf, Edward Yang, Zachary DeVito, Martin Raison,
  Alykhan Tejani, Sasank Chilamkurthy, Benoit Steiner, Lu~Fang, Junjie Bai, and
  Soumith Chintala.
\newblock {PyTorch}: An imperative style, high-performance deep learning
  library.
\newblock In \emph{Advances in Neural Information Processing Systems 32}, pages
  8024--8035, 2019.

\bibitem[Mirzaei et~al.(2012)Mirzaei, Schaback, and Dehghan]{Mirzaei2012}
Davoud Mirzaei, Robert Schaback, and Mehdi Dehghan.
\newblock On generalized moving least squares and diffuse derivatives.
\newblock \emph{IMA Journal of Numerical Analysis}, 32\penalty0 (3):\penalty0
  983--1000, 2012.
\newblock \doi{10.1093/imanum/drr030}.

\bibitem[Doody et~al.(2023)Doody, Rodgers, Afanasiev, Boehm, Krischer, Chiang,
  and Simmons]{Doody2023}
Claire Doody, Arthur Rodgers, Michael Afanasiev, Christian Boehm, Lion
  Krischer, Andrea Chiang, and Nathan Simmons.
\newblock {CANVAS}: An adjoint waveform tomography model of {California} and
  {Nevada}.
\newblock \emph{Journal of Geophysical Research: Solid Earth}, 128\penalty0
  (12):\penalty0 e2023JB027583, 2023.
\newblock \doi{10.1029/2023JB027583}.

\bibitem[Yeh et~al.(2026)Yeh, Ben-Zion, and Olsen]{YehBenZion2026}
Te-Yang Yeh, Yehuda Ben-Zion, and Kim Olsen.
\newblock {MUSCAL}---the multi-scale {P} and {S} seismic velocity models of
  {California} and {Western Nevada}.
\newblock Zenodo, 2026.
\newblock URL \url{https://doi.org/10.5281/zenodo.19243476}.
\newblock MUSCAL version 1 data release.

\bibitem[Sethian(1996)]{Sethian1996}
James~A. Sethian.
\newblock A fast marching level set method for monotonically advancing fronts.
\newblock \emph{Proceedings of the National Academy of Sciences}, 93\penalty0
  (4):\penalty0 1591--1595, 1996.
\newblock \doi{10.1073/pnas.93.4.1591}.

\bibitem[Giroux(2021)]{Giroux2021}
Bernard Giroux.
\newblock ttcrpy: A python package for traveltime computation and raytracing.
\newblock \emph{SoftwareX}, 16:\penalty0 100834, 2021.
\newblock \doi{10.1016/j.softx.2021.100834}.

\bibitem[He et~al.(2014)He, Shang, Yu, Zhang, and Van~der Hilst]{HeEtAl2014}
Rizheng He, Xuefeng Shang, Chunquan Yu, Haijiang Zhang, and Robert~D. Van~der
  Hilst.
\newblock A unified map of {Moho} depth and {$V_p/V_s$} ratio of continental
  {China} by receiver function analysis.
\newblock \emph{Geophysical Journal International}, 199\penalty0 (3):\penalty0
  1910--1918, 2014.
\newblock \doi{10.1093/gji/ggu365}.

\bibitem[Laske et~al.(2013)Laske, Masters, Ma, and Pasyanos]{LaskeEtAl2013}
Gabi Laske, Guy Masters, Zhitu Ma, and Michael Pasyanos.
\newblock Update on {CRUST1.0}: A 1-degree global model of {Earth}'s crust.
\newblock Geophysical Research Abstracts, 15, EGU2013-2658, 2013.
\newblock URL \url{https://igppweb.ucsd.edu/~gabi/crust1.html}.

\bibitem[{Natural Earth}(2026)]{NaturalEarth}
{Natural Earth}.
\newblock 1:110m physical vectors: Land, 2026.
\newblock URL
  \url{https://www.naturalearthdata.com/downloads/110m-physical-vectors/110m-land/}.
\newblock Accessed 7 August 2026.

\bibitem[Jost and Herrmann(1989)]{JostHerrmann1989}
M.~L. Jost and R.~B. Herrmann.
\newblock A student's guide to and review of moment tensors.
\newblock \emph{Seismological Research Letters}, 60\penalty0 (2):\penalty0
  37--57, 1989.
\newblock \doi{10.1785/gssrl.60.2.37}.

\bibitem[{U.S. Geological Survey Earthquake Hazards Program}(2017)]{USGSComCat}
{U.S. Geological Survey Earthquake Hazards Program}.
\newblock {Advanced National Seismic System} ({ANSS}) comprehensive catalog of
  earthquake events and products.
\newblock U.S. Geological Survey, 2017.
\newblock URL \url{https://doi.org/10.5066/F7MS3QZH}.

\bibitem[Hardebeck and Shearer(2002)]{HardebeckShearer2002}
Jeanne~L. Hardebeck and Peter~M. Shearer.
\newblock A new method for determining first-motion focal mechanisms.
\newblock \emph{Bulletin of the Seismological Society of America}, 92\penalty0
  (6):\penalty0 2264--2276, 2002.
\newblock \doi{10.1785/0120010200}.

\end{thebibliography}

\clearpage
\suppressfloats[t]
\setcounter{figure}{0}
\setcounter{table}{0}
\setcounter{equation}{0}
\setcounter{section}{0}
\renewcommand{\theHfigure}{S.\arabic{figure}}
\renewcommand{\theHtable}{S.\arabic{table}}
\renewcommand{\theHequation}{S.\arabic{equation}}
\renewcommand{\theHsection}{S.\arabic{section}}
\section*{Supplementary Information}
The public release and its verified scope are described in the main Data and Code availability statements. Historical archive paths below preserve experiment provenance; domestic catalogue and geographic reproduction materials are not part of the public package.

\renewcommand{\thefigure}{S\arabic{figure}}
\renewcommand{\thetable}{S\arabic{table}}
\renewcommand{\theequation}{S\arabic{equation}}

\section{Data Partitions and Provenance}

Observed records were partitioned by event identifier and synthetic records by source coordinate. The three partitions are disjoint for every experiment reported in the main text. The fixed-external-mechanism partition assigns all 26 exact USGS ComCat associations to test before training; the resulting train/validation/test counts are 6,369/1,124/26 events and 143,657/26,467/2,410 records. The earlier broad observational benchmark contains 5,423/1,156/940 events and 124,254/26,386/21,894 records and is retained only for secondary diagnostics. The primary dense controlled-medium partition contains 397/85/85 train/validation/test source groups; the original MUSCAL benchmark used for secondary transfer controls contains 397/92/70.

The China four-phase cache contains 2,903,562 retained values from 791,404 in-domain events. Deterministic event hashing assigns 2,322,094/291,562/289,906 records to train/validation/test before competing-branch removal. Phase counts are 1,001,527 Pg, 990,955 Sg, 673,362 Pn and 237,718 Sn. The split-index SHA-256 digests and every cleaning threshold are stored with the experiment outputs.

The synthetic validation groups correspond to a test partition used during an initial model-development stage. The held-out test groups were selected deterministically from the remaining source pool and excluded from every training, validation, architecture, coverage and endpoint-eikonal run reported in the paper. Because this test geometry was fixed across the planned comparison matrix, the paper interprets it as a common benchmark rather than a source of model-selection feedback.

\begin{figure}[t]
  \centering
  \includegraphics[width=0.96\textwidth]{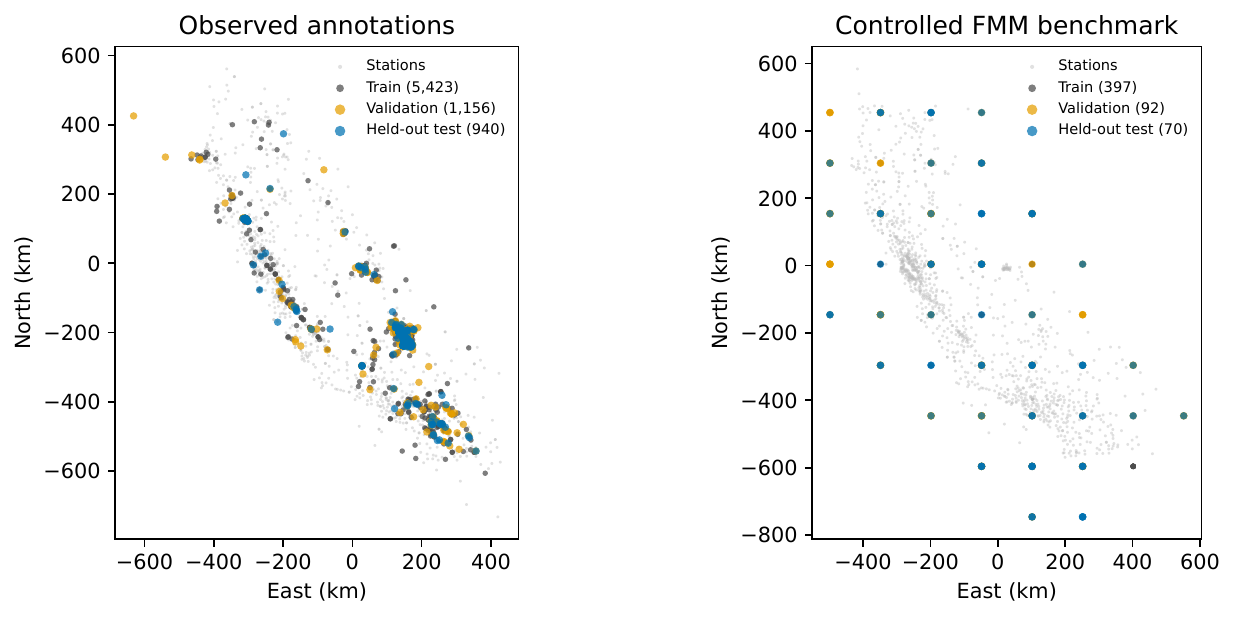}
  \caption{\textbf{Disjoint event and source partitions.} The earlier broad observational partition is shown at left; the controlled MUSCAL/FMM benchmark is split by source coordinate at right. The external-mechanism experiment uses a stricter event-disjoint manifest that assigns every exact ComCat tensor association to test. Stations are shown in grey.}
  \label{si:fig:splits}
\end{figure}

Machine-readable manifests store all group identifiers. The integrity package in
\path{supplementary/integrity_stage2_5/} records cache hashes, split intersections,
expected run counts, summary recomputation and citation checks.

\section{Cross-Architecture Generality}

The conventional tanh and fixed-Fourier fields use raw normalized source and
receiver coordinates, no reference time, no causal head and approximately the
same parameter count as the residual minimal-prior field. Their equal training
budgets are summarized in Table~\ref{si:tab:architecture_generality}. The tanh
field recovers geometry at full coverage but its scalar error increases
at 10\% coverage. The fixed-Fourier field has large scalar and gradient errors even at full coverage. Because
only one fixed frequency distribution was tested, this is a null result for the
specified representation rather than for Fourier features as a class.

A separate scalar-matched control adds the fixed smooth mode
$0.05\sin(64\pi\vect{d}^{\mathsf T}\widetilde{\vect{x}}_s)$ to each of the five
selected primary fields, with
$\vect{d}\propto(1,\sqrt{2},\sqrt{3})$. The perturbation changes mean test MAE
from $0.7171$ to $0.7174$~s while changing the five-seed mean median angle from
$5.34\degree$ to $38.94\degree$ and relative-vector error from 0.108 to 0.987.
Because the value amplitude is bounded before evaluation, this is a constructive
matched-scalar control rather than an additional trained architecture.

\begin{table}[ht]
\centering
\caption{Single-seed equal-budget architecture generality. Entries give travel-time MAE (s), median FMM angular error (degrees) and relative-vector error, in that order.}
\label{si:tab:architecture_generality}
\small
\begin{adjustbox}{max width=\linewidth}\begin{tabular}{lcccc}
\toprule
Field & 100\% & 25\% & 10\% & 5\% \\
\midrule
Residual minimal & 0.660/5.45/0.116 & 1.004/8.17/0.184 & 1.966/24.95/0.575 & 9.737/69.04/1.116 \\
Plain tanh & 0.917/6.18/0.131 & 1.442/10.02/0.249 & 9.738/57.66/0.992 & 9.867/58.57/0.937 \\
Fixed Fourier & 4.206/71.84/3.078 & 8.566/81.37/5.667 & 9.457/87.62/2.237 & 10.227/85.97/2.004 \\
\bottomrule
\end{tabular}\end{adjustbox}
\end{table}

\section{Controlled-Medium Generality}

Two independent benchmarks use an $81\times81\times17$ grid with 5-km spacing,
147 sources, 81 stations and a common 103/22/22 source split. The layered P
velocity is $4.6+0.03z$~km~s$^{-1}$. The heterogeneous medium adds smooth
sinusoidal and Gaussian anomalies with an approximately 50-km scale and a P
range of 3.8--7.4~km~s$^{-1}$. Source-centred FMM supplies scalar times;
station-centred reciprocal FMM supplies gradients independently.

\begin{table}[ht]
\centering
\caption{Value-only medium generality. Entries give travel-time MAE (s), median FMM angular error (degrees) and relative-vector error, in that order.}
\label{si:tab:medium_generality}
\small
\begin{adjustbox}{max width=\linewidth}\begin{tabular}{lcccc}
\toprule
Medium & 100\% & 25\% & 10\% & 5\% \\
\midrule
Layered & 0.044/2.22/0.042 & 0.095/2.23/0.042 & 0.178/3.17/0.060 & 0.378/8.77/0.164 \\
3-D heterogeneous & 0.270/3.58/0.075 & 0.870/7.35/0.152 & 1.082/9.56/0.192 & 1.208/11.66/0.229 \\
\bottomrule
\end{tabular}\end{adjustbox}
\end{table}

The same architecture is jointly adequate to 5\% in the layered medium but only
to 25\% in the heterogeneous medium. Dividing fill distance by the nominal
50-km anomaly scale does not establish a common relation because the layered
medium has no finite lateral correlation scale and only two media are available.
No cross-medium scaling law is claimed.

\begin{figure}[t]
  \centering
  \includegraphics[width=0.96\textwidth]{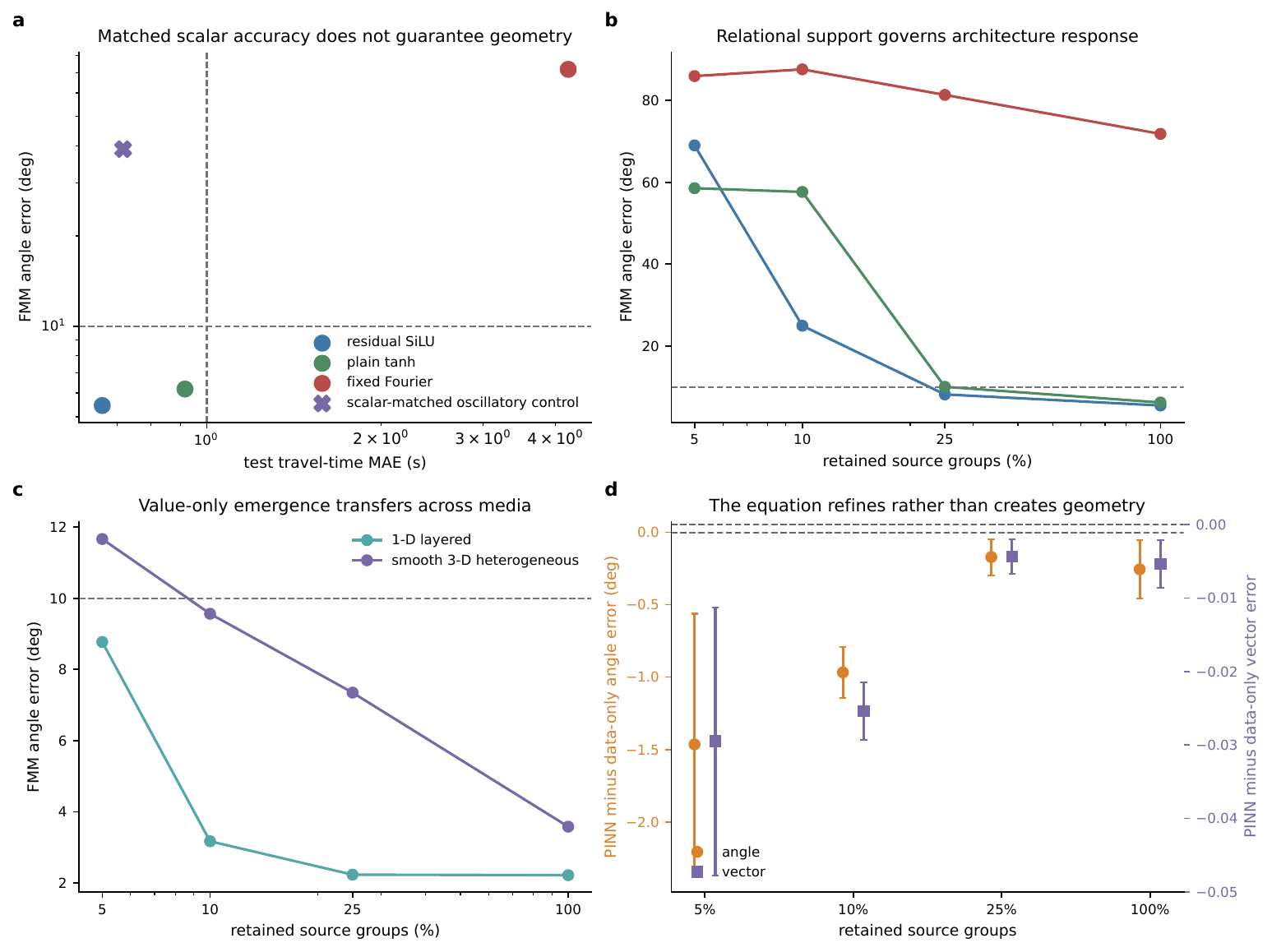}
  \caption{\textbf{Extended representation, medium and equation response.} (a) Residual-SiLU and tanh fields recover gradients, whereas the scalar-matched oscillatory control does not. (b) Relational support controls derivative recovery across parameterizations. (c) Value-only geometry transfers between layered and smooth three-dimensional heterogeneous media. (d) Endpoint-eikonal supervision refines geometry already identified at high coverage and partially compensates at 5\% support. Direct $V_\mathrm{P}/V_\mathrm{S}$ recovery is shown in Fig.~3 of the main text.}
  \label{si:fig:extended_generality}
\end{figure}

\section{Local Identifiability and the Sampling Law}

For a differentiable scalar field observed locally at coordinate offsets
$\Delta\vect{x}_i$, the linearized inverse problem is
\begin{equation}
\Delta\vect{T}=\vect{X}\vect{g}+\vect{r},\qquad
\vect{X}_{i:}=\Delta\vect{x}_i^{\mathsf T},\qquad
\vect{g}=\nabla T(\vect{x}_0).
\end{equation}
With Gaussian weights $w_i$, the normal matrix is
$\vect{G}=\vect{X}^{\mathsf T}\vect{W}\vect{X}$. Its rank determines whether
the three components of $\vect{g}$ are identifiable in the local linearized problem. The eigenvalues
of $\vect{G}/\operatorname{tr}(\vect{G})$ separate directional diversity from
the absolute offset scale. Small fill distance controls local truncation,
whereas the smallest eigenvalue and normalized determinant penalize nearly
coplanar support. Distance outside the support box distinguishes interpolation
from extrapolation.

The weighted local-linear estimate is
\begin{equation}
\widehat{\vect{g}}=(\vect{X}^{\mathsf T}\vect{W}\vect{X})^{-1}
\vect{X}^{\mathsf T}\vect{W}\Delta\vect{T},
\label{si:eq:supp_local_gradient_estimator}
\end{equation}
and, under the smoothness conditions below, obeys
\begin{equation}
\|\widehat{\vect{g}}-\vect{g}\|_2\leq
\frac{\|\vect{W}^{1/2}\vect{\epsilon}\|_2+
\frac{M}{2}\left(\sum_iw_i\|\Delta\vect{x}_i\|_2^4\right)^{1/2}}
{\sqrt{\lambda_{\min}(\vect{G})}}.
\label{si:eq:supp_gradient_error_bound}
\end{equation}

\paragraph{Local gradient proposition.}
Suppose $T$ is twice continuously differentiable on the neighbourhood of
$\vect{x}_0$, $\|\nabla^2T\|_2\leq M$, and the observed differences satisfy
$\Delta\vect{T}=\vect{X}\vect{g}+\vect{r}+\vect{\epsilon}$.  If
$\vect{G}=\vect{X}^{\mathsf T}\vect{W}\vect{X}$ is positive definite, the
weighted local-linear estimate in Eq.~\ref{si:eq:supp_local_gradient_estimator} is
unique and obeys Eq.~\ref{si:eq:supp_gradient_error_bound}.  If $\vect{G}$ is singular,
there is a non-zero direction $\vect{u}$ for which
$\vect{X}\vect{u}=0$, so the component of $\vect{g}$ along $\vect{u}$ cannot be
identified from these local differences.

To prove the bound, let $\vect{A}=\vect{W}^{1/2}\vect{X}$.  Then
$\widehat{\vect{g}}-\vect{g}=\vect{A}^{\dagger}
\vect{W}^{1/2}(\vect{r}+\vect{\epsilon})$ and
$\|\vect{A}^{\dagger}\|_2=1/\sqrt{\lambda_{\min}(\vect{G})}$.
Taylor's theorem gives
$|r_i|\leq(M/2)\|\Delta\vect{x}_i\|_2^2$; the triangle inequality yields
Eq.~\ref{si:eq:supp_gradient_error_bound}.  The result makes the sampling trade-off
explicit: shrinking a neighbourhood reduces its curvature remainder but may
also reduce the weakest design eigenvalue and amplify observational noise.
It is a first-order moving-least-squares/local-polynomial derivative bound,
not a claim that neural differentiation creates information absent from the
sample design \citep{LancasterSalkauskas1981,Levin1998,DeBrabanter2013}.

\paragraph{Learned-field corollary.}
The same design matrix bounds the automatic derivative of any smooth learned
field directly. Define the scalar relation mismatch
\begin{equation}
\delta_i=\{T_\theta(\vect{x}_i)-T_\theta(\vect{x}_0)\}
-\{T(\vect{x}_i)-T(\vect{x}_0)\}.
\end{equation}
If the Hessian norms of $T_\theta$ and $T$ are bounded by $M_\theta$ and $M$,
subtracting their two Taylor expansions and applying the same pseudoinverse
argument gives
\begin{equation}
\|\nabla T_\theta(\vect{x}_0)-\nabla T(\vect{x}_0)\|_2\leq
\frac{\|\vect{W}^{1/2}\vect{\delta}\|_2+
\frac{M_\theta+M}{2}\left(\sum_iw_i\|\Delta\vect{x}_i\|_2^4\right)^{1/2}}
{\sqrt{\lambda_{\min}(\vect{G})}}.
\label{si:eq:supp_learned_gradient_bound}
\end{equation}
If measured values contain noise, the weighted norm of the corresponding
noise differences adds to the numerator. Equation~\ref{si:eq:supp_learned_gradient_bound}
is the direct bridge from scalar fit to neural-field gradient: small relational
error determines a small gradient error only when the learned and physical
fields remain locally smooth and the observation design has no weak direction.
The scalar-matched oscillatory control violates the useful-curvature regime
while retaining nearly identical values, exactly as the bound predicts.

Source-coordinate tensors hold station identity fixed because
$\partial T(\vect{x}_r,\vect{x}_s)/\partial\vect{x}_s$ varies the source while
holding the receiver constant. Receiver tensors use the retained station
coordinates. For requested neighbourhood sizes $k=8,16,32,64$, the Gaussian
bandwidth is half the $k$th-neighbour distance. Thirty-three scalar candidates
were evaluated over 17 validation-only conditions: six random source fractions,
six structured source or path restrictions and five station fractions.
The selection score is the mean of the median within-condition Spearman
correlations with validation-FMM angle and vector error. Locked-test outcomes
were not loaded during candidate selection.

The frozen descriptor is the path-conditioned source boundary penalty
\begin{equation}
b(\vect{x})=d_{\rm outside}(\vect{x},\mathcal{B})+
\frac{1}{d_{\rm inside}(\vect{x},\partial\mathcal{B})+1},
\end{equation}
where $\mathcal{B}$ is the station-specific source-support box. Its validation
selection score is 0.255 (angle 0.198; vector 0.312). Its locked-test results
are retained in Table~\ref{si:tab:identifiability_descriptor} as a secondary
structured-field diagnostic. They do not enter the primary minimal-coordinate
Fig.~2. No locked-test derivative outcome selected the descriptor or any subset.

\begin{table}[H]
\centering
\caption{Secondary validation-frozen descriptor associations in the structured-field diagnostic matrix. These fits contain engineered pair features and are kept separate from the primary minimal-coordinate evidence. Locked-test run-level coefficients use the median descriptor and error from each of 30 fits; within-run coefficients are computed over paths and then summarized across fits.}
\label{si:tab:identifiability_descriptor}
\small
\begin{adjustbox}{max width=\linewidth}\begin{tabular}{lcc}
\toprule
Evaluation & Angular error $\rho$ & Relative-vector error $\rho$ \\
\midrule
Validation selection, median within condition & 0.198 & 0.312 \\
Locked test, run level ($n=30$) & 0.697 & 0.692 \\
Locked test, median within run & 0.148 & 0.214 \\
\bottomrule
\end{tabular}\end{adjustbox}
\end{table}

This structured-field matrix records the original validation-frozen descriptor
development but is not used as primary evidence in Fig.~2 or in the Results.
The minimal-coordinate subset ensemble, endpoint-rank intervention and
role-completion experiment provide the primary sampling-law tests.

We evaluated the design tensor at every FMM gradient query in the 30 frozen
25\%, 10\% and 5\% source-subset runs of the primary minimal-coordinate field.
To isolate directional completeness from neighbourhood scale, the analysis uses
$\widetilde{\lambda}_{\min}=\lambda_{\min}(G)/\operatorname{tr}(G)$ for the
path-conditioned $k=32$ source design. Across the 30 run medians,
$\widetilde{\lambda}_{\min}$ has Spearman $\rho=-0.911$ with angular error and
$-0.928$ with relative-vector error. This ordering is not solely a comparison
between coverage levels. With source count fixed, the angular/vector
coefficients are $-0.636/-0.661$ across the ten 10\% subsets and
$-0.648/-0.733$ across the ten 5\% subsets; the 25\% fields occupy a saturated
low-error regime ($-0.091/-0.285$). Architecture, scalar targets, optimization
seed and approximately matched update count remain fixed within each comparison.

\begin{figure}[H]
  \centering
  \includegraphics[width=0.86\textwidth]{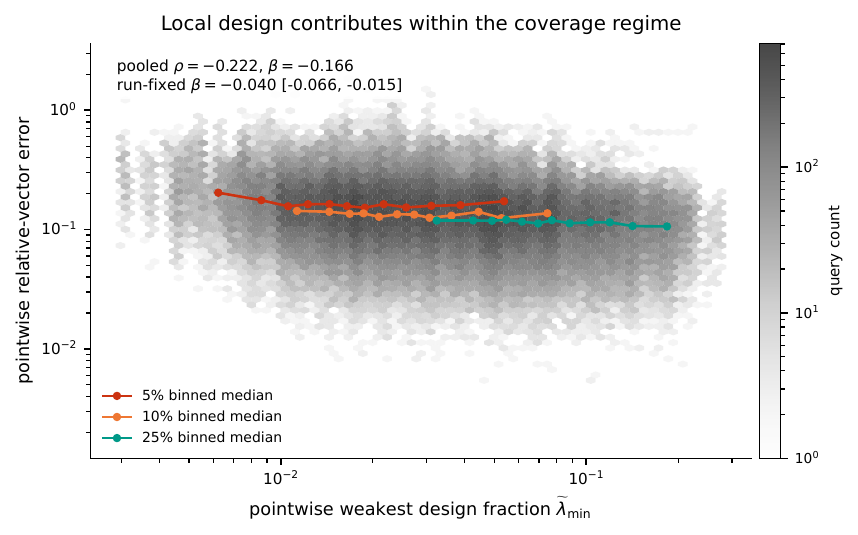}
  \caption{\textbf{Local design modulates gradient error inside the coverage regime.} Hexagonal density shows all 206,550 locked point queries from the 30 minimal-coordinate fields; coloured curves are equal-count bin medians within each coverage level. The pooled exponent combines the coverage transition, whereas the run-fixed exponent removes every run mean before fitting and remains negative under complete-run bootstrap resampling.}
  \label{si:fig:eq5_pointwise}
\end{figure}

The pooled pointwise association over 206,550 queries is weaker
($\rho=-0.222$), and coverage-stratified pointwise coefficients are $-0.065$,
$-0.042$ and $-0.059$ at 25\%, 10\% and 5\%. We do not use this pooled value as
a pointwise prediction claim. Equation~\ref{si:eq:supp_learned_gradient_bound}
contains relational mismatch, curvature and neighbourhood scale in its numerator;
$\widetilde{\lambda}_{\min}$ alone is therefore not expected to order every
query. After removing the mean log design and log error of every run, the
pointwise exponent is $-0.040$ (run-cluster 95\% interval $-0.066$ to
$-0.015$), with 24 of 30 individual slopes negative
(Supplementary Fig.~\ref{si:fig:eq5_pointwise}). The controlled result is that
weakening the run-level design, including among equal-count sparse subsets,
systematically weakens the learned gradient, while local directional support
adds a smaller resolved contribution inside each coverage regime.

Surface receivers have identically zero vertical directional isotropy, whereas
the three-dimensional source designs retain a positive smallest eigenvalue.
This endpoint rank asymmetry is geometric, not an outcome of optimization.

\section{Independent and Structured Source Subsets}

Ten independent training-source subsets were run at each key sparse fraction.
Coverage seed 20260806 is the original nested subset; the other seeds are
31001--31009. All use optimization seed 101. The original subset at each level
also has optimization seeds 101--105. Table~\ref{si:tab:repeated} separates
sampling-realization variability from optimization variability.

\begin{table}[ht]
\centering
\caption{Independent-subset results. Values are means $\pm$ sampling SD across ten subsets; the last column is SD across five optimization seeds for the original subset.}
\label{si:tab:repeated}
\small
\begin{adjustbox}{max width=\linewidth}\begin{tabular}{ccccc}
\toprule
Sources & Time MAE (s) & Angle (deg) & Relative vector & Angle SD, optimization \\
\midrule
25\% & $0.735\pm0.011$ & $5.75\pm0.22$ & $0.116\pm0.004$ & 0.053 \\
10\% & $0.803\pm0.015$ & $6.69\pm0.61$ & $0.135\pm0.011$ & 0.152 \\
5\%  & $0.983\pm0.121$ & $8.28\pm1.16$ & $0.167\pm0.023$ & 0.851 \\
\bottomrule
\end{tabular}\end{adjustbox}
\end{table}

The joint adequacy fractions are 1.0, 1.0 and 0.5 at 25, 10 and 5\%.
At a fixed fraction, which sources are retained produces more variability than
optimization seed in median angle and relative-vector error. This conclusion is
conditional on the tested optimizer and five optimization replicates.

Structured screens probe different rank and extrapolation failures. At 25\%,
shallow-only sampling produces 2.559~s/$30.72\degree$, restricted spatial
extent 2.350~s/$11.28\degree$, a spatial cluster 1.269~s/$10.18\degree$, a
depth gap 0.939~s/$9.61\degree$, and boundary-biased sampling
1.224~s/$8.68\degree$. The 25\% azimuth wedge is a path filter rather than a
source subset and gives 3.310~s/$12.08\degree$. Their selected penalty, fill
distance and local spectra do not produce a common error curve. The absence of
a scaling collapse is retained as a constraint on the main claim.

\section{Annotation Composition}

The historical mixed-pick cache contains 172,534 event--station records associated with 7,519 events. Among selected P targets, 102,204 are manual and 50,904 are automatic, so automatic picks contribute 33.25\% of P values. The 121,630 selected S targets are manual. Those historical travel-time targets are therefore described as annotated arrivals. The new manual-only tomography experiment rebuilds targets before training, excludes every automatic pick, and is documented separately below. The U/D labels used in the focal experiment are a separate set of manual first-motion polarities.

For repeated event--station phase records, the selection order is highest score, manual status and earliest valid arrival. Records with invalid origin time or travel times outside 0--120~s are excluded. Catalogue hypocentres and origin times are upstream seismological products and may retain velocity-model dependence even though the neural-network objective uses no velocity field.

\section{Architecture Parameterization Details}

The primary evidence model contains 532,994 trainable parameters and maps only
the six normalized receiver/source coordinates to two unbounded outputs.  It
has no engineered pair geometry, reference travel time, causal output transform
or finite output range.  An 80-s multiplicative output scale is used only as a
numerical parameterization; unlike a hyperbolic-tangent bound, it does not
restrict the predicted time.  The implementation retains seven identically
zero input columns so that the residual field has the same parameter count as
the structured ablations; these constants contain no pair information.

All residual-MLP variants in the following architectural ablation contain 532,994 trainable parameters and share the same optimizer and data. The full model uses 13 inputs: normalized receiver and source coordinates, coordinate difference, midpoint and distance. It adds learned corrections bounded by 10~s to homogeneous P and S--P reference times and applies a smooth causal output head.

\begin{table}[ht]
\centering
\caption{Architectural information retained by each equal-parameter variant.}
\small
\begin{adjustbox}{max width=\linewidth}\begin{tabular}{lcccc}
\toprule
Variant & Coordinates & Engineered geometry & Homogeneous reference & Causal head \\
\midrule
Full & yes & yes & yes & yes \\
Coordinates only & yes & no & yes & yes \\
No reference & yes & yes & no & yes \\
No causal head & yes & yes & yes & no \\
Legacy bounded minimal & yes & no & no & no \\
Primary coordinate field & yes & no & no & no \\
\bottomrule
\end{tabular}\end{adjustbox}
\end{table}

Reference-free and legacy minimal-prior variants use an 80-s bounded learned-output scale instead of the full model's 10-s correction scale. The primary minimal coordinate field removes that bound. The distinction preserves exact reproducibility of the earlier equal-parameter ablation while making the main evidence independent of a prespecified travel-time range.

In the full model, the median homogeneous-reference gradient norm is 1.094 times the complete P-gradient norm and the learned-correction norm is 0.694 times the complete norm. Their vector sum determines the final source gradient. No minimal-prior test prediction violates $T_P>0$ or $T_S>T_P$, even though these inequalities are not imposed by its head.

\begin{figure}[t]
  \centering
  \includegraphics[width=0.99\textwidth]{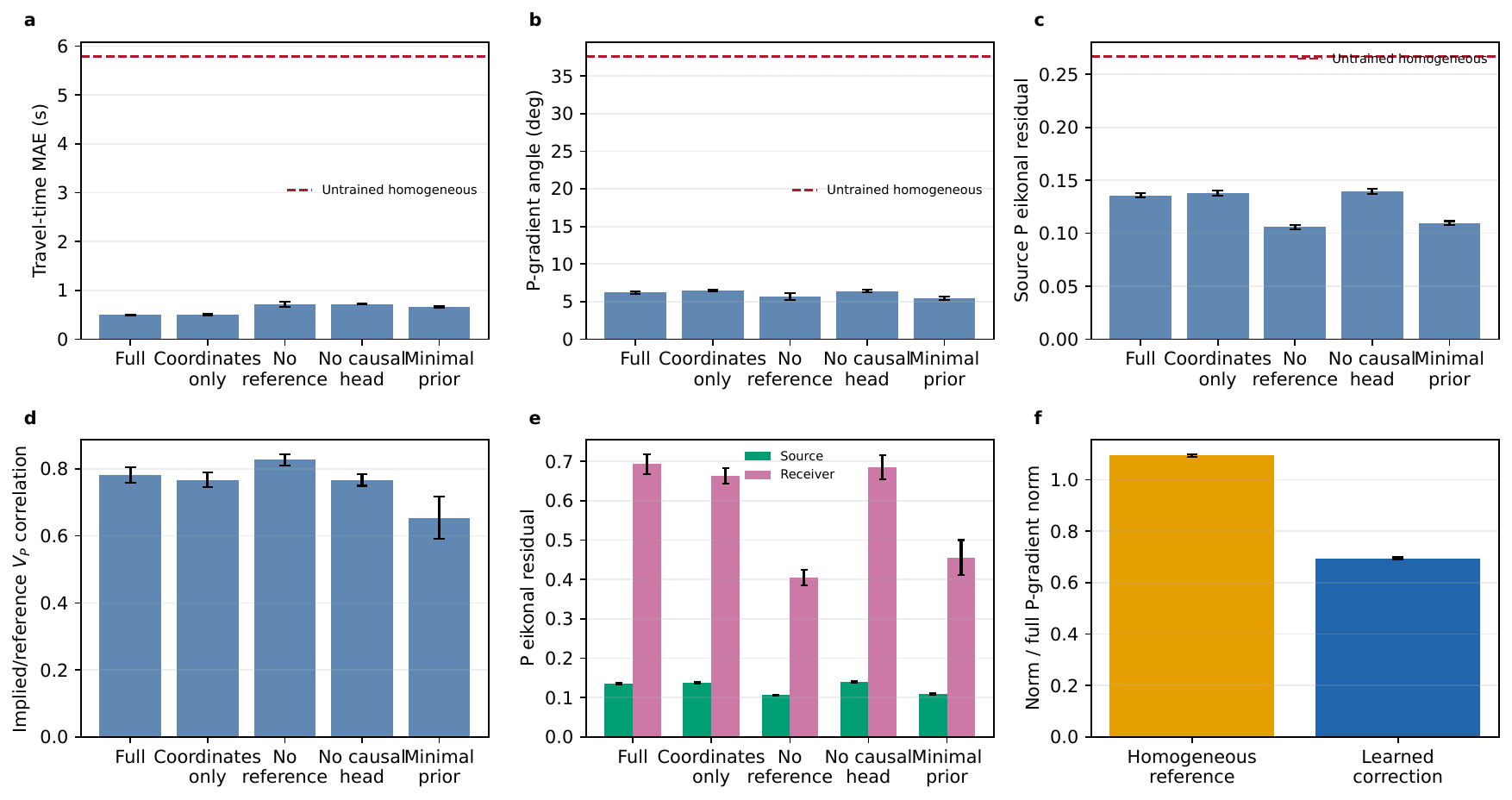}
  \caption{\textbf{Extended architecture and endpoint diagnostics.} Five-seed value, source-gradient, eikonal, velocity-correlation, source--receiver and gradient-decomposition metrics for the full and equal-parameter architectural variants. Reference-free variants also use a larger learned-output scale, as described above.}
  \label{si:fig:extended_architecture}
\end{figure}

\section{Endpoint-eikonal Selection and Optimization Diagnostics}

The endpoint-eikonal weight was selected independently at each source coverage from
$\{0.03,\allowbreak0.1,\allowbreak0.3,\allowbreak1,\allowbreak3,\allowbreak10,\allowbreak30,\allowbreak100\}$ using validation combined MAE. For the primary
minimal-coordinate field, selected values were 10, 10, 100 and 10 at 100, 25,
10 and 5\% coverage, respectively. The selection run used optimization seed 101. The final five-seed summaries
also contain a separately launched seed-101 fit at the selected weight, so the reported
intervals summarize optimization variability but are not fully independent of
hyperparameter development.

\begin{table}[ht]
\centering
\caption{Primary minimal-coordinate value-only and endpoint-eikonal results. Entries give seed-averaged travel-time MAE (s), median FMM angular error (degrees) and relative-vector error, followed by the fraction of five seeds satisfying the joint rule.}
\label{si:tab:primary_pinn}
\small
\begin{adjustbox}{max width=\linewidth}\begin{tabular}{lcc}
\toprule
Coverage & Value only & Endpoint eikonal \\
\midrule
100\% & 0.717/5.34/0.108 (5/5) & 0.721/5.09/0.102 (5/5) \\
25\%  & 0.725/5.64/0.113 (5/5) & 0.726/5.47/0.108 (5/5) \\
10\%  & 0.824/6.68/0.138 (5/5) & 0.770/5.71/0.112 (5/5) \\
5\%   & 1.076/8.63/0.172 (1/5) & 0.971/7.17/0.142 (4/5) \\
\bottomrule
\end{tabular}\end{adjustbox}
\end{table}

The dimensionless loss places P and S endpoint residuals on a common scale.
At 100 and 25\% coverage, paired time intervals cross zero and the angle/vector
changes are small; the joint qualification is unchanged. At 10\%, the equation
refines all three quantities, but the value-only field already qualifies in
every seed. At 5\%, angle and vector differences are resolved and qualification
rises from one to four of five seeds, whereas the time interval crosses zero.
This is partial derivative compensation without a resolved scalar gain. Thus no tested coverage shows that the equation is
required for the physical geometry established in the supported data regime.
Run-level traces and selection JSON files are stored with the revision outputs.

\section{Independent Same-Data Confirmation of Metric Recovery}

The scalar-relational result was discovered after the reciprocal-augmentation
screen, so it was not promoted directly from that test. Instead, its complete
configuration was frozen before generating a new random three-dimensional
velocity realization, a new scrambled-Sobol source set and a jittered surface
array. The relational-only and PINN fields then used exactly the same 184,469
P/S scalar records, 1,434/307/307 source-disjoint split, minimal-coordinate
residual-SiLU architecture, 100-epoch budget and seeds 101--103. The relational
field received same-station finite differences constructed from measured scalar
values, with frozen weight 10 and no derivative labels; the PINN
received exact endpoint velocities through a weight-1 eikonal term evaluated on
10\% of each batch. No velocity result selected a checkpoint or parameter.

\begin{table}[ht]
\centering
\caption{Prospectively locked independent same-data confirmation. Velocity metrics use the strict common support of 307 held-out events and aggregate inverse source-gradient norms over receivers and three optimization seeds.}
\label{si:tab:metric_confirmation}
\small
\begin{adjustbox}{max width=\linewidth}\begin{tabular}{lrrrrrr}
\toprule
Method & Time MAE & Vp MAE & Vp Pearson & Vp detrended $\rho$ & Vs MAE & Vs detrended $\rho$ \\
 & (s) & (km~s$^{-1}$) & & & (km~s$^{-1}$) & \\
\midrule
Scalar relational only & 0.0871 & 0.0537 & 0.9638 & 0.8861 & 0.0274 & 0.9034 \\
Endpoint-eikonal PINN & 0.0913 & 0.0575 & 0.9598 & 0.8735 & 0.0306 & 0.8876 \\
\bottomrule
\end{tabular}\end{adjustbox}
\end{table}

The event-paired PINN-minus-relational Vp MAE difference is
0.00378~km~s$^{-1}$, with a 20,000-replicate bootstrap 95\% interval of
0.00066--0.00696~km~s$^{-1}$. The corresponding Vs difference is
0.00320~km~s$^{-1}$ (0.00184--0.00456~km~s$^{-1}$). All three frozen Vp
conditions pass: lower relational MAE, a bootstrap lower bound above zero and
depth-detrended rank correlation no lower than the PINN value. Independent
verification reproduces the headline errors from the stored event fields and
confirms the protocol, data and checkpoint hashes, the matched training inputs
and the absence of test access during training.

After that prospectively locked comparison was opened, its architecture,
training budget, seeds and loss weights were held fixed and the two missing
factorial cells were completed: scalar values alone, and scalar relations plus
endpoint eikonal supervision. These cells are a fixed post-test confirmation,
not part of the original prospective success gate. All twelve checkpoints are
selected by scalar validation MAE and evaluated on the same 307-event support.

\begin{table}[ht]
\centering
\caption{Matched four-group metric-identifiability comparison. Every group uses identical scalar data, architecture, 100-epoch budget and seeds. Relations are same-station finite differences of measured scalar values; eikonal groups additionally receive true endpoint velocity.}
\label{si:tab:metric_factorial}
\small
\begin{adjustbox}{max width=\linewidth}\begin{tabular}{lccccc}
\toprule
Training information & Time MAE & Vp MAE & Vp detrended $\rho$ & Vs MAE & Vs detrended $\rho$ \\
 & (s) & (km~s$^{-1}$) & & (km~s$^{-1}$) & \\
\midrule
Values only & 0.0905 & 0.0574 & 0.877 & 0.0306 & 0.886 \\
Values + scalar relations & 0.0871 & 0.0537 & 0.886 & 0.0274 & 0.903 \\
Values + endpoint eikonal & 0.0913 & 0.0575 & 0.874 & 0.0306 & 0.888 \\
Values + relations + eikonal & 0.1130 & 0.0636 & 0.864 & 0.0366 & 0.862 \\
\bottomrule
\end{tabular}\end{adjustbox}
\end{table}

Relative to values alone, scalar relations reduce Vp/Vs absolute error by
0.00363/0.00313~km~s$^{-1}$, with paired 95\% intervals
0.00140--0.00586 and 0.00191--0.00439~km~s$^{-1}$. The corresponding eikonal
changes are $-0.00015/-0.00007$~km~s$^{-1}$, with intervals spanning zero
($-0.00403$--0.00380 and $-0.00165$--0.00152~km~s$^{-1}$). Adding both
auxiliary terms degrades rather than improves this fixed setting. The factorial
therefore shows that the physical metric readout succeeds from scalar-value
training, that scalar secants improve its accuracy, and that the tested
endpoint-eikonal loss has no statistically resolved standalone benefit at full
coverage.

\subsection{Prospectively Locked Four-Cell Confirmation}

A second confirmation prospectively froze the complete four-cell factorial
before a new velocity realization was generated. The protocol, generator
family, realization and optimization seeds,
acquisition, split, loss weights, readout, 20,000-replicate hierarchical
bootstrap and success gates were written and hashed first (protocol SHA-256
\texttt{1c709736a3a9da4dde4d87ab77ab6aea87bd43eafbad631922cd92213c07137f}).
All four cells use the identical scalar travel-time cache and the same
minimal-coordinate architecture, 100-epoch budget and seeds 101--103. Test
velocity truth was opened once, after all 12 scalar-validation-selected
checkpoints had been completed and hashed. The common test contains 307
events and 27,239 event--station paths. This is internal test sealing within the
author project, using a new realization of the previously specified generator
family. Protocol hashing establishes the identity of the archived specification;
the accompanying execution and test-access records document its order of use.
No public registration or independent external-team replication is claimed.

\begin{table}[ht]
\centering
\caption{Test-sealed four-cell confirmation on a new heterogeneous velocity realization. All methods use identical scalar values, acquisition, split, architecture, budget and optimization seeds.}
\label{si:tab:blind_metric_factorial}
\small
\begin{adjustbox}{max width=\linewidth}\begin{tabular}{lccccc}
\toprule
Training information & Time MAE & Vp MAE & Vp detrended $\rho$ & Vs MAE & Vs detrended $\rho$ \\
 & (s) & (km~s$^{-1}$) & & (km~s$^{-1}$) & \\
\midrule
Values only & 0.0835 & 0.0606 & 0.791 & 0.0258 & 0.862 \\
Values + scalar relations & 0.0889 & 0.0567 & 0.803 & 0.0220 & 0.890 \\
Values + endpoint eikonal & 0.0877 & 0.0534 & 0.787 & 0.0245 & 0.851 \\
Values + relations + eikonal & 0.0818 & 0.0526 & 0.804 & 0.0208 & 0.896 \\
\bottomrule
\end{tabular}\end{adjustbox}
\end{table}

Values alone pass every predeclared emergence gate: Vp/Vs MAE are below
0.080/0.050~km~s$^{-1}$ and both depth-detrended rank correlations exceed
0.75. Scalar relations reduce Vp and Vs absolute error by 0.00392
(hierarchical-bootstrap 95\% interval 0.00089--0.00882) and
0.00375~km~s$^{-1}$ (0.00271--0.00763), respectively. These resolved
refinements use measured scalar secants and contain neither velocity nor
equation supervision.

Endpoint-eikonal supervision changes Vp/Vs MAE relative to values alone by
$-0.00727/-0.00130$~km~s$^{-1}$; the corresponding intervals,
$[-0.02871,0.00046]$ and $[-0.01453,0.00540]$~km~s$^{-1}$, both cross zero.
The experiment therefore resolves neither an eikonal advantage nor formal
equivalence. For the prospectively specified values-only-minus-eikonal contrasts, the
upper limits are 0.02871 and 0.01453~km~s$^{-1}$, exceeding the
0.010/0.006~km~s$^{-1}$ margins; non-inferiority is therefore unresolved. The
combined cell gives the smallest point
errors; relative to the eikonal cell, its Vp gain is unresolved and its Vs gain
is 0.00371~km~s$^{-1}$ (0.00053--0.01063). The affirmative blind result is the
one tested directly by the absolute gates: an untouched value-only field
recovers both local metric structures before any governing equation enters its
objective. Scalar relations resolve a data-only refinement. The eikonal relation is used
for all post-training velocity readouts; adding its residual to training has
no statistically resolved standalone advantage at full support.

\begin{figure}[t]
  \centering
  \includegraphics[width=0.94\textwidth]{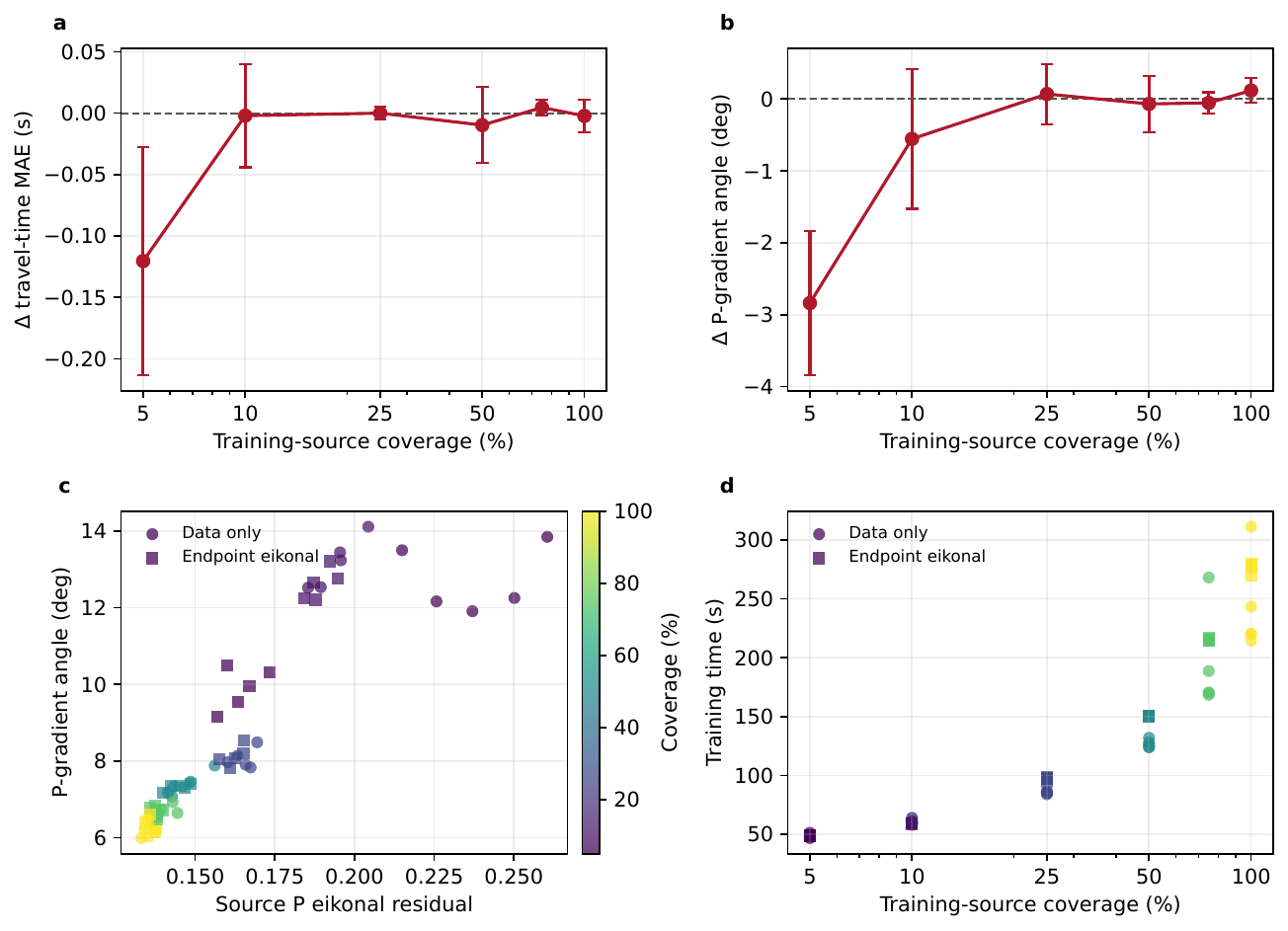}
  \caption{\textbf{Secondary structured-field endpoint-eikonal comparison.} Paired time and angular changes, eikonal-residual--direction relation and equal-epoch training cost in the earlier structured transfer field. The primary minimal-coordinate coverage comparison is reported in Table~\ref{si:tab:primary_pinn}; Fig.~3d of the main text shows the separate same-data metric comparison.}
  \label{si:fig:extended_pinn}
\end{figure}

For completeness, Fig.~\ref{si:fig:extended_pinn} retains the earlier structured
transfer-field screen. Its architecture and selected weights differ from the
primary minimal-coordinate comparison and it is not used to support the central
claim.

\section{Event-Correlated Catalogue-Error Stress}

The controlled stress uses the same medium, records and source-disjoint split as
the independent metric confirmation. The 184,469 event--station records, 225
stations, architecture, scalar-relational objective, 100-epoch budget,
checkpoint rule and thresholds remain fixed. One horizontal, depth and
origin-time offset is drawn for an event and copied to every one of its station
records. The same-station relation graph is deterministically reconstructed
from each perturbed catalogue coordinate set, because those are the coordinates
available to the learner; no FMM-gradient result selects a neighbour or
checkpoint. Three perturbation realizations are crossed with three optimization
seeds. All 18 checkpoints are complete before the fixed 96-receiver gradient
evaluation is generated.

Neither the annotation JSON nor the China phase archive provides formal
per-event uncertainty fields. The 2/5-km horizontal and 5/10-km depth levels
are therefore labelled stresses. The controlled 0.04/0.10-s origin-time scales
equal the median and rounded 90th percentile absolute origin disagreement among
18 exact California catalogue--ComCat associations passing the fixed 2-s/15-km
audit; they are empirical association scales, not formal standard errors.
Table~\ref{si:tab:catalogue_stress_absolute} gives the absolute truth-known
results. Every run retains all 307 test events and all nine runs at both stress
levels satisfy the pooled angle/vector criterion. Event-level and path-level
qualification distinguish moderate stability from the strong-error transition:
they are 0.958/0.811 at moderate stress and 0.619/0.503 at severe stress.

\begin{table}[ht]
\centering
\caption{Truth-known event-correlated catalogue stress. Values are medians across three clean or nine perturbed runs; physical MAE is evaluated against unperturbed FMM times.}
\label{si:tab:catalogue_stress_absolute}
\small
\setlength{\tabcolsep}{4pt}
\begin{adjustbox}{max width=\linewidth}\begin{tabular}{lrrrrrr}
\toprule
Stress (horizontal/depth/time) & Physical MAE & Catalogue-fit MAE & Angle & Vector error & Event qualified & Vp MAE \\
 & (s) & (s) & (deg) & & fraction & (km~s$^{-1}$) \\
\midrule
None & 0.093 & 0.093 & 1.64 & 0.034 & 1.000 & 0.053 \\
2/5~km/0.04~s & 0.316 & 0.317 & 5.27 & 0.105 & 0.958 & 0.159 \\
5/10~km/0.10~s & 0.669 & 0.672 & 9.14 & 0.187 & 0.619 & 0.236 \\
\bottomrule
\end{tabular}\end{adjustbox}
\end{table}

Paired event/realization/seed resampling quantifies degradation from the
same-seed clean field (Table~\ref{si:tab:catalogue_stress_paired}). Moderate
stress leaves the pooled geometry inside the joint threshold but
raises Vp error by 0.104~km~s$^{-1}$. Severe stress approaches the angular
threshold and reduces event qualification by 0.391. Catalogue error therefore
sets a measurable scale for recovery rather than changing the origin of the
geometry identified in the clean truth-known experiment.

\begin{table}[ht]
\centering
\caption{Paired change from the unperturbed field under event-correlated stress. Intervals use 20,000 hierarchical bootstrap replicates over events, perturbation realizations and optimization seeds.}
\label{si:tab:catalogue_stress_paired}
\small
\begin{adjustbox}{max width=\linewidth}\begin{tabular}{lrr}
\toprule
Metric & 2/5~km/0.04~s & 5/10~km/0.10~s \\
\midrule
Physical time MAE (s) & 0.219 (0.204--0.236) & 0.586 (0.554--0.619) \\
P-gradient angle (deg) & 3.63 (2.95--4.17) & 8.10 (7.47--8.80) \\
Relative-vector error & 0.071 (0.060--0.081) & 0.161 (0.149--0.173) \\
Event qualification & $-0.049$ ($-0.089$--$-0.013$) & $-0.391$ ($-0.456$--$-0.325$) \\
Vp absolute error (km~s$^{-1}$) & 0.104 (0.090--0.120) & 0.182 (0.153--0.214) \\
\bottomrule
\end{tabular}\end{adjustbox}
\end{table}

\begin{figure}[t]
  \centering
  \includegraphics[width=0.98\textwidth]{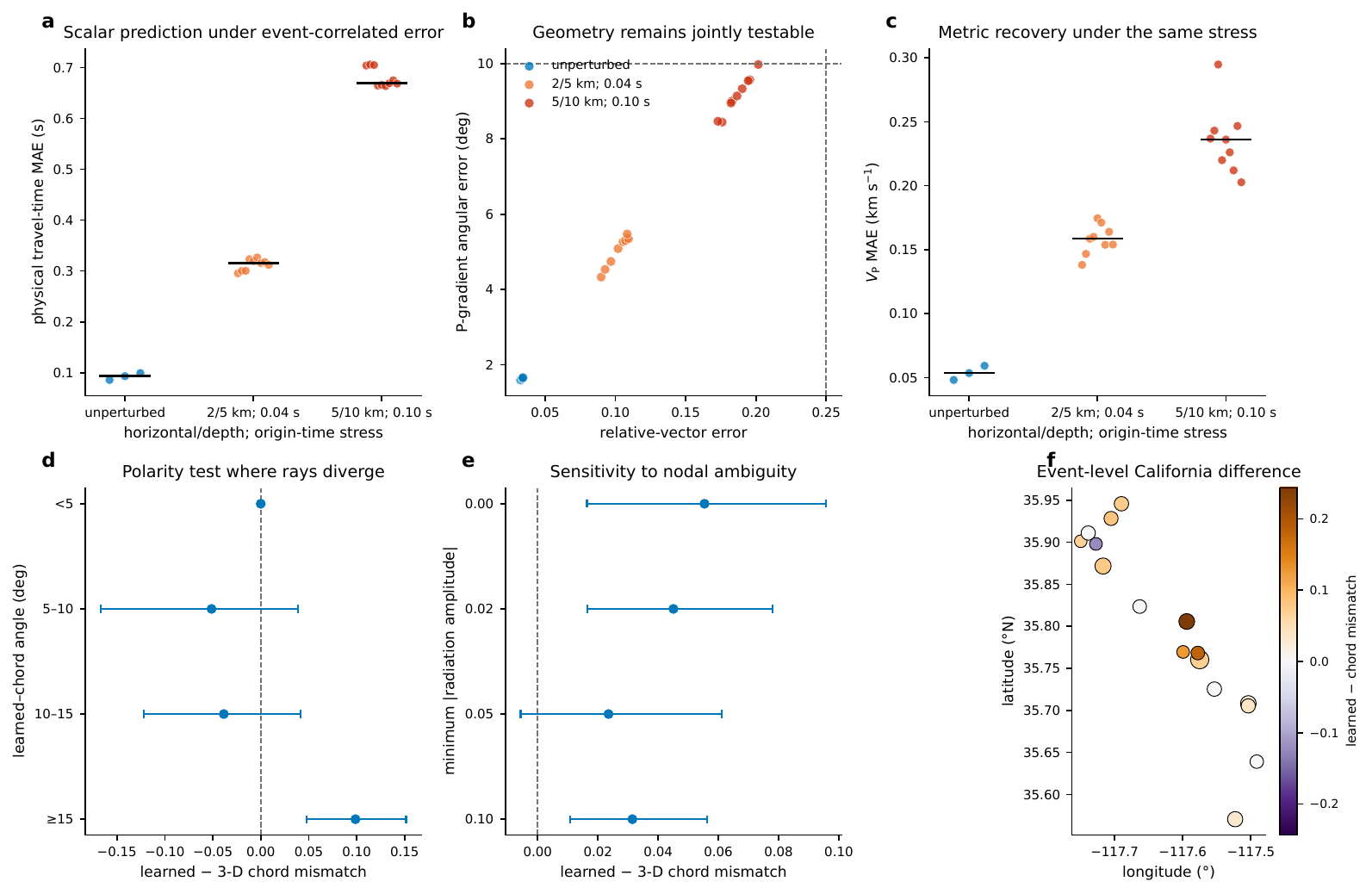}
  \caption{\textbf{Catalogue-error robustness and California attribution.} (a) Physical P/S time MAE for three unperturbed and nine runs at each event-correlated stress level; black segments mark medians. (b) P-gradient angle and relative-vector error for the same runs, with the frozen $10\degree$/0.25 joint thresholds. (c) Event-aggregated $V_\mathrm{P}$ error. The stress labels give horizontal radial RMS/depth standard deviation and origin-time standard deviation. (d) Fixed-tensor California learned-minus-three-dimensional-chord mismatch in prespecified learned--chord angle bins. (e) The same difference as low-radiation-amplitude polarities are removed under every ray. Error bars in (d,e) are 95\% event-bootstrap intervals. (f) Event-level learned-minus-chord mismatch; symbol area scales with common polarity count. Positive differences favour the chord.}
  \label{si:fig:catalogue_robustness}
\end{figure}

\section{Full Numerical Comparisons}

The generated tables below retain the complete five-seed results for the secondary
structured transfer model and its architectural ablations, including scalar
MAE, angular error, relative-vector error and source eikonal residual. They are
reported as transfer controls rather than as the primary evidence model.

\begin{table}[t]
\centering
\caption{Coverage-controlled synthetic benchmark. Entries are mean $\pm$ standard deviation across five optimization seeds. Gradient metrics use the out-of-loss reciprocal-FMM source-gradient set.}
\label{si:tab:coverage}
\small
\begin{adjustbox}{max width=\linewidth}\begin{tabular}{lrrrr}
\toprule
Method and source coverage & MAE (s) & Angle ($^\circ$) & Vector error & Source eikonal \\
\midrule
Data only, 100\% & 0.503 $\pm$ 0.008 & 6.20 $\pm$ 0.14 & 0.135 $\pm$ 0.005 & 0.136 $\pm$ 0.002 \\
Data only, 75\% & 0.564 $\pm$ 0.005 & 6.73 $\pm$ 0.12 & 0.148 $\pm$ 0.005 & 0.141 $\pm$ 0.003 \\
Data only, 50\% & 0.661 $\pm$ 0.029 & 7.38 $\pm$ 0.31 & 0.168 $\pm$ 0.011 & 0.146 $\pm$ 0.006 \\
Data only, 25\% & 0.883 $\pm$ 0.013 & 8.07 $\pm$ 0.26 & 0.183 $\pm$ 0.006 & 0.165 $\pm$ 0.004 \\
Data only, 10\% & 1.079 $\pm$ 0.018 & 13.17 $\pm$ 0.67 & 0.290 $\pm$ 0.018 & 0.194 $\pm$ 0.007 \\
Data only, 5\% & 1.322 $\pm$ 0.053 & 12.73 $\pm$ 0.87 & 0.308 $\pm$ 0.023 & 0.238 $\pm$ 0.018 \\
\addlinespace
Endpoint eikonal, 100\% & 0.501 $\pm$ 0.008 & 6.32 $\pm$ 0.23 & 0.136 $\pm$ 0.005 & 0.136 $\pm$ 0.001 \\
Endpoint eikonal, 75\% & 0.568 $\pm$ 0.008 & 6.67 $\pm$ 0.16 & 0.145 $\pm$ 0.003 & 0.138 $\pm$ 0.001 \\
Endpoint eikonal, 50\% & 0.651 $\pm$ 0.010 & 7.32 $\pm$ 0.08 & 0.163 $\pm$ 0.003 & 0.145 $\pm$ 0.003 \\
Endpoint eikonal, 25\% & 0.883 $\pm$ 0.014 & 8.14 $\pm$ 0.27 & 0.182 $\pm$ 0.006 & 0.162 $\pm$ 0.003 \\
Endpoint eikonal, 10\% & 1.077 $\pm$ 0.023 & 12.61 $\pm$ 0.41 & 0.277 $\pm$ 0.010 & 0.189 $\pm$ 0.004 \\
Endpoint eikonal, 5\% & 1.201 $\pm$ 0.034 & 9.89 $\pm$ 0.55 & 0.222 $\pm$ 0.013 & 0.164 $\pm$ 0.006 \\
\bottomrule
\end{tabular}\end{adjustbox}
\end{table}

\begin{table}[t]
\centering
\caption{Matched architecture-parameterization controls at full synthetic coverage. Neural entries are mean $\pm$ standard deviation across five optimization seeds; the homogeneous baseline is untrained. Reference-free rows use an 80-s learned-output scale rather than the full model's 10-s correction scale.}
\label{si:tab:architecture}
\small
\begin{adjustbox}{max width=\linewidth}\begin{tabular}{lrrr}
\toprule
Model & MAE (s) & Angle ($^\circ$) & Vector error \\
\midrule
Full & 0.503 $\pm$ 0.008 & 6.20 $\pm$ 0.14 & 0.135 $\pm$ 0.005 \\
Coordinates only & 0.507 $\pm$ 0.013 & 6.50 $\pm$ 0.15 & 0.142 $\pm$ 0.003 \\
No homogeneous reference & 0.714 $\pm$ 0.054 & 5.71 $\pm$ 0.46 & 0.119 $\pm$ 0.008 \\
No causal head & 0.723 $\pm$ 0.006 & 6.41 $\pm$ 0.15 & 0.137 $\pm$ 0.003 \\
Minimal prior & 0.661 $\pm$ 0.017 & 5.42 $\pm$ 0.25 & 0.117 $\pm$ 0.003 \\
Homogeneous, untrained & 5.791 & 37.62 & 0.728 \\
\bottomrule
\end{tabular}\end{adjustbox}
\end{table}

\begin{figure}[t]
  \centering
  \includegraphics[width=0.94\textwidth]{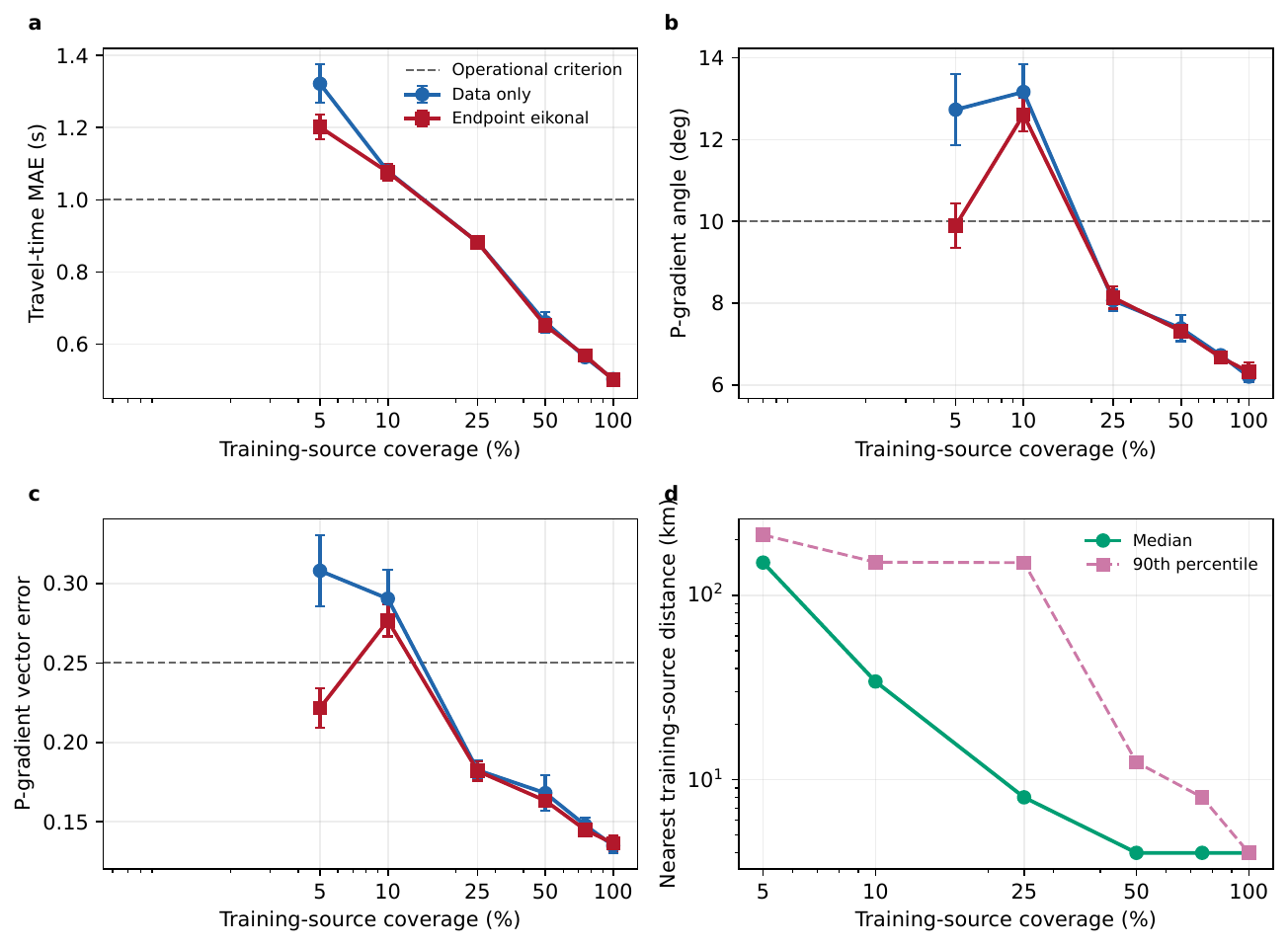}
  \caption{\textbf{Extended coverage comparison.} Travel-time, angular, vector and nearest-training-source metrics for data-only and validation-selected endpoint-eikonal fields.}
  \label{si:fig:extended_coverage}
\end{figure}

\section{Exactly Matched Annotation and FMM Targets}

Matched synthetic targets were computed for every annotated event--station row while retaining row order, coordinates, identifiers, missing-phase masks, partitions and seeds. The matched cache differs only in its travel-time values. Sixty event depths and 171,749 receiver depths lie above the MUSCAL grid surface and were evaluated at the surface node. This convention affects target comparison near the upper boundary but is identical across the paired runs.

\begin{figure}[t]
  \centering
  \includegraphics[width=0.86\textwidth]{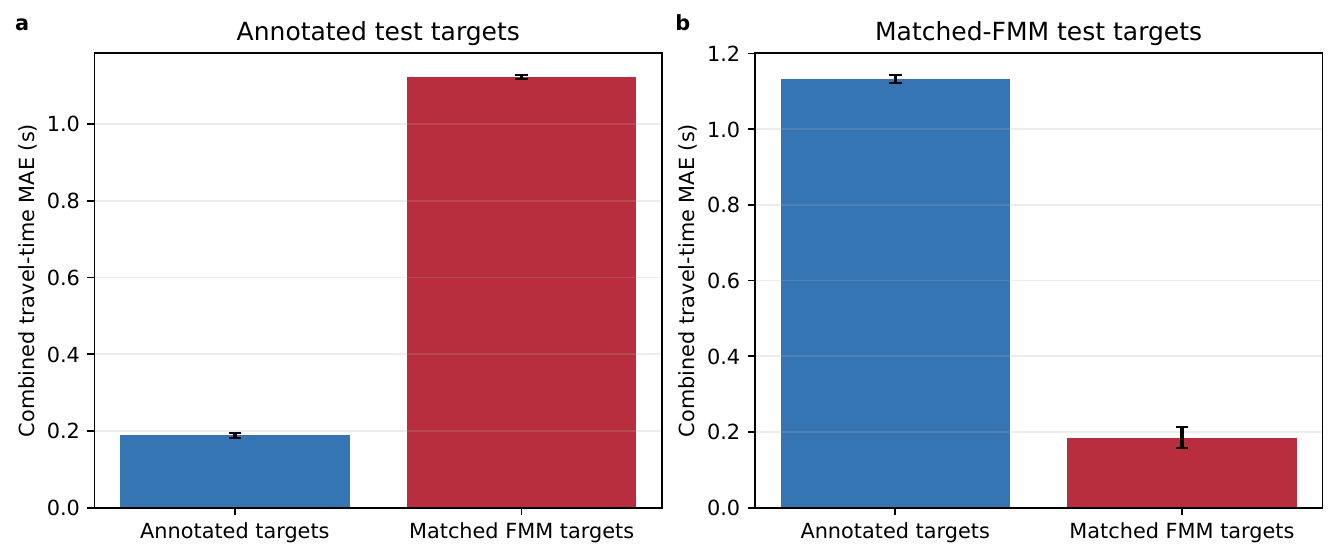}
  \caption{\textbf{Matched-target matrix.} Networks trained on annotated or MUSCAL/FMM targets use identical event--station inputs, phase masks, partitions, sample count, seeds and optimization. Bars show five-seed test MAE on annotated (a) and matched-FMM (b) targets.}
  \label{si:fig:matched}
\end{figure}

Annotation-trained fields reach $0.189\pm0.006$~s MAE on annotations, and FMM-trained
fields reach $0.185\pm0.027$~s on FMM targets. Cross-target MAE is $1.132\pm0.011$~s
for annotation training evaluated on FMM values and $1.122\pm0.005$~s in the reverse
direction. Both target systems are learnable; their difference reflects catalogue,
velocity, gridding, station-correction and picking conventions.

\section{Coverage Stress Tests}

Station thinning was evaluated with one fixed seed while retaining all training sources.
Source-gradient angular error rises from $6.41\degree$ at 75\% stations to
$6.83\degree$, $7.03\degree$, $8.42\degree$ and $9.37\degree$ at 50, 25, 10 and
5\%, respectively. Travel-time MAE crosses 1~s at 10\% station coverage.

Single-seed source-selection screens retained 99 of 397 source groups for random,
shallow-depth and spatial-extent 25\% subsets. Their MAE/angular errors were
0.897~s/$7.83\degree$, 2.559~s/$30.72\degree$ and 2.350~s/$11.28\degree$.
An azimuth-wedge path filter retained 132,954 paths from 346 source groups and produced
3.310~s/$12.08\degree$. The wedge is a path-direction restriction rather than a
25\% source subset. These screens show that depth, spatial extent and path azimuth affect
derivative recovery beyond sample count.

\section{China Four-Phase Gradient and Depth-Support Audit}

The depth-balanced China field is selected without derivative or crustal-thickness targets. Table~\ref{si:tab:china_phase} reports its natural-distribution locked test errors. All four heads improve on the frozen robust scalar regression, with the largest relative gain for Pn.

\begin{table}[ht]
\centering
\caption{Event-disjoint China four-phase scalar accuracy.}
\label{si:tab:china_phase}
\small
\begin{adjustbox}{max width=\linewidth}\begin{tabular}{lrrrr}
\toprule
Phase & Test records & Coordinate-field MAE (s) & Baseline MAE (s) & Improvement \\
\midrule
Pg & 100,194 & 0.398 & 0.436 & 8.7\% \\
Sg & 99,038  & 0.578 & 0.630 & 8.2\% \\
Pn & 65,651  & 0.934 & 1.204 & 22.4\% \\
Sn & 23,822  & 1.631 & 1.838 & 11.2\% \\
\bottomrule
\end{tabular}\end{adjustbox}
\end{table}

The fixed component audit keeps the 62,086 derivative-qualified
held-out relations, 194-cell comparison mask, 2$\degree$ grid and smoothing
weight fixed. Table~\ref{si:tab:china_components} reports the resulting separation
between scalar-time and gradient-magnitude contributions. The fixed-magnitude
variant uses regional slownesses $p_c=0.16910$ and $p_m=0.12443$~s~km$^{-1}$
from robust Pg/Pn moveout fits on training arrivals. Specifically, the analysis
reads the reciprocals of \texttt{crustal\_velocity\_km\_s} and
\texttt{mantle\_velocity\_km\_s} in the training moveout record; it does not
take medians of the neural gradient magnitudes. The common observation mask
remains derivative-qualified for all components, including fixed slowness.
The shuffled result summarizes 100 deterministic joint permutations of learned
$(p_c,p_m)$ pairs among paths.

\begin{table}[ht]
\centering
\caption{China Pg--Pn component audit on one common supported-cell mask. Shuffled entries are permutation medians with 5th--95th percentiles.}
\label{si:tab:china_components}
\small
\begin{adjustbox}{max width=\linewidth}\begin{tabular}{lccc}
\toprule
Readout & Spearman $\rho$ & Linear-detrended Pearson $r$ & MAE (km) \\
\midrule
Learned time + learned magnitudes & 0.778 & 0.409 & 9.21 \\
Learned time + fixed magnitudes & 0.910 & 0.732 & 8.93 \\
Learned time + shuffled magnitudes & 0.897 (0.890--0.904) & 0.706 (0.688--0.727) & 10.75 (10.68--10.83) \\
Catalogue time + fixed magnitudes & 0.909 & 0.759 & 8.69 \\
Global moveout null & $-0.022$ & $-0.078$ & 12.78 \\
\bottomrule
\end{tabular}\end{adjustbox}
\end{table}

The locked component intervention localizes the continental signal in the
scalar Pn term: fixed regional slowness gives the highest correlation among learned-time readouts,
and learned magnitudes do not exceed it or the upper shuffled-correlation
percentile. The continental map therefore establishes spatial Moho information
in scalar Pn relations, while the independent synthetic confirmation supplies
the direct test of learned local metric structure. Spatially varying learned
magnitudes modify the absolute readout, including lower error than shuffled
magnitudes, rather than generating the continental correlation.

\begin{figure}[H]
  \centering
  \includegraphics[width=0.99\textwidth]{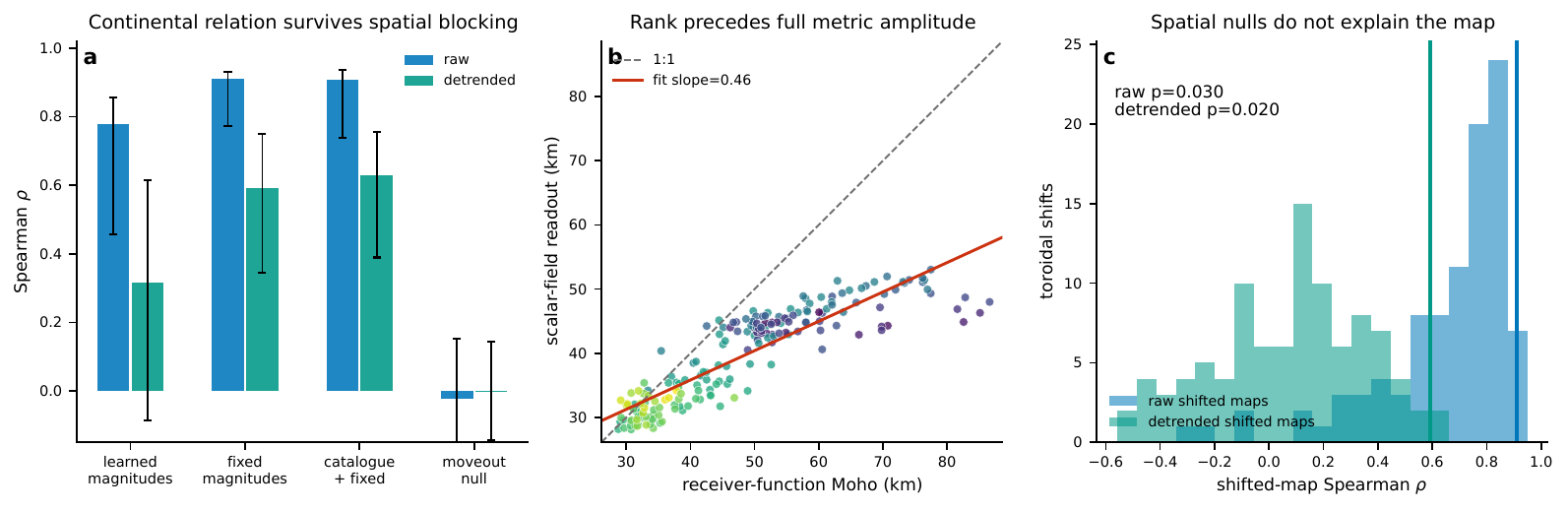}
  \caption{\textbf{The continental relation survives spatial inference.} (a) Spearman association for the learned-magnitude, fixed-magnitude, catalogue-time and global-moveout readouts; bars show 10$\degree$ spatial-block 95\% intervals for raw and quadratic-trend-removed maps. (b) The scalar-field/fixed-slowness readout against the independent receiver-function reference; the fitted slope quantifies amplitude compression without erasing rank structure. (c) Correlations obtained from all admissible non-zero toroidal translations of the prediction and support mask. Vertical lines mark the observed raw and detrended correlations.}
  \label{si:fig:china_spatial_inference}
\end{figure}

The fixed-slowness map remains associated with the independent reference after
geographical structure is made part of the null hypothesis
(Supplementary Fig.~\ref{si:fig:china_spatial_inference}). Its raw
$\rho=0.910$ has 6$\degree$ and 10$\degree$ spatial-block 95\% intervals of
0.838--0.933 and 0.772--0.931. Removing separate quadratic
longitude--latitude surfaces from prediction and reference leaves
$\rho=0.592$, with corresponding intervals of 0.401--0.729 and
0.345--0.750. Among all 98 admissible non-zero toroidal translations of the
prediction and its support mask, only one reaches the observed detrended
correlation (plus-one one-sided $p=0.020$); two reach the raw correlation
($p=0.030$). The fitted
prediction-on-reference slope is 0.457 and its standard-deviation ratio is
0.516, showing that the field identifies continental rank and pattern before
it reproduces the full reference amplitude.

The fixed-regional-slowness readout is also stable when the frozen China field
is queried at event-correlated perturbed source coordinates. One horizontal
and depth offset is shared by every record of an event, while the original
62,086-observation support mask, 194-cell evaluation mask and smoothing weight
remain fixed. Table~\ref{si:tab:china_catalogue_stress} reports three perturbation
realizations and 200 event-cluster bootstrap replicates. The China phase archive
does not contain a formal origin-time uncertainty or an independently grounded
time-error scale, so this observational test perturbs space only without
assigning an unsupported time-error distribution.

\begin{table}[ht]
\centering
\caption{Fixed-field China Moho robustness to event-correlated catalogue-coordinate stress. Horizontal values are radial RMS; intervals resample events and perturbation realization.}
\label{si:tab:china_catalogue_stress}
\small
\begin{adjustbox}{max width=\linewidth}\begin{tabular}{lccc}
\toprule
Horizontal/depth stress & Spearman $\rho$ & MAE (km) & Support retained \\
\midrule
None & 0.910 & 8.93 & 100\% \\
2/5~km & 0.907 (0.899--0.913) & 8.88 (8.79--8.96) & 98.85--98.98\% \\
5/10~km & 0.900 (0.888--0.908) & 8.50 (8.38--8.63) & 97.62--97.75\% \\
\bottomrule
\end{tabular}\end{adjustbox}
\end{table}

Literal integration of the Pn source gradient toward a deep turning point was audited separately from the endpoint-gradient readout used in the main text. Although the training partition contains approximately 528,000 branch-qualified Pn values, only ten 2$\degree$ source cells contain at least 30 Pn observations in every 10-km depth bin from the surface to 50~km. Eight support-qualified rays were traced from each cell. Twenty-one of 80 reach an apparent gradient turn before leaving catalogue-supported cell--depth bins; 58 of the remaining 59 lose support first, and one reaches the integration stop without either condition. The resolved turns cluster geographically rather than reproducing the independent receiver-function thicknesses. This confirms the sampling law: endpoint queries at observed coordinates are identified by thousands of held-out relations, whereas literal deep turning paths require new deep-coordinate coverage.

\section{Reciprocal Endpoint Role Completion}

\subsection{Truth-known role completion on identical scalar pairs}

The controlled source--receiver asymmetry provides a direct causal test of
endpoint-role support. Five no-exchange fields and five 50\%-exchange fields
use the identical medium, source-disjoint split, scalar pairs, architecture,
epoch budget and optimization seeds. Exchange reverses presentation order; it
neither duplicates a pair nor supplies a derivative, velocity or equation
target. All 6,885 test paths are paired across arms and form 85 source clusters
of 81 stations. Table~\ref{si:tab:synthetic_reciprocity_confirmation} reports
seed variability and paired source-cluster inference.

\begin{table}[H]
\centering
\caption{Truth-known synthetic endpoint-role confirmation. Values are mean $\pm$ sample SD across five seeds. The final column gives the paired swap-minus-control effect and its 20,000-replicate source-cluster 95\% interval.}
\label{si:tab:synthetic_reciprocity_confirmation}
\small
\begin{adjustbox}{max width=\linewidth}\begin{tabular}{lrrr}
\toprule
Diagnostic & No exchange & 50\% exchange & Paired effect (95\% interval) \\
\midrule
Direct P MAE (s) & $0.5231\pm0.0063$ & $0.5266\pm0.0052$ & $+0.0035$ (0.0002, 0.0070) \\
Exchanged-role P MAE (s) & $14.433\pm0.483$ & $0.5276\pm0.0053$ & $-13.905$ ($-15.441$, $-12.345$) \\
P reciprocity median (s) & $13.703\pm0.655$ & $0.1189\pm0.0278$ & $-13.585$ ($-15.883$, $-10.816$) \\
P reciprocity 95th percentile (s) & $32.837\pm1.823$ & $0.3390\pm0.0560$ & $-32.498$ ($-33.494$, $-31.298$) \\
Role-gradient angle ($\degree$) & $72.52\pm1.51$ & $1.503\pm0.159$ & $-71.02$ ($-75.37$, $-67.63$) \\
Direct role vs FMM ($\degree$) & $5.342\pm0.075$ & $5.563\pm0.165$ & $+0.222$ (0.127, 0.313) \\
Exchanged role vs FMM ($\degree$) & $72.70\pm1.37$ & $5.603\pm0.127$ & $-67.09$ ($-71.11$, $-63.95$) \\
Direct eikonal residual, mean & $0.03897\pm0.00054$ & $0.04158\pm0.00161$ & $+0.00260$ (0.00104, 0.00416) \\
Exchanged eikonal residual, mean & $5.577\pm0.377$ & $0.04212\pm0.00266$ & $-5.535$ ($-6.282$, $-4.815$) \\
\bottomrule
\end{tabular}\end{adjustbox}
\end{table}

Endpoint-role completion reduces the median reciprocity gap by 99.13\%
(95\% interval 98.92--99.25\%), the role-gradient angle by 97.93\%
(97.74--98.11\%) and exchanged-role FMM error by 92.29\%
(91.86--92.74\%). The direct query remains intact: its P-time change is
0.0035~s, with a crossed seed--source sensitivity interval of
$-0.0036$ to 0.0114~s, and its gradient remains near $5.5\degree$. Thus the
same scalar information becomes differentiable in both endpoint roles once
both ordered relations are presented. Reciprocity is explicitly supplied as a
symmetry-informed scalar augmentation; the experiment tests global field
closure and is not labelled minimal-prior.

The design is a fixed same-medium replication performed after the seed-101
endpoint asymmetry was observed. Its cluster interval addresses new sources
under the fixed 81-station network, while the five-seed SD records optimization
repeatability. The independent-medium experiments test representation and
metric transfer separately.

\subsection{Catalogue-scale role completion}

The value-only field observes an ordered relation graph: catalogue stations
occupy the receiver slot and earthquakes occupy the source slot. Reciprocity is
therefore not automatically identified at the unsupported reversed role. We
tested this mechanism with a protocol frozen before the new ablation results
were opened. Three matched fields use a shared endpoint normalization and no
exchange; three use the same architecture, batches, 30 epochs and seeds but
reverse a deterministic 50\% of training presentations. A reversed presentation
retains the original phase and scalar travel time, so exchange replaces
orientation rather than adding observations or optimization steps. No velocity,
ray, derivative, Moho reference or equation enters the loss. Because the
operation explicitly supplies source--receiver symmetry, it is a
reciprocity-informed relational augmentation and is not labelled minimal-prior.

\begin{table}[H]
\centering
\caption{Matched China source--receiver role-completion experiment. Values are mean $\pm$ sample standard deviation across three identical-seed pairs. Original-orientation test errors use the natural catalogue ordering.}
\label{si:tab:reciprocity_ablation}
\small
\begin{adjustbox}{max width=\linewidth}\begin{tabular}{lrr}
\toprule
Diagnostic & No exchange & 50\% exchange \\
\midrule
Pg test MAE (s) & $0.3974\pm0.0023$ & $0.3980\pm0.0031$ \\
Pn test MAE (s) & $0.9278\pm0.0073$ & $0.9392\pm0.0076$ \\
Pn reciprocity gap, median (s) & $1.763\pm0.499$ & $0.094\pm0.006$ \\
Pg reciprocity gap, median (s) & $2.321\pm0.520$ & $0.143\pm0.004$ \\
Pn reciprocal-gradient angle & $52.28\pm4.42\degree$ & $2.49\pm0.37\degree$ \\
Pg reciprocal-gradient angle & $52.78\pm2.78\degree$ & $2.85\pm0.37\degree$ \\
Pn relative-vector discrepancy & $1.026\pm0.125$ & $0.050\pm0.007$ \\
Direct Moho $\rho$ (193 cells) & $0.791\pm0.025$ & $0.789\pm0.013$ \\
Direct Moho MAE (km) & $9.45\pm0.50$ & $9.37\pm0.41$ \\
Two-end vector-average MAE (km) & $13.40\pm1.68$ & $8.72\pm0.20$ \\
\bottomrule
\end{tabular}\end{adjustbox}
\end{table}

Exchange collapses the reciprocal scalar and vector discrepancies while
changing original-orientation Pn MAE by only 0.011~s. It leaves the direct
continental map unchanged within optimization variability, demonstrating that
the Moho pattern was already carried by the scalar Pn field. Its effect instead
appears in the two-end query, whose absolute thickness error decreases by
4.67~km. The intervention therefore separates three objects that a scalar MAE
alone cannot distinguish: the supported endpoint geometry, reciprocal closure
and the downstream regional signal.

\begin{figure}[H]
  \centering
  \includegraphics[width=0.94\textwidth]{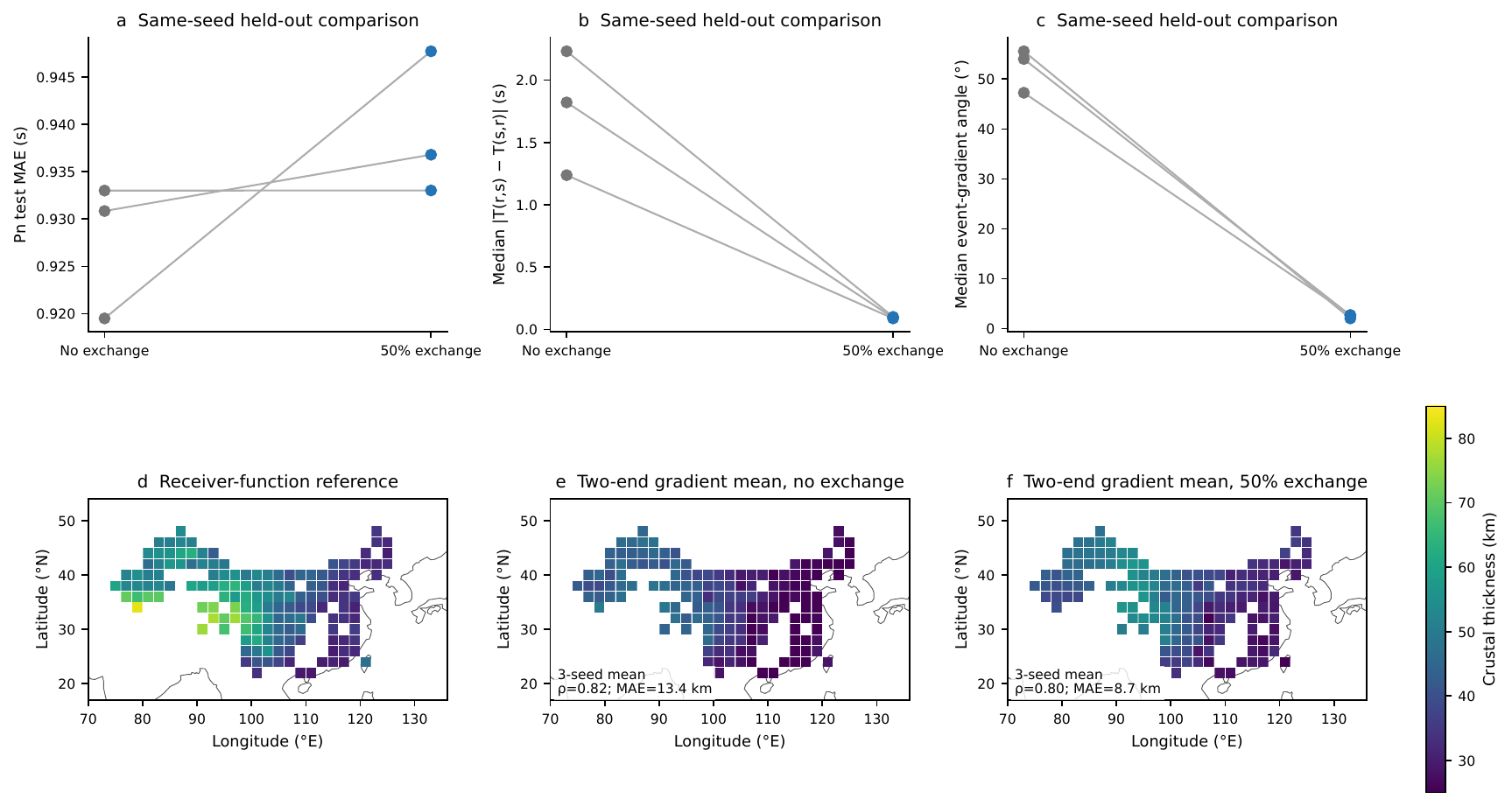}
  \caption{\textbf{Scalar role completion closes reciprocal geometry.} Matched three-seed comparisons show that exchanging existing source--receiver pairs leaves original-orientation prediction and the direct Moho map essentially unchanged, while sharply reducing scalar reciprocity error, reciprocal endpoint-gradient disagreement and symmetric-readout thickness error. Exchange supplies symmetry but no new scalar target or governing equation.}
  \label{si:fig:reciprocity_ablation}
\end{figure}

\section{Pathwise Data Support for Direct Geometry}

Reciprocal MUSCAL FMM fields use a local domain extending 40~km beyond the
evaluated endpoint geometry. Reference gradients are sampled at source endpoints
by centred grid differences and trilinear interpolation on a 2-km grid. Five
observed events contribute 78 event--station pairs with catalogue sources above
the model surface; those pairs are evaluated after surface clipping and do not
define derivatives with respect to the original negative depths.

The direct trajectory experiment uses the dense controlled heterogeneous medium
on $z\in[0,80]$~km.  Training-source coordinates span
$[-160,160]^2\times[0,60]$~km at 40-km horizontal and 10-km vertical spacing.
For a fixed receiver, integration changes the \emph{source-coordinate} query
continuously; this dense volume therefore supplies support along the interior
path rather than only at three source-depth planes.

Forty-eight deterministic held-out source--receiver paths span 100--300~km
endpoint distances.  All learned and FMM integrations reach the 7.5-km receiver
neighbourhood and none contacts the 80-km computational bottom.  Forty paths
remain entirely inside the sampled source-coordinate box.  Their median initial
learned--FMM angle is $5.77\degree$ and their median-of-ray path separation is
4.70~km.  The remaining eight cross a lateral or depth face of the sampled box
and are retained as support-monitoring cases rather than main-text evidence.

The primary minimal-coordinate controlled field has a P reciprocity-error median
of 14.08~s and a 95th percentile of 34.07~s; its S median is 24.75~s. These
values are reported alongside its accurate source derivative because they mark
the unobserved reversed endpoint role, not failure of the supported
source-conditioned query. For comparison, the secondary structured MUSCAL
field has a P reciprocity-error median of $1.269\pm0.316$~s and a 95th percentile
of $4.411\pm0.917$~s, and full-coverage endpoint-eikonal training gives
$1.390\pm0.172$~s median reciprocity error. These secondary values describe a different
architecture from the primary field. The matched exchange intervention in
Table~\ref{si:tab:synthetic_reciprocity_confirmation} tests reciprocal closure directly.

\section{Truth-Known Synthetic Focal Mechanisms}

One hundred held-out double-couple mechanisms span 73 source groups, with dip
restricted to 15--85$\degree$. Thirty-six stations are selected deterministically
per event, polarities are generated from the heterogeneous 3-D FMM vectors, and
two of 36 signs are flipped. The common inversion uses a 5$\degree$ global grid
and 1$\degree$ refinement. Table
\ref{si:tab:synthetic_focal} reports known-truth performance.

\begin{table}[ht]
\centering
\caption{Truth-known focal-mechanism recovery across 100 events at 73 held-out source locations. Errors use the Frobenius-inner-product tensor angle defined in Methods; the last two columns give recovery fractions.}
\label{si:tab:synthetic_focal}
\small
\begin{adjustbox}{max width=\linewidth}\begin{tabular}{lcccc}
\toprule
Ray model & Median angle & 90th percentile & $\leq15\degree$ & $\leq30\degree$ \\
\midrule
3-D FMM oracle & 11.34$\degree$ & 32.67$\degree$ & 0.680 & 0.860 \\
Value-learned & 15.38$\degree$ & 33.22$\degree$ & 0.490 & 0.890 \\
1-D layered FMM & 13.38$\degree$ & 31.23$\degree$ & 0.560 & 0.890 \\
Homogeneous straight & 52.87$\degree$ & 99.03$\degree$ & 0.010 & 0.120 \\
\bottomrule
\end{tabular}\end{adjustbox}
\end{table}

The paired learned-minus-oracle mean angular-error difference is $3.26\degree$ (source-cluster
bootstrap 95\% interval 1.20--5.90$\degree$). Learned minus 1-D layered is
$1.57\degree$ (0.24--3.75$\degree$), and learned minus straight is
$-40.19\degree$ ($-47.82$ to $-34.15\degree$). The scalar-trained geometry is
therefore decisively more informative than straight rays and approaches, but
does not equal, the explicit 1-D and 3-D model-based comparators.

\section{Fixed External-Mechanism Validation}

The ComCat query returned 429 events carrying a moment-tensor product in the
specified date and geographic window. Twenty-six have an exact associated
identifier in the annotation catalogue. All 26 were assigned to the final test
partition, including events later removed by mechanism or polarity quality
control \citep{USGSComCat}. Eighteen products pass the reviewed waveform-tensor rules. Six events
lack an accepted reviewed product and two reviewed Mwr products exceed the
frozen origin/centroid match tolerances. Two of the 18 accepted mechanisms have
only 14 common valid manual polarities and are excluded by the prespecified
minimum of 15. The final 16 events contain 447 manual U/D observations, all with
valid reciprocal-FMM pairs and zero HASH status flags.

The chord baseline is the normalized full three-dimensional
source--receiver vector, not a horizontal or zero-elevation construction. All
447 station coordinates have non-zero elevation, spanning $-0.038$ to 2.795~km.
Direct reconstruction from the source and station coordinates reproduces the
stored chord to $1.1\times10^{-16}$ maximum absolute component difference.
Setting station elevation to zero would change the chord by a median
$1.18\degree$ (95th percentile $3.41\degree$; maximum $9.15\degree$).

\begin{table}[ht]
\centering
\caption{Fixed-mechanism observational validation. Event summaries give equal weight to each of 16 events; pooled values give equal weight to each of 447 polarities.}
\label{si:tab:external_fixed}
\small
\begin{adjustbox}{max width=\linewidth}\begin{tabular}{lccccc}
\toprule
Ray model & Event mean & Event median & Event IQR & Pooled & Mismatches \\
\midrule
Value-learned & 0.158 & 0.146 & 0.087--0.204 & 0.154 & 69 \\
MUSCAL 3-D & 0.219 & 0.120 & 0.059--0.364 & 0.201 & 90 \\
HASH 1-D & 0.145 & 0.116 & 0.073--0.190 & 0.136 & 61 \\
3-D chord including elevation & 0.103 & 0.083 & 0.038--0.147 & 0.092 & 41 \\
\bottomrule
\end{tabular}\end{adjustbox}
\end{table}

\begin{table}[ht]
\centering
\caption{Paired event-bootstrap differences in fixed-mechanism mismatch. Negative values favour learned rays. The non-inferiority rule requires the upper interval bound to be below 0.05.}
\label{si:tab:external_bootstrap}
\small
\begin{adjustbox}{max width=\linewidth}\begin{tabular}{lccc}
\toprule
Comparator & Mean difference & 95\% interval & Non-inferior \\
\midrule
MUSCAL 3-D & $-0.061$ & $-0.186$ to 0.048 & yes \\
HASH 1-D & 0.014 & $-0.039$ to 0.062 & no \\
3-D chord including elevation & 0.055 & 0.017 to 0.096 & no \\
\bottomrule
\end{tabular}\end{adjustbox}
\end{table}

The learned--chord angle directly tests whether the polarities distinguish rays
that are geometrically different. Table~\ref{si:tab:external_separation} reports
event-weighted differences within the four frozen bins. The difference is zero
where rays are within $5\degree$, unresolved in the two intermediate bins and
positive for the 306 polarities separated by at least $15\degree$. The
observational ranking therefore does not arise solely from polarities for which
the two ray constructions are indistinguishable.

\begin{table}[ht]
\centering
\caption{California fixed-tensor learned-minus-three-dimensional-chord mismatch by ray separation. Positive values favour the chord; intervals resample events.}
\label{si:tab:external_separation}
\small
\begin{adjustbox}{max width=\linewidth}\begin{tabular}{lrrr}
\toprule
Learned--chord angle & Polarities & Events & Difference (95\% interval) \\
\midrule
$<5\degree$ & 24 & 7 & 0.000 (0.000--0.000) \\
5--$10\degree$ & 58 & 13 & $-0.051$ ($-0.167$--0.038) \\
10--$15\degree$ & 59 & 16 & $-0.039$ ($-0.122$--0.042) \\
$\geq15\degree$ & 306 & 16 & 0.099 (0.048--0.152) \\
\bottomrule
\end{tabular}\end{adjustbox}
\end{table}

Radiation-amplitude filtering reduces nodal ambiguity without changing the
fixed external tensor. Requiring every ray to have normalized absolute
amplitude at least 0.02, 0.05 and 0.10 retains respectively 400, 335 and 261
polarities from 16, 15 and 14 events. Learned-minus-chord differences are 0.045
(0.016--0.078), 0.024 ($-0.006$--0.061) and 0.031 (0.011--0.056); the 0.05
threshold makes the difference unresolved but does not establish a learned-ray
advantage. At that threshold learned-minus-HASH is 0.007 ($-0.019$--0.034) and
learned-minus-MUSCAL is $-0.038$ ($-0.134$--0.045).

No tested event attribute resolves the learned--chord difference in this
16-event sample. Spearman correlations are 0.375 ($-0.197$--0.782) with source
depth, $-0.046$ ($-0.600$--0.514) with median epicentral distance, $-0.157$
($-0.586$--0.393) with azimuthal gap, $-0.134$ ($-0.563$--0.372) with
magnitude, 0.218 ($-0.333$--0.682) with polarity count and 0.249
($-0.291$--0.622) with waveform fit. The map and all event-level values are
shown in Fig.~\ref{si:fig:catalogue_robustness} and supplied in machine-readable
form. No geological subclass is assigned post hoc.

A fixed-field catalogue-coordinate stress applies one shared source offset to
every polarity of an event while leaving stations and external tensors fixed.
Under the 2-km horizontal/5-km depth stress, learned mismatch changes by 0.029
($-0.002$--0.071) and chord mismatch by 0.022 ($-0.0004$--0.043). Under the
5-km/10-km stress, the changes are 0.047 (0.010--0.089) and 0.046
(0.016--0.079). All 16 events retain valid learned gradients in all three
realizations. Origin-time shifts are recorded but do not affect a spatial
gradient or polarity sign. The comparable degradation of both rays supplies a
coordinate-sensitivity boundary, not a learned-over-chord ranking.

Integrity checks confirm zero event overlap, kilometre coordinate inputs,
s~km$^{-1}$ source gradients, unit ray vectors, manual-only U/D codes, symmetric
normalized tensors and agreement between tensor eigenvectors and reported P/T
axes. The primary event bootstrap treats events as the sampling unit but does not
propagate uncertainty in external waveform tensors, polarity codes, catalogue
locations, the neural training seed or the velocity models used by MUSCAL and
HASH; the dedicated source-coordinate stress above quantifies only the
catalogue-location component. ComCat Mwr products also use regional velocity models in their Green's
functions; fixed mechanism means independent of the scored manual polarities,
not independent of all propagation assumptions.

\begin{figure}[t]
  \centering
  \includegraphics[width=0.96\textwidth]{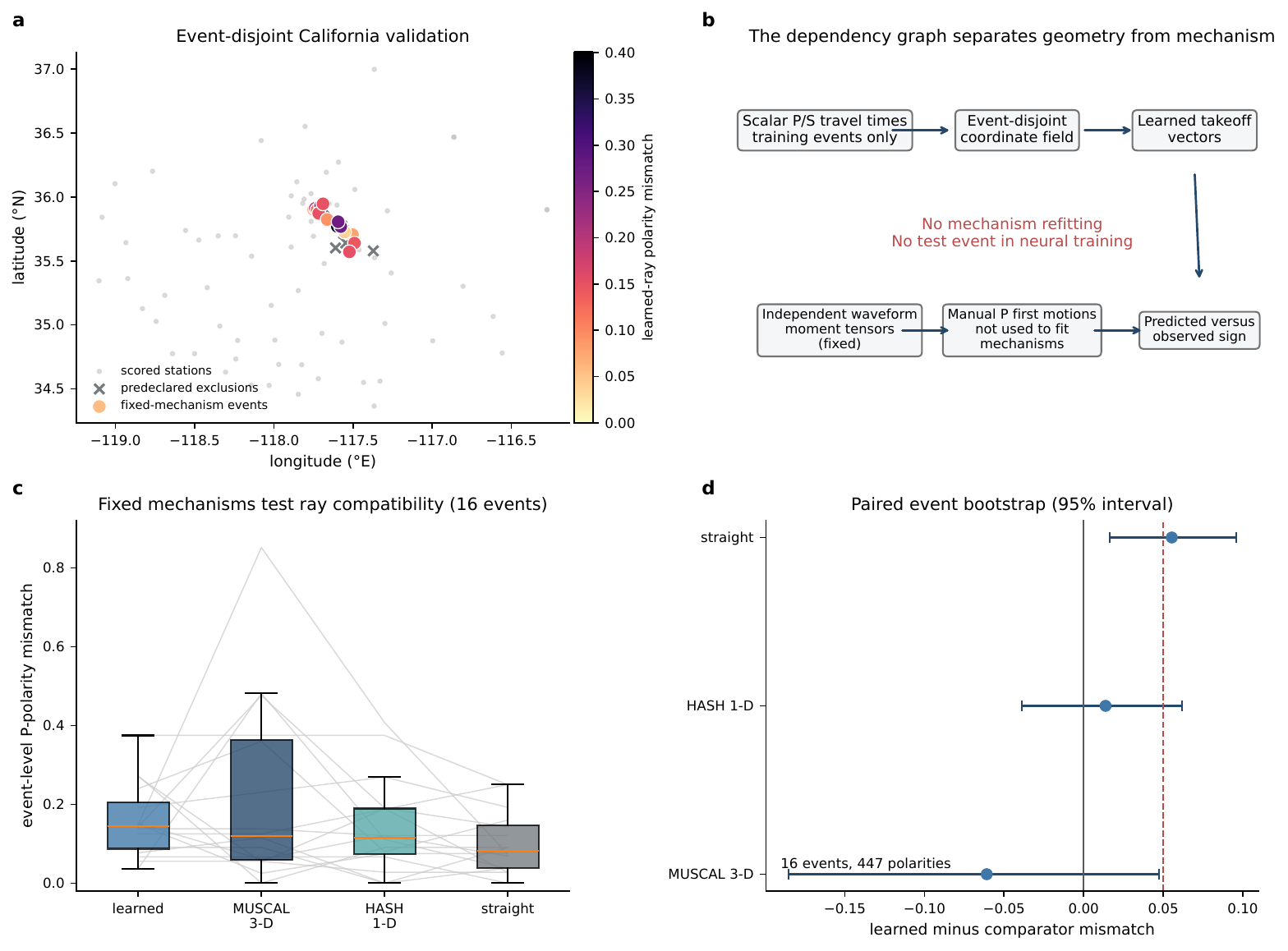}
  \caption{\textbf{Event-disjoint California observational transfer.} (a) Sixteen accepted events and 447 common manual P first motions. (b) The dependency graph separates catalogue travel-time learning from fixed external mechanisms. (c) Event-level mismatch for learned, MUSCAL 3-D, HASH 1-D and the full three-dimensional source--receiver chord including station elevation. (d) Paired event-bootstrap differences. This experiment tests observational compatibility and event-disjoint transfer; the main-text synthetic experiment supplies known mechanism truth.}
  \label{si:fig:external_fixed_figure}
\end{figure}

\section{Legacy Refitted-Mechanism Compatibility Test}

The downstream experiment contains 93 events, 1,548 fitting polarities and 664 held-out
polarities. The learned-vector result uses the seed-101 observed-arrival checkpoint and
one deterministic stratified polarity partition. The 20,000-replica event bootstrap
quantifies sampling across events. It does not propagate neural-training, hypocentral,
velocity-model or polarity-partition uncertainty. A production extension could ensemble
these sources of uncertainty in trials analogous to the multiple velocity tables and
hypocentral perturbations used by HASH.

\begin{figure}[t]
  \centering
  \includegraphics[width=0.98\textwidth]{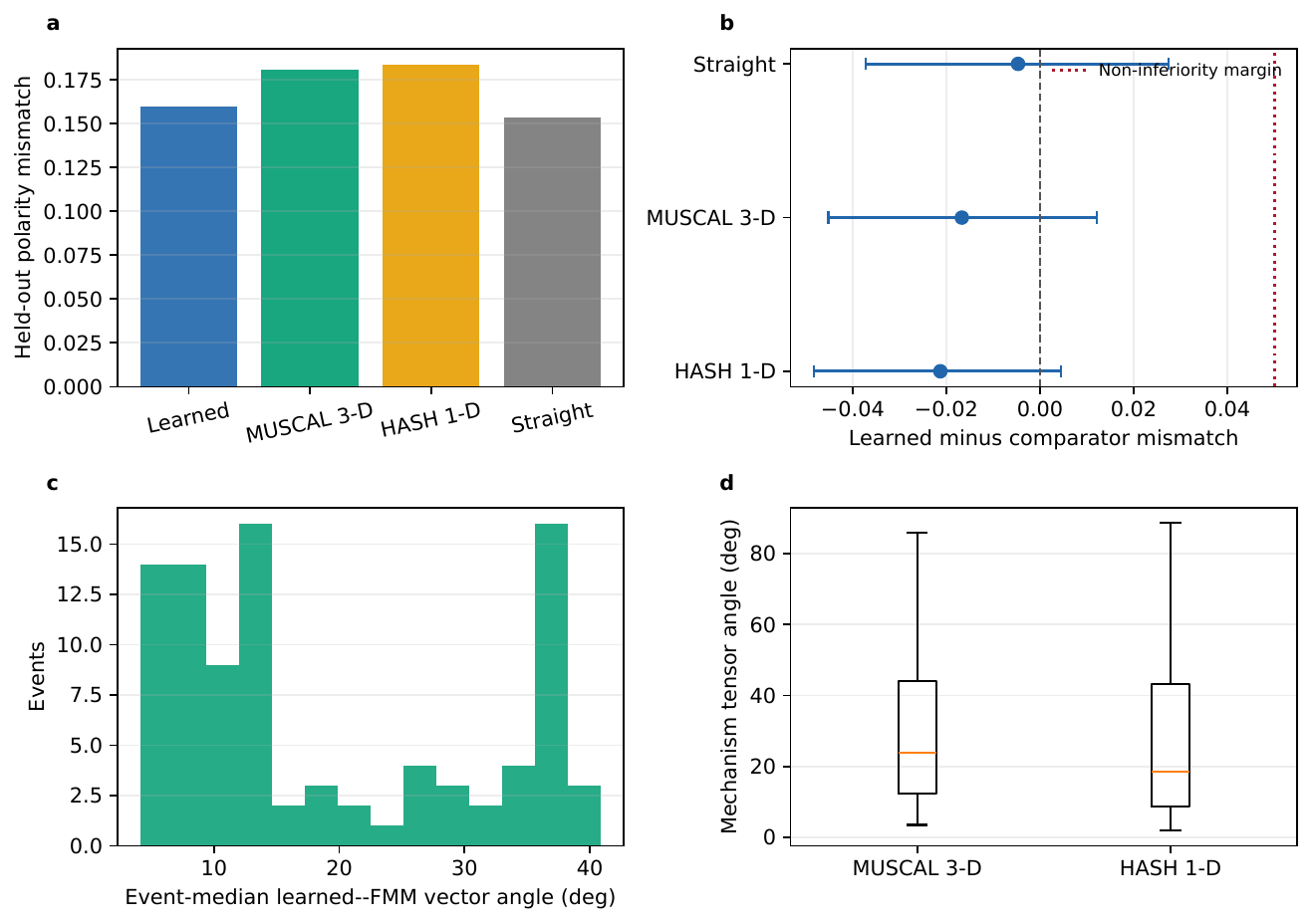}
  \caption{\textbf{Legacy California held-out-polarity compatibility comparison.} (a) Pooled held-out mismatch after a separate mechanism is fitted for each ray model. (b) Event-bootstrap learned-minus-comparator differences and the 0.05 non-inferiority margin. (c) Learned--MUSCAL takeoff-vector differences. (d) Moment-tensor differences among the separately fitted mechanisms. Because the mechanism changes with the ray model, this analysis is retained only as a compatibility diagnostic and is not the independent real-data result.}
  \label{si:fig:observed_focal}
\end{figure}

\FloatBarrier
\section{Same-Observation Classical Relational Controls}
\label{si:sec:classical_relational}

Geometric recovery extends beyond neural parameterizations. Local-polynomial and moving-least-squares (MLS) estimates~\citep{LancasterSalkauskas1981,Levin1998,Mirzaei2012} read propagation geometry from the same scalar observations without velocity, gradient or equation targets. We compare them with the primary coordinate-only field on the same dense heterogeneous medium at 100\% and 25\% training-source coverage, and on the dense layered medium at 100\%. Both media contain 567 sources and 81 surface receivers, with the identical 397/85/85 source partition. The 25\% subset contains 99 training sources, selected with the original coverage seed 20260806. Every method is scored on all 85 held-out sources and 6,885 source--receiver rows.

For a receiver and phase, let $\vect{u}_i=(\vect{x}_i-\vect{q})/\vect{h}$ and let $\vect{p}_d(\vect{u}_i)$ contain monomials through total degree $d$. The coefficient vector minimizes
\begin{equation}
\sum_{i\in\mathrm{train}} e^{-\|\vect{u}_i\|^2/2}
\left[T_i-\vect{p}_d(\vect{u}_i)^{\mathsf T}\vect{c}(\vect{q})\right]^2
+\lambda\sum_{j>0}c_j(\vect{q})^2.
\end{equation}
The intercept $c_0(\vect{q})$ is the predicted time. The first-order coefficients divided by the coordinate bandwidths estimate the gradient directly (LP/GMLS). The full MLS derivative is $\nabla_{\vect{q}}c_0(\vect{q})$ and also differentiates the moving weights and basis. These are two readouts of the same fit, not independent training algorithms. All training sources enter the Gaussian sum; a fixed bandwidth avoids nearest-neighbour stencil changes. This receiver-conditioned comparison tests source-coordinate interpolation at observed stations, not interpolation to unseen receiver coordinates.

The frozen grid contains degrees 1/2/3, horizontal bandwidths 20/40/80~km, vertical bandwidths 10/20/40~km and ridge penalties 0/$10^{-6}$. Validation scalar MAE selects the winner within each degree and across degrees. Neither gradient truth nor any downstream label enters selection. The full-coverage heterogeneous winner is cubic with $(h_{xy},h_z,\lambda)=(20,40,10^{-6})$; the 25\% winner is quadratic with $(20,40,0)$; the layered winner is cubic with $(20,20,0)$. The complete 54-candidate searches and rejected ill-conditioned candidates are archived. The ten original heterogeneous neural runs are unchanged; three layered seeds 101--103 use the same minimal-coordinate architecture, scalar objective and 60-epoch budget.

\begin{table}[ht]
\centering
\caption{\textbf{Classical and neural queries on identical observations.} LP1--3 use fitted polynomial slopes; MLS1--3 differentiate the moving scalar field. The asterisk marks the order selected by validation scalar MAE. Neural entries are means across five heterogeneous or three layered seeds. Angle and relative-vector error are path medians against reciprocal-FMM P gradients. Velocity MAEs (km~s$^{-1}$) use the median inverse gradient norm across 81 receivers at each of 85 test sources, followed by mean absolute error over sources. These are the dense directional benchmarks, distinct from the 307-source metric-confirmation experiments.}
\label{si:tab:classical_controls}
\small
\setlength{\tabcolsep}{3pt}
\begin{adjustbox}{max width=\linewidth}\begin{tabular}{llrrrrr}
\toprule
Medium / coverage & Readout & MAE (s) & Angle ($\degree$) & Vector error & $V_\mathrm{P}$ MAE & $V_\mathrm{S}$ MAE \\
\midrule
Heterogeneous, 100\% & LP1 & 0.493 & 5.265 & 0.117 & 0.320 & 0.183 \\
 & MLS1 & 0.493 & 4.386 & 0.094 & 0.174 & 0.099 \\
 & LP2 & 0.084 & 2.519 & 0.054 & 0.083 & 0.047 \\
 & MLS2 & 0.084 & 2.200 & 0.048 & 0.068 & 0.039 \\
 & LP3 & 0.055 & 2.734 & 0.055 & 0.329 & 0.188 \\
 & MLS3$^{*}$ & 0.055 & 2.734 & 0.055 & 0.326 & 0.186 \\
 & Neural mean & 0.717 & 5.342 & 0.108 & 0.163 & 0.092 \\
\midrule
Heterogeneous, 25\% & LP1 & 0.908 & 8.794 & 0.202 & 0.520 & 0.297 \\
 & MLS1 & 0.908 & 10.379 & 0.219 & 0.380 & 0.217 \\
 & LP2 & 0.497 & 5.573 & 0.117 & 0.181 & 0.103 \\
 & MLS2$^{*}$ & 0.497 & 6.555 & 0.144 & 0.270 & 0.154 \\
 & LP3 & 0.576 & 6.716 & 0.142 & 0.255 & 0.146 \\
 & MLS3 & 0.576 & 7.051 & 0.154 & 0.253 & 0.145 \\
 & Neural mean & 0.725 & 5.639 & 0.113 & 0.167 & 0.095 \\
\midrule
Layered, 100\% & LP1 & 0.458 & 4.679 & 0.104 & 0.318 & 0.182 \\
 & MLS1 & 0.458 & 3.601 & 0.081 & 0.156 & 0.089 \\
 & LP2 & 0.047 & 1.007 & 0.022 & 0.038 & 0.022 \\
 & MLS2 & 0.047 & 1.029 & 0.023 & 0.033 & 0.019 \\
 & LP3 & 0.023 & 0.725 & 0.017 & 0.023 & 0.013 \\
 & MLS3$^{*}$ & 0.023 & 0.725 & 0.018 & 0.023 & 0.013 \\
 & Neural mean & 0.121 & 1.359 & 0.029 & 0.039 & 0.021 \\
\bottomrule
\end{tabular}\end{adjustbox}

\end{table}

Full-coverage cubic MLS recovers more accurate scalar times and P directions than the current neural field in both media (Table~\ref{si:tab:classical_controls}). In the heterogeneous medium, the paired neural-minus-MLS source-mean angular difference is $1.99\degree$ (95\% source-bootstrap interval $1.49$--$2.46\degree$). At 25\% coverage, its sign reverses to $-4.34\degree$ ($-5.89$ to $-2.91\degree$), although MLS retains lower scalar MAE. These mean-angle comparisons differ from the path medians reported in the table. Local-polynomial slopes can behave differently from full MLS derivatives: the selected quadratic slope gives $5.57\degree$ median error at 25\%, compared with $6.56\degree$ for the full derivative of the identical scalar fit.

The velocity query separates scalar approximation accuracy from metric recovery. In the full heterogeneous medium, source-median $V_\mathrm{P}$ MAE is 0.326~km~s$^{-1}$ for scalar-selected cubic MLS and 0.163 for the neural field, whereas the quadratic polynomial slope gives 0.083. Thus scalar-optimal polynomial order is not necessarily metric-optimal; all orders remain reported rather than being reselected on test velocity. The layered medium has no resolved depth-detrended anomaly, so anomaly correlation is not interpreted there.

\begin{figure}[p]
\centering
\includegraphics[width=\textwidth]{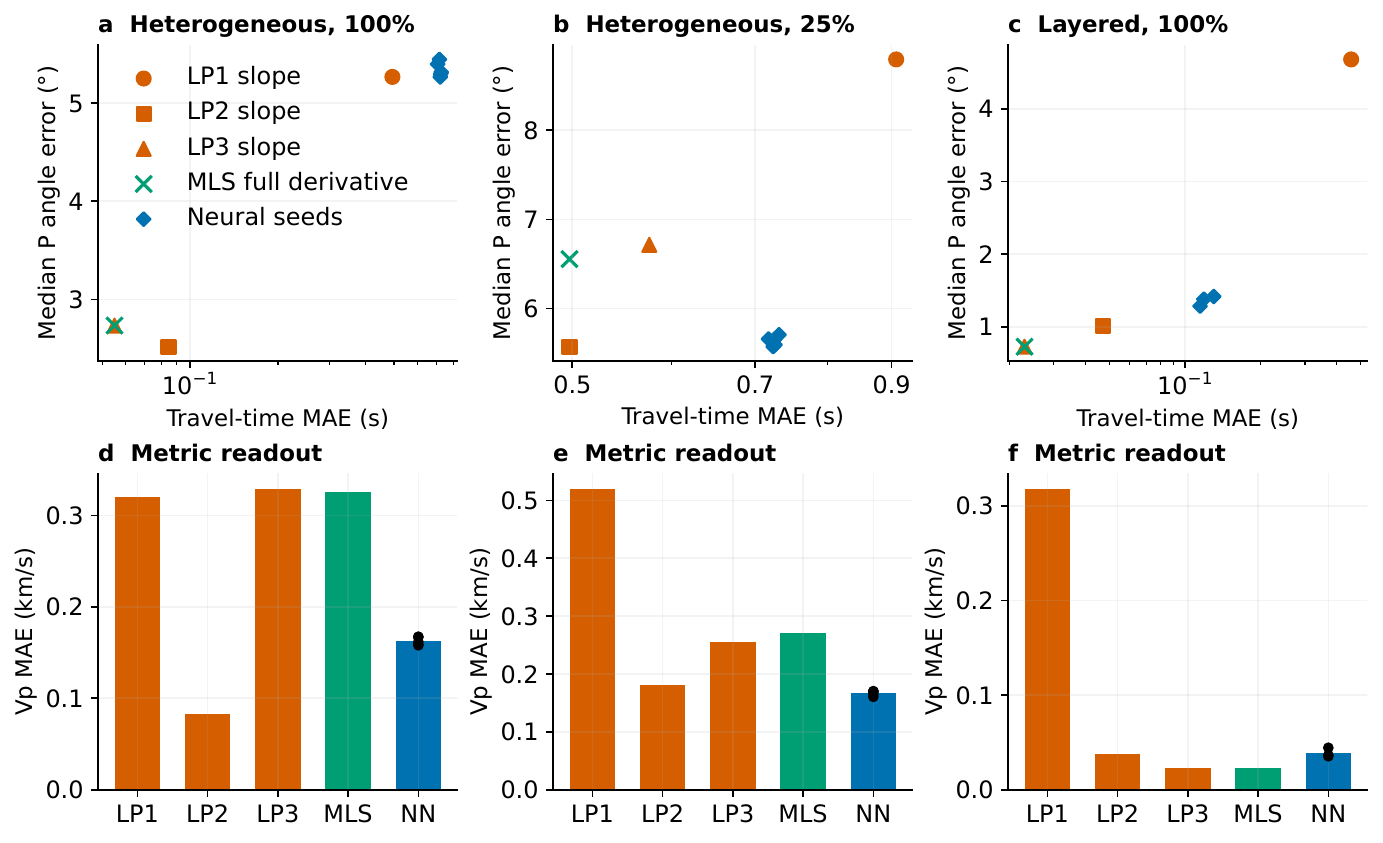}
\caption{\textbf{Scalar relations support geometry through classical and neural representations.} (a--c) Test scalar MAE and median P-gradient angular error for each validation-selected polynomial order, the full derivative of the overall scalar-selected MLS field and individual neural seeds. LP and MLS markers can overlap because the readouts share fitted values. (d--f) Local $V_\mathrm{P}$ errors after median aggregation over receivers at each source. Neural bars are seed means; black points are individual seeds. LP1--3 denote polynomial-slope readouts and NN denotes the neural field. No test errors were used to select the plotted settings. All tested orders and both derivative definitions appear in Table~\ref{si:tab:classical_controls}.}
\label{si:fig:classical_values_geometry}
\end{figure}

Downstream evaluation preserves the archived 48 ray starts and 100 mechanisms exactly. For all 48 paths, the median separation from reciprocal FMM is 4.44~km for the neural field, 2.06~km for the cubic polynomial slope and 2.10~km for cubic MLS. Receiver-neighbourhood arrival counts are respectively 48, 44 and 42. These medians include non-arriving paths and are therefore reported alongside convergence counts, not as evidence of complete-ray success. On the original fixed 40-path support subset, separations are 4.70/1.86/2.00~km and arrivals are 40/37/36. The mask is not changed to remove classical failures.

The mechanism queries reuse all 73 source groups, 36 stations per mechanism, two polarity flips and the original inversion settings. Median tensor-angle errors are $15.38/13.31/13.00\degree$ for neural/polynomial-slope/MLS geometry; recovery within $30\degree$ is 89/88/86\%. Neural-minus-classical source-mean error differences are $1.56\degree$ (95\% interval $-1.05$ to $4.23\degree$) for polynomial slopes and $0.73\degree$ ($-2.54$ to $3.79\degree$) for MLS. Neither difference is statistically resolved. The intervals use 20,000 paired resamples of source groups, averaging repeated mechanisms within each group first. Scalar and gradient intervals average neural seeds before source resampling; they describe source sampling conditional on those fitted models and are not adjusted for multiple comparisons.

\begin{figure}[p]
\centering
\includegraphics[width=\textwidth]{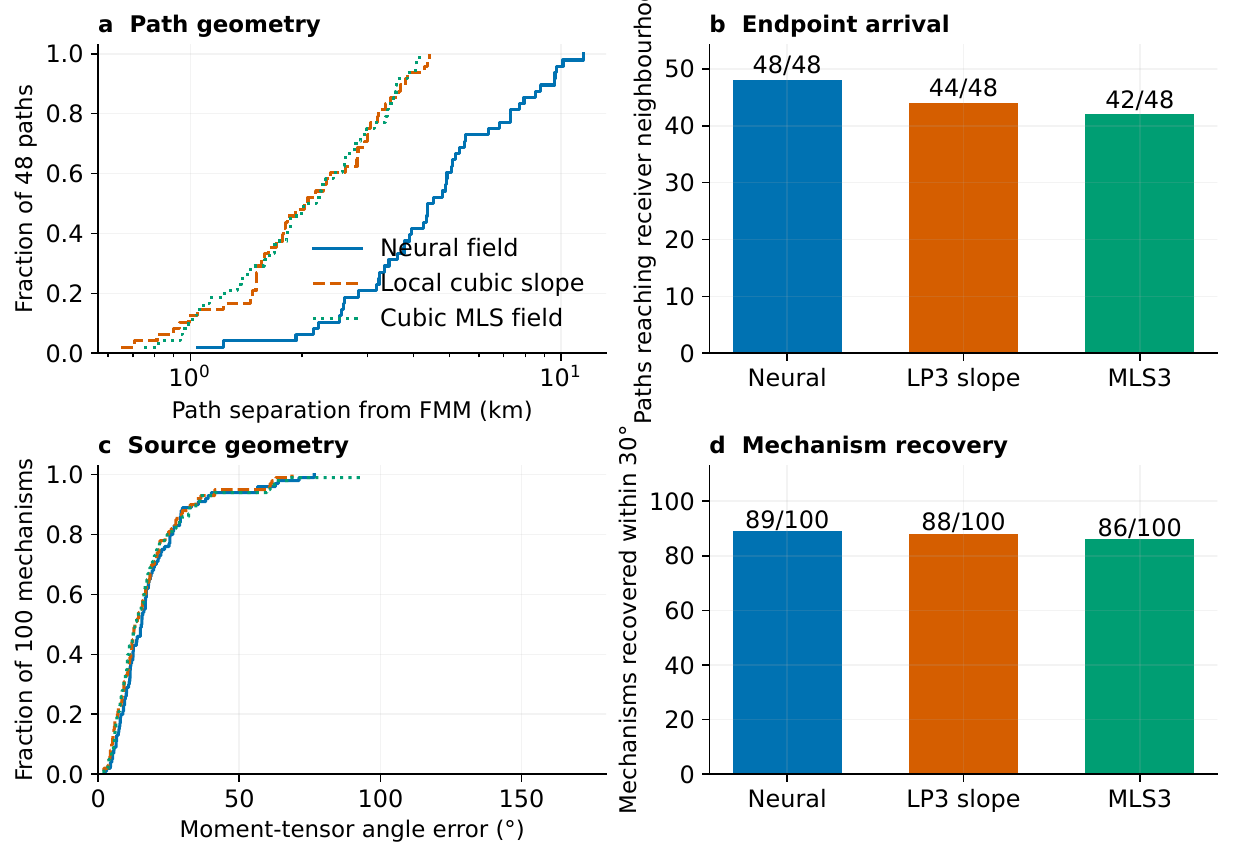}
\caption{\textbf{Classical scalar fields support the same downstream physical queries.} (a) Path-separation distributions for all 48 archived starts, including paths that do not reach the receiver neighbourhood. (b) Arrival counts using identical 1-km midpoint integration, 700-step limit and 7.5-km receiver radius. (c) Tensor-angle distributions for all 100 paired mechanisms, shown over the complete 0--180$\degree$ range. (d) Counts recovered within $30\degree$. Classical methods use the cubic fit selected only by validation travel-time error; the neural comparator is the archived seed-101 field. Different path-accuracy and convergence rankings are retained.}
\label{si:fig:classical_downstream}
\end{figure}

Protocol hashes, source-row identity checks, predictions, gradients, ray paths, mechanism observations and figure source data are archived in \path{supplementary/revision/classical_relational_baselines/run_1/}. Unit tests verify polynomial reproduction through degree three and physical-unit transformations. At the first ten validation sources, 0.001-km centred differences agree with analytic full MLS gradients: median relative vector differences are below $3.3\times10^{-8}$ across the three cases and median norm ratios are within $4\times10^{-9}$ of one. The maximum relative difference is $1.8\times10^{-4}$. One archived neural mechanism is independently replayed to verify the polarity and inversion alignment. The protocol was frozen before fitting these new classical controls to the existing benchmarks; this comparison is not a fresh blinded realization.

\FloatBarrier
\section{China Source-Geographic Holdout}
\label{si:sec:china_source_geographic}

This experiment asks whether arrival prediction transfers to geographically withheld sources, separately from the continental event-disjoint query in the main figures. The design was fixed before any new fitting: source blocks are 4$\degree$ wide, anchored at 70$\degree$E, 15$\degree$N, and assigned by the first eight SHA-256 bytes of \texttt{china-spatial-v1:ix:iy}, interpreted as a big-endian integer modulo ten. Buckets zero and one are test and validation, respectively; other buckets are training. Training events within 25~km horizontal AEQD distance of any held-out source are excluded. The minimum retained separation is 25.0007~km. Four event identifiers have inconsistent source coordinates; their 13 rows are excluded as ambiguous rather than repaired. Neither the original cache nor its older experimental results is altered.

\begin{figure}[p]
\centering
\includegraphics[width=\textwidth]{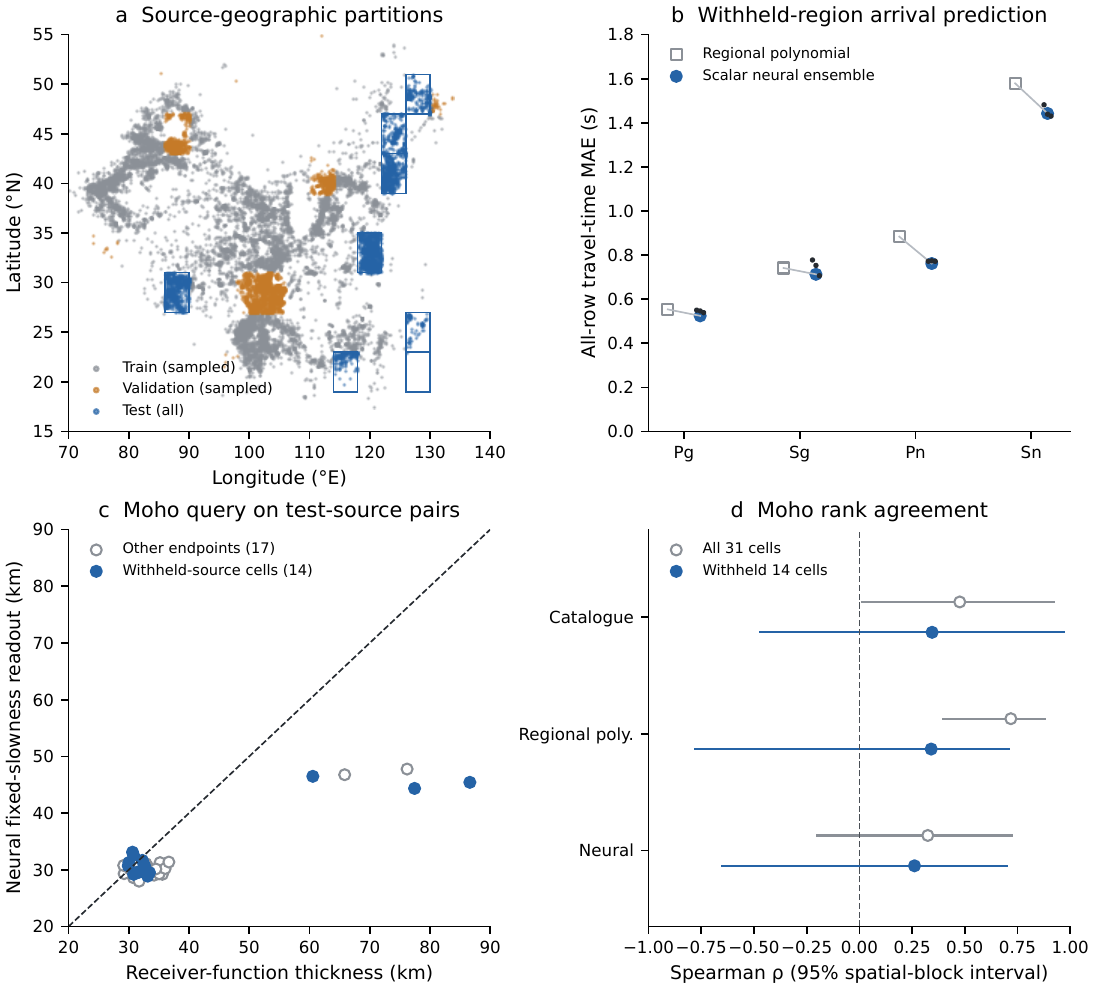}
\caption{\textbf{Source-geographic arrival transfer and its Moho query.} (a) Spatial split after the 25-km buffer, with all 14,109 test sources and deterministic samples of up to 20,000 training and validation sources; outlines mark the eight occupied test blocks. (b) MAE for all 75,843 test arrivals. Filled circles denote the median-of-three neural ensemble, squares the regional polynomial, and small black points the individual neural seeds. (c) Every supported common reference cell, with the 14 withheld-source cells filled and the other 17 endpoints open. The dashed line is identity; neither high-thickness values nor compressed predictions are clipped. (d) Spearman correlations and 95\% 10$\degree$ spatial-block intervals for neural, regional-polynomial and direct-catalogue fixed-slowness readouts, shown on the complete correlation domain. The 31-cell endpoint map must not be read as 31 withheld-source cells.}
\label{si:fig:china_source_geographic}
\end{figure}

After buffering, the split contains 517,388 training events in 112 occupied blocks, 235,383 validation events in 15 blocks and 14,109 test events in eight blocks. The unequal event fractions follow the fixed geographic hashing and highly uneven seismicity; no blocks were exchanged to improve balance or results. A further 24,520 events occupy the buffer. The test contains 75,843 arrivals. Receivers and training paths may sample the withheld source regions, so this is a source-geographic test, not a completely unobserved-volume experiment (Fig.~\ref{si:fig:china_source_geographic}).

The raw-coordinate four-phase field retains width 256, five residual blocks and the original depth-balanced loader. Each of seeds 101--103 trains for 30 epochs with 160,000 presentations per phase per epoch, batch size 8,192, AdamW learning rate 0.002 and weight decay $10^{-6}$, cosine decay to $10^{-5}$, normalized smooth-L1 parameter 0.1 and gradient clipping at five. No endpoint exchange, derivative loss or equation loss is applied. China-centred AEQD inputs are in kilometres and scaled by $(1000,1000,50)$~km for each endpoint. Role-specific centres and phase means/scales are re-estimated on new training rows. Training-only moveout fits define the competing-branch screen. A deterministic, depth-stratified validation subset selects the checkpoint; test scoring begins only after all three selected weights are hashed.

The phase-specific regional polynomial comparator uses the original 12 scalar features, ten robust reweighting iterations and at most 300,000 training rows per phase, chosen with seed 20260909. This regional baseline is not the receiver-conditioned MLS estimator tested on the separate controlled media. The three neural predictions are combined by their pointwise median. Table~\ref{si:tab:china_spatial_scalar} reports every phase and seed on all test observations, including Pn values rejected by the training-derived branch screen. Paired intervals use 2,000 resamples of complete 4$\degree$ source blocks and are conditional on these fitted seeds, without multiple-comparison adjustment. Pg/Sg/Pn/Sn occupy 6/6/8/7 test blocks, respectively.

\begin{table}[ht]
\centering
\small
\caption{\textbf{Arrival prediction at withheld source regions.} All values are MAE in seconds. The final column is ensemble minus regional-polynomial absolute error with a 95\% whole-source-block bootstrap interval. Negative differences favour the scalar neural ensemble. No held-out residual screen is applied to these primary rows.}
\label{si:tab:china_spatial_scalar}
\setlength{\tabcolsep}{4pt}
\begin{adjustbox}{max width=\linewidth}\begin{tabular}{lrrrrrrl}
\toprule
Phase & Rows & Seed 101 & Seed 102 & Seed 103 & Ensemble & Regional & Paired difference [95\% CI] \\
\midrule
Pg & 21,265 & 0.550 & 0.547 & 0.539 & 0.524 & 0.554 & $-0.030$ [$-0.096,-0.013$] \\
Sg & 23,372 & 0.778 & 0.753 & 0.707 & 0.713 & 0.742 & $-0.029$ [$-0.124,-0.009$] \\
Pn & 25,222 & 0.772 & 0.774 & 0.772 & 0.762 & 0.884 & $-0.122$ [$-0.291,-0.048$] \\
Sn & 5,984 & 1.482 & 1.439 & 1.430 & 1.442 & 1.579 & $-0.137$ [$-0.461,0.315$] \\
\bottomrule
\end{tabular}\end{adjustbox}
\end{table}

Pn MAE decreases by 13.8\% relative to regional regression. The three seed MAEs are 0.772, 0.774 and 0.772~s. On the separately reported branch-screened Pn subset (24,841 rows), ensemble and regional MAEs are 0.745 and 0.862~s; that subset does not replace the all-row outcome. Sn's paired difference is not statistically resolved. These results establish transfer of scalar arrival prediction under the specified geographic separation, without using a Moho reference in fitting or selection.

The secondary thickness query uses 24,789 nominal Pn test pairs at $\geq100$~km that pass the training-derived branch screen. The held-out observed times are used only for this common quality qualification and the direct-catalogue comparator. Fixed crustal and mantle slownesses are re-estimated from new training arrivals. All methods share the same pair incidence matrix, 2$\degree$ grid, robust four-iteration solver, smoothing weight 0.1 and ridge coefficient 0.002. Each prior centre is the median predicted pair-thickness sum divided by two. Unlike the original component audit, this test uses no learned-gradient qualification and requires 200 incident endpoints per supported cell. Thus its numerical map scores must not be interpreted as changes on the original 194-cell comparison.

\begin{table}[ht]
\centering
\small
\caption{\textbf{Fixed-slowness Moho queries from spatial-test pairs.} Each map is locked before reading the existing receiver-function reference. Correlation intervals resample entire 10$\degree$ cell blocks 2,000 times (seven blocks for all supported endpoints; four for withheld-source cells). The final column is Spearman correlation after separate quadratic longitude--latitude detrending of prediction and reference.}
\label{si:tab:china_spatial_moho}
\setlength{\tabcolsep}{5pt}
\begin{adjustbox}{max width=\linewidth}\begin{tabular}{llrrlr}
\toprule
Scope & Readout & Cells & MAE (km) & $\rho$ [95\% CI] & Detrended $\rho$ \\
\midrule
All endpoints & Neural ensemble & 31 & 6.82 & 0.325 [$-0.199,0.722$] & $-0.021$ \\
 & Regional polynomial & 31 & 6.28 & 0.719 [0.398,0.879] & 0.368 \\
 & Catalogue arrivals & 31 & 5.61 & 0.476 [0.012,0.922] & 0.350 \\
\midrule
Withheld sources & Neural ensemble & 14 & 7.93 & 0.262 [$-0.652,0.700$] & $-0.086$ \\
 & Regional polynomial & 14 & 8.02 & 0.341 [$-0.781,0.710$] & $-0.411$ \\
 & Catalogue arrivals & 14 & 7.18 & 0.345 [$-0.472,0.972$] & 0.073 \\
\bottomrule
\end{tabular}\end{adjustbox}
\end{table}

There are 31 supported common endpoint cells, of which only 14 lie inside withheld source blocks. The neural ensemble has 6.82~km MAE and $\rho=0.325$ across all 31; the regional polynomial has 6.28~km and $\rho=0.719$. Within the 14 source-geographic cells, the ensemble gives 7.93~km MAE and $\rho=0.262$, with a 95\% spatial-block interval of $-0.652$ to 0.700. The broad interval does not resolve rank agreement. Pearson correlation on these 14 cells is 0.925, but it is driven by broad thickness contrast and is not substituted for the predeclared rank statistic. The ensemble reproduces a Tibetan contrast of 14.71~km against 45.86~km on this particular mask. Better scalar prediction therefore does not certify the spatial thickness ranking. The test adds source-geographic arrival transfer, not another confirmed continental Moho reconstruction.

The protocol hash is \path{9f8467d667cb823d64aa7ae8a7b82f23594696106e0cffdb0f88aa04387248a4}. Frozen settings, complete source split, all per-seed predictions, raw map outputs, figure source data and audits are archived under \path{supplementary/revision/china_spatial_holdout/}. The companion reviewer archive includes all three selected weights. No architecture, block assignment, threshold or smoothing parameter was changed after test access.

\section{Portable Checkpoint Reproduction}

The historical reviewer archive contains executable code and data sufficient to recompute the primary 6,885-path gradient result, all twelve selected weights and training inputs of the blinded metric confirmation, the original China four-phase checkpoint/cache, and the new three-seed geographic test. Original protocol and provenance files retain their original hashes; a separate portable manifest maps archive-relative files to byte sizes and SHA-256 digests. Runtime paths are derived from the extraction directory, not historical author directories.

The complete inference check was executed on CPU after extracting the archive into a separate temporary directory. It verifies every content hash before loading weights, differentiates times with respect to physical-km coordinates, aggregates inverse gradient norms over the original 20--70-km receiver selection, and reproduces all four blinded objectives on the same 307 test sources. The maximum per-seed velocity discrepancy from the archived computation is below $3.9\times10^{-6}$~km~s$^{-1}$. The original China phase MAEs agree within $10^{-6}$~s; the new geographic predictions agree within $1.7\times10^{-4}$~s. The primary seed-101 gradient angle is $5.4431\degree$ and relative-vector error is 0.10926, consistent with its archived result and distinct from the five-seed mean in the abstract. The tested environment is Python 3.14.4, NumPy 2.4.2, SciPy 1.16.2, PyTorch 2.11.0 and pyproj 3.7.2 on macOS arm64.

This verification reruns checkpoint inference and the specified derived velocity metrics, not the entire training matrix. Portable commands for the twelve blinded optimization runs are supplied with new output directories and final-test scoring disabled during fitting. Full historical regional-matrix retraining and FMM target regeneration have additional dependencies identified in the code README. The archive and reproduction report make this verified scope explicit.

\section{Manual-Only Learning and Non-PINN Velocity Extraction}
\label{si:sec:manual_tomography}

This experiment implements the sequence of real-arrival learning, frozen-field sampling at new endpoints, and conventional velocity inversion. The manual-only California fields are newly trained. Each uses 100 initial epochs followed by a validation-motivated 200-epoch scalar-only refinement; both stages precede final-test evaluation. The initial query fits are archived as development runs and are not selected by reference-velocity performance. Historical fields trained on the mixed manual/automatic P cache are neither renamed nor reused as manual-only fields. All 102,204 P and 121,630 S targets in the new cache have explicit manual status; there is no automatic fallback. No velocity or equation residual enters the neural objective. The separate controlled readout uses only the three value-only checkpoints from the second, previously blinded metric realization; no PINN checkpoint contributes to the generated times or to the inverse problem.

\begin{table}[htbp]
\centering
\small
\caption{Manual-only data partitions and the regional subset used for tomography. Regional selection uses source/receiver geometry and training-station support. The whole-catalogue teacher test and the regional tomography test have different denominators.}
\begin{adjustbox}{max width=\linewidth}\begin{tabular}{llrrr}
\toprule
Scope & Partition & Events & P targets & S targets / records \\
\midrule
Whole manual cache & Training & 5,423 & 73,777 & 87,697 \\
 & Validation & 1,156 & 15,513 & 18,427 \\
 & Test & 940 & 12,914 & 15,506 \\
Regional tomography & Training & 4,177 & 33,241 & 39,537 \\
 & Validation & 891 & 6,995 & 8,354 \\
 & Test & 740 & 5,820 & 7,003 \\
\bottomrule
\end{tabular}\end{adjustbox}
\end{table}

The controlled regional subset contains 1,434/307/307 training/validation/test sources and 9,525/2,051/2,004 records, each with both phases. Both experiments generate 1,024 new source locations and 24 new surface locations. After the 20--70-km pair restriction, 1,024 California sources contribute 14,103 pairs and 1,012 controlled sources contribute 4,768 pairs. Query geometry depends only on training coordinates. No generated value is described as a manual pick. Random seeds are 202609161 for California and 202609162 for the controlled readout. All three neural seeds (101--103) enter the arithmetic mean scalar field.

The observed-pick inversion comparator uses the selected regional source--station paths, whereas each teacher is trained on its full catalogue or controlled cache. Query-only versus observed-pick inversion is therefore a comparison of these specified readouts, not a claim of algorithmic superiority under identical information. The principal matched intervention compares unperturbed and oscillatory versions of the same frozen scalar field on identical random coordinates and with the same inverse-model family.

The four inversion arms and four spatial weights are specified in Methods. All candidate models, including those that perform less well, are retained. The homogeneous comparator uses each arm's own robust scalar distance--time fit. The zero-spatial-roughness comparator retains the 3,328-node grid, weak amplitude damping and positivity bounds; it is therefore not an absence-of-prior control. Increasing $\lambda$ changes spatial roughness and additional amplitude damping together, so this sweep does not isolate their individual contributions. Grid spacing is a representation choice, not a measured resolving length. The ray solver uses fixed source and receiver positions and a common phase intercept; it does not jointly relocate earthquakes. Original validation picks select the inverse-model iteration and spatial weight. Within each dataset, all selected P/S readouts are hashed before its reference velocities and final-test scores enter evaluation. Manual fitting settings are locked before the completed controlled experiment is scored; the continuing manual fits use only their own original validation arrivals.

Numerical development identified a row-order inconsistency when the ray-tracing library's kernel was requested for interleaved origins. Scalar times retained input order whereas kernel rows were grouped. Explicit sorting and inverse sorting fixes their alignment. The discarded development fits are not scientific results. The final solver verifies $F(m)=L(m)m$ at every kernel evaluation; shuffled-origin homogeneous and heterogeneous tests agree to approximately $4\times10^{-15}$~s. The homogeneous path-time error is at most 0.0091~s on the verification paths. Smooth directional finite-difference tests differ from the linearized kernel by 0.29\% in the homogeneous field and 0.62\% in a heterogeneous field. A rough S-wave trial in the mixed-data arm also produced nonterminating library ray backtracking. A 120-s worker watchdog, added before test access, rejects such numerical trial evaluations and lets the same training-objective line search reduce its step. Failed linearizations retain the previous valid checkpoint. Rejection scores never enter final physical predictions. Diagnostic models and failure logs are archived. These checks separate numerical-solver failure from limitations of the learned scalar representation.

\subsection{Frozen-test results and extraction limits}

The whole-catalogue manual test contains 940 events and 15,506 event--station records. Ensemble P/S MAEs are 0.12938/0.23381~s. Individual-seed P MAEs are 0.13094, 0.12975 and 0.13051~s; S MAEs are 0.23493, 0.23496 and 0.23598~s. The separate regional geometry restriction retains 740 events and 7,003 records, including 5,820 P targets. The strict common velocity support contains 622 P and 694 S events, occupying 11 and 12 horizontal 20-km blocks. The controlled support contains 306 events in 108 blocks for each phase. Bootstrap intervals are conditional on these sampled regions and trained fields; California's few occupied blocks limit spatial inference.

The controlled direct-gradient summary differs from the earlier metric-confirmation tables because this comparison differentiates the ensemble-mean scalar field on its own regional paths and common event support, instead of taking the median velocity across individual fields. The weights are unchanged; the new definition matches exactly the field supplying the generated values.

California query-only tomography has P/S MUSCAL biases of +0.334/+0.183~km~s$^{-1}$. Its MAEs exceed the unperturbed gradient readout by 0.0949/0.0358~km~s$^{-1}$ (block 95\% intervals 0.0509--0.1115/0.0202--0.0443). Its depth-detrended velocity standard deviations are 0.0392/0.0226~km~s$^{-1}$, compared with 0.1624/0.1125~km~s$^{-1}$ in MUSCAL. The selected regularizer therefore compresses lateral amplitude; a high raw rank correlation largely reflects depth structure. The direct-gradient depth-detrended correlations are also weak and negative ($-0.129/-0.108$). The zero-roughness query control has stronger residual rank correlations (0.484/0.371) but worse scalar validation error and velocity MAE; it is retained as a sensitivity result, not selected post hoc using MUSCAL. On the 622 common P/S events, query-model fifth/median/95th-percentile $V_P/V_S$ ratios are 1.702/1.710/1.720, with no $V_P\leq V_S$ values. This phase ordering was not imposed during neural training or the separate inversions.

The oscillation raises California gradient MAEs by 1.785/0.516~km~s$^{-1}$, but changes query-model MAEs by $-0.00107/-0.000001$~km~s$^{-1}$; block intervals for the latter changes include zero. In the known medium, query-model degradation is small but resolved: +0.00288/+0.00124~km~s$^{-1}$, with block intervals 0.00132--0.00448/0.00051--0.00200. It is therefore stable, not exactly invariant. Removing explicit spatial roughness increases unperturbed known-medium MAEs from 0.0579/0.0255 to 0.0926/0.0467~km~s$^{-1}$, and perturbed MAEs from 0.0607/0.0268 to 0.1154/0.0513~km~s$^{-1}$. The corresponding California changes are 0.3948/0.2454 to 0.4775/0.2989 and 0.3938/0.2454 to 0.4775/0.2990~km~s$^{-1}$. Every one of these paired MAE reductions has a block interval excluding zero. This supports regularization as one contributor to stable extraction, while the finite representation and bounds also constrain the solution.

Median receiver-to-receiver relative velocity interquartile ranges are 4.20\%/5.01\% in California and 1.70\%/1.67\% in the controlled medium. An exact isotropic travel-time field on a differentiable branch would give the same endpoint speed for each receiver. The observed dispersion diagnoses an imperfect local metric readout, but does not by itself distinguish catalogue errors, representation bias and propagation-model mismatch. The new intervention perturbs the frozen function, so it cannot retrospectively identify the cause of every earlier noisy-training failure.

\begin{table}[htbp]
\centering\small
\caption{Controlled held-out scalar errors and validation-selected inverse settings. MAE is in seconds; each pair of settings is $(\lambda,\mathrm{iteration})$. Selection uses original validation arrivals only.}
\begin{adjustbox}{max width=\linewidth}\begin{tabular}{lrrrr}
\toprule
Readout & P MAE & S MAE & P setting & S setting \\
\midrule
Frozen scalar field & 0.0601 & 0.0882 & --- & --- \\
Perturbed scalar field & 0.0682 & 0.0942 & --- & --- \\
Query tomography & 0.0623 & 0.0902 & (10, 3) & (10, 3) \\
Perturbed query tomography & 0.0639 & 0.0919 & (10, 4) & (10, 4) \\
Observed-pick tomography & 0.0272 & 0.0455 & (1, 5) & (1, 4) \\
Picks + queries & 0.0330 & 0.0535 & (0.1, 4) & (1, 5) \\
Queries, zero roughness & 0.1018 & 0.1426 & (0, 5) & (0, 5) \\
Perturbed, zero roughness & 0.1082 & 0.1461 & (0, 5) & (0, 5) \\
Homogeneous from picks & 0.1802 & 0.2898 & --- & --- \\
Homogeneous from queries & 0.1791 & 0.2894 & --- & --- \\
\bottomrule
\end{tabular}\end{adjustbox}
\end{table}

\begin{table}[htbp]
\centering\small
\caption{Manual held-out scalar errors and validation-selected inverse settings. MAE is in seconds; each pair of settings is $(\lambda,\mathrm{iteration})$. Selection uses original validation arrivals only.}
\begin{adjustbox}{max width=\linewidth}\begin{tabular}{lrrrr}
\toprule
Readout & P MAE & S MAE & P setting & S setting \\
\midrule
Frozen scalar field & 0.1208 & 0.2201 & --- & --- \\
Perturbed scalar field & 0.1244 & 0.2225 & --- & --- \\
Query tomography & 0.1519 & 0.2635 & (10, 3) & (10, 3) \\
Perturbed query tomography & 0.1521 & 0.2637 & (10, 3) & (10, 3) \\
Observed-pick tomography & 0.1203 & 0.2119 & (1, 5) & (1, 5) \\
Picks + queries & 0.1300 & 0.2245 & (0.1, 5) & (1, 5) \\
Queries, zero roughness & 0.1781 & 0.3134 & (0, 1) & (0, 1) \\
Perturbed, zero roughness & 0.1781 & 0.3133 & (0, 1) & (0, 1) \\
Homogeneous from picks & 0.1804 & 0.3187 & --- & --- \\
Homogeneous from queries & 0.1811 & 0.3200 & --- & --- \\
\bottomrule
\end{tabular}\end{adjustbox}
\end{table}

\begin{table}[htbp]
\centering
\small
\caption{Known-medium accuracy for every frozen readout. Velocity MAE is in km~s$^{-1}$; $\rho_z$ is Spearman correlation after separate cubic depth detrending. Homogeneous models have no lateral structure. The zero-roughness controls retain weak amplitude damping and the finite grid.}
\begin{adjustbox}{max width=\linewidth}\begin{tabular}{lrrrr}
\toprule
Readout & P MAE & P $\rho_z$ & S MAE & S $\rho_z$ \\
\midrule
Direct gradient & 0.0688 & 0.759 & 0.0335 & 0.813 \\
Perturbed gradient & 3.0956 & -0.012 & 1.2767 & -0.007 \\
Query tomography & 0.0579 & 0.844 & 0.0255 & 0.922 \\
Perturbed query tomography & 0.0607 & 0.815 & 0.0268 & 0.905 \\
Observed-pick tomography & 0.0285 & 0.968 & 0.0167 & 0.960 \\
Picks + queries & 0.0345 & 0.944 & 0.0174 & 0.964 \\
Queries, zero roughness & 0.0926 & 0.720 & 0.0467 & 0.819 \\
Perturbed, zero roughness & 0.1154 & 0.605 & 0.0513 & 0.759 \\
Homogeneous from picks & 0.1975 & --- & 0.1103 & --- \\
Homogeneous from queries & 0.1995 & --- & 0.1108 & --- \\
\bottomrule
\end{tabular}\end{adjustbox}
\end{table}

\begin{table}[htbp]
\centering
\small
\caption{California consistency with the external MUSCAL reference for every frozen readout. Velocity MAE is in km~s$^{-1}$; $\rho_z$ is Spearman correlation after separate cubic depth detrending. Homogeneous models have no lateral structure. The zero-roughness controls retain weak amplitude damping and the finite grid.}
\begin{adjustbox}{max width=\linewidth}\begin{tabular}{lrrrr}
\toprule
Readout & P MAE & P $\rho_z$ & S MAE & S $\rho_z$ \\
\midrule
Direct gradient & 0.2999 & -0.129 & 0.2097 & -0.108 \\
Perturbed gradient & 2.0852 & -0.068 & 0.7256 & -0.042 \\
Query tomography & 0.3948 & 0.231 & 0.2454 & -0.174 \\
Perturbed query tomography & 0.3938 & 0.183 & 0.2454 & -0.182 \\
Observed-pick tomography & 0.3674 & 0.241 & 0.2347 & 0.191 \\
Picks + queries & 0.4375 & 0.285 & 0.2490 & 0.036 \\
Queries, zero roughness & 0.4775 & 0.484 & 0.2989 & 0.371 \\
Perturbed, zero roughness & 0.4775 & 0.449 & 0.2990 & 0.345 \\
Homogeneous from picks & 0.4846 & --- & 0.3009 & --- \\
Homogeneous from queries & 0.4842 & --- & 0.3023 & --- \\
\bottomrule
\end{tabular}\end{adjustbox}
\end{table}

\newpage
\subsection{Why value information can survive a derivative failure}

For $\delta T=\epsilon\sin(\vect{k}^{\mathsf T}\vect{x})$, the value error is bounded by $\epsilon$, whereas $\|\nabla\delta T\|$ can reach $\epsilon\|\vect{k}\|$. Small scalar error therefore does not control the local inverse-gradient norm. Conventional extraction uses a different inverse map. At fixed ray paths and fixed robust weights, one linearized update for the slowness and intercept vector $\vect{p}$ has perturbation
\begin{equation}
\delta\vect{p}=
\left(\vect{A}^{\mathsf T}\vect{W}\vect{A}+
\vect{R}_\lambda^{\mathsf T}\vect{R}_\lambda\right)^{-1}
\vect{A}^{\mathsf T}\vect{W}\,\delta\vect{t},
\end{equation}
where $\vect{R}_\lambda$ contains the roughness, amplitude and intercept penalties in the main-text objective. Its error is controlled by the regularized inverse operator acting on scalar perturbations, without an explicit differentiation factor $\|\vect{k}\|$. The finite representation, damping and cross-path consistency can therefore stabilize a readout of values whose pointwise derivatives are unreliable. This local argument does not prove unique recovery, remove ray-path nonlinearity, or eliminate bias from the chosen representation and regularizer. It also does not equate the fixed oscillatory intervention with random observation noise or hypocentre errors.

\begin{figure}[htbp]
\centering
\includegraphics[width=0.99\textwidth]{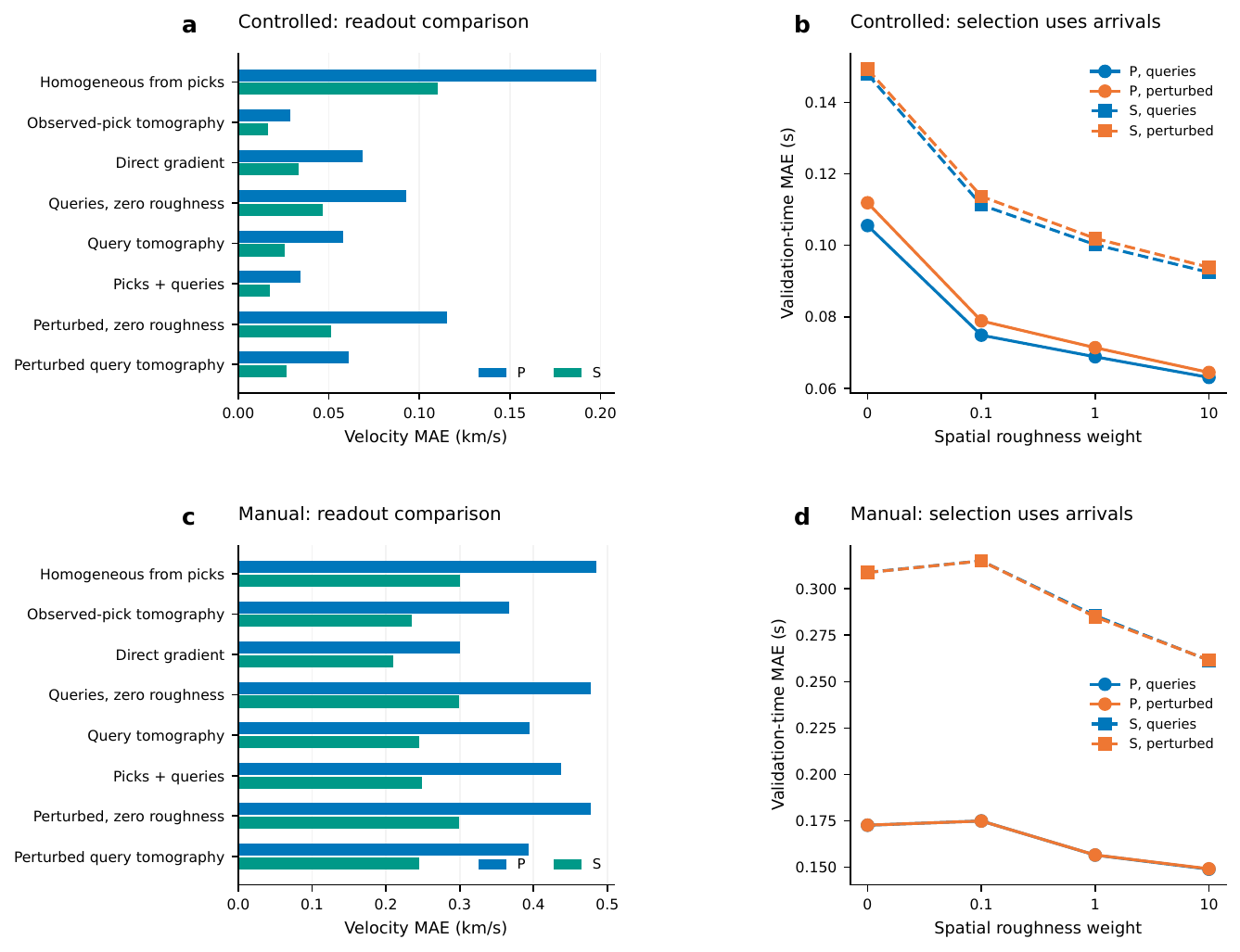}
\caption{\textbf{Conventional readout and regularization controls.} Left: velocity MAE for the observed-pick homogeneous baseline, unperturbed direct gradients and six conventional readouts in the known medium and the manual-only California experiment. Right: original validation-arrival MAE for the four spatial-roughness weights, with minimum validation error selecting the frozen inverse model. P and S use separate inversions. The California reference is external MUSCAL consistency rather than true-Earth accuracy. Tables report every comparator, including perturbed gradients and the query-based homogeneous baseline.}
\label{si:fig:manual_tomography_controls}
\end{figure}

\begin{figure}[htbp]
\centering
\includegraphics[width=0.95\textwidth]{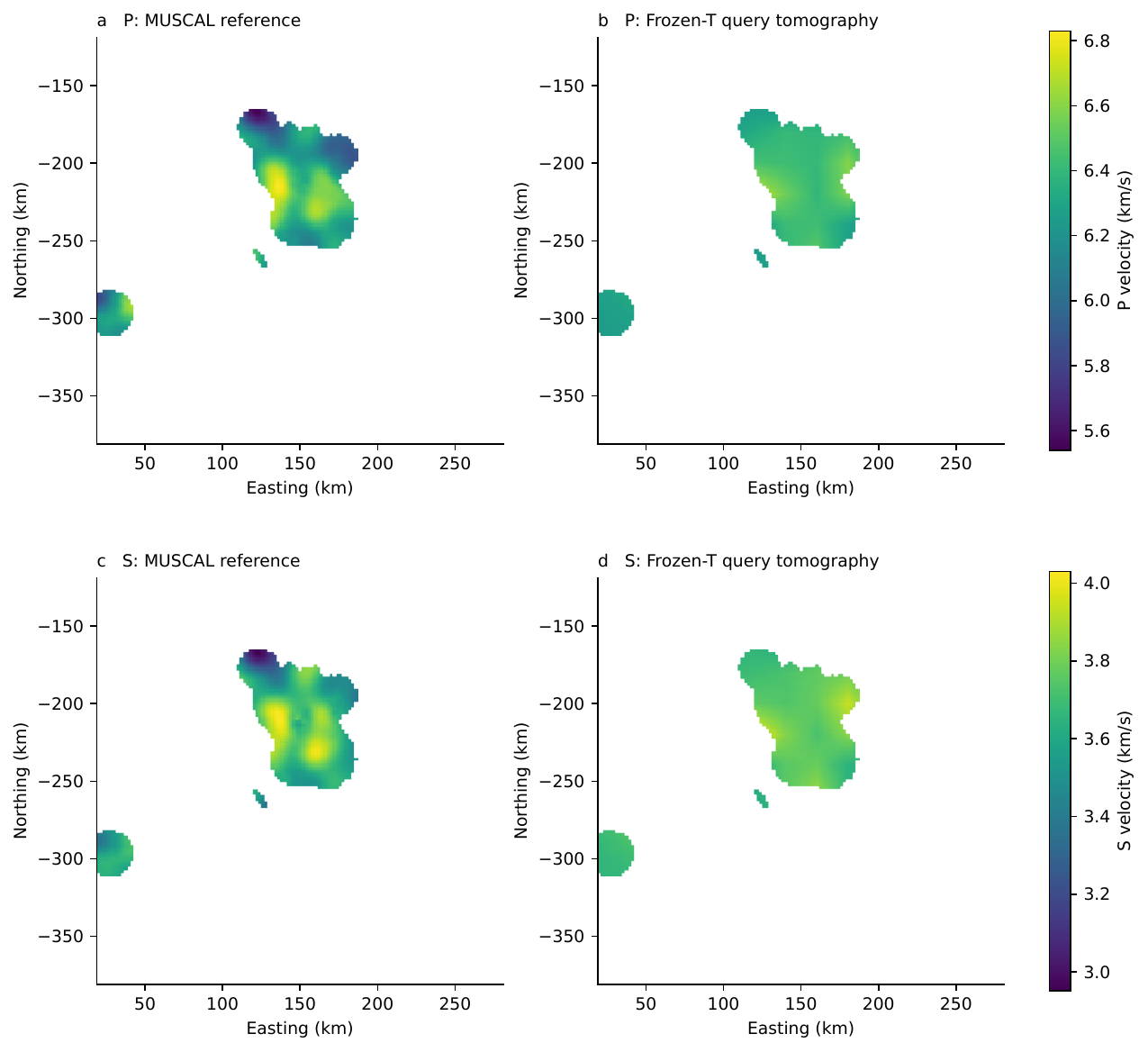}
\caption{\textbf{California velocity sections extracted from generated scalar times.} External MUSCAL and query-only tomography at a fixed 8-km depth. White regions fail the same training-source-neighbour support condition used for random sampling or lack three eligible virtual receivers at 20--70~km. Both maps for each phase use the full joint velocity range. The mask is geometric and is not selected by agreement. All velocity metrics in the tables use held-out hypocentres over 2--20~km depth, not these display pixels. Fine map sampling does not imply resolution below the inversion grid.}
\label{si:fig:manual_tomography_slice}
\end{figure}

\begin{figure}[htbp]
\centering
\includegraphics[width=0.95\textwidth]{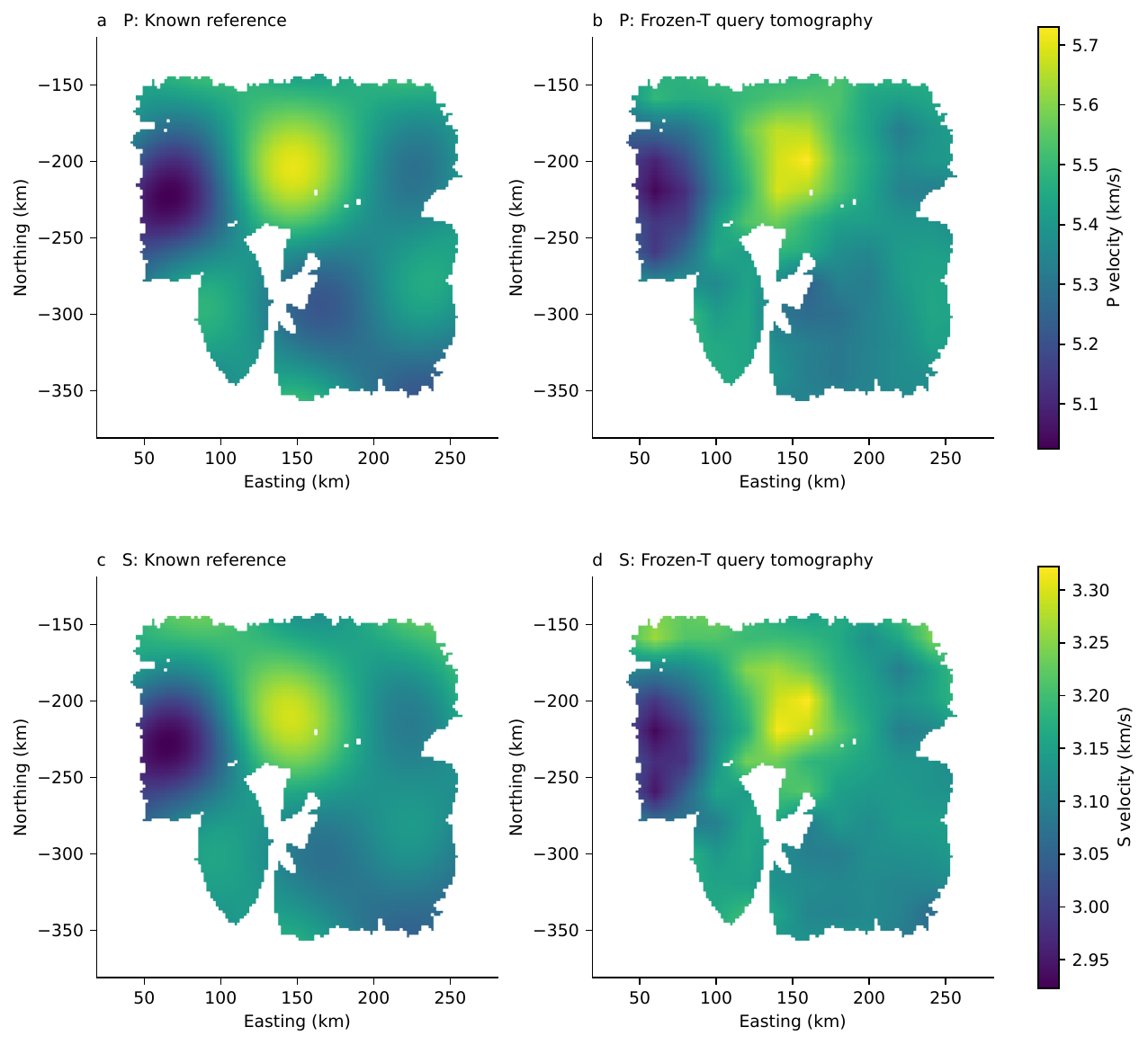}
\caption{\textbf{Known-medium sections from the same scalar-query extraction.} The true velocity and frozen-T query tomography at 8-km depth, with the same geometric support and colour-range rules as Fig.~\ref{si:fig:manual_tomography_slice}. This controlled comparison tests structural extraction where velocity truth is available; it does not replace the real-arrival experiment.}
\label{si:fig:controlled_tomography_slice}
\end{figure}

The companion archive retains the full manual cache, immutable event identifiers, generated coordinates and scalar times, three manual teachers, three controlled value-only teachers, every inverse candidate, final-test arrays, scripts and SHA-256 manifests. CPU checkpoint verification recomputes teacher values and derivatives and independently interpolates the conventional fields. A separate command reruns the inverse fits from the frozen training/validation and generated-query files. Archived inference verification is distinguished from complete neural retraining and from acquisition of the original annotation or external velocity archives.

\clearpage
\subsection{Fixed-receiver averaging and training-time norm regularization}

A further diagnostic tests whether restricting epicentral distance, locally averaging a frozen field, or regularizing its gradient magnitude improves the readout. This follows inspection of the preceding tests and is explicitly a sensitivity study on the same event partitions. All distances, bandwidths, regularization weights and training budgets are fixed before these new comparisons are scored. No PINN or velocity label is introduced.

The 10--50-km window contains 48,386 of the 87,697 manual training records (55.2\%), with the largest 10-km bin at 30--40~km. We compare it with 20--70~km using the same original 32 training-selected surface stations, including their measured elevations. Sources retain the regional horizontal domain and 2--20-km depth restriction. Local queries require the fourth training-source neighbour to lie within scaled distance 1.5 using 10/10/3-km scales. Every offset at every bandwidth must satisfy this rule; all methods share the resulting event support. The controlled cache starts at 20-km epicentral distance, so its nominal 10--50-km subset actually spans 20--50~km. Window comparisons additionally use only events with at least three finite phase paths in both windows.

For each fixed receiver, we draw 1,024 scrambled-Sobol normal-quantile offsets capped at $\pm2.5$ and include their negatives. Horizontal bandwidths are $h=0.5,1,2$~km, with vertical width $h/4$. The readout is
\begin{equation}
\overline{\vect g}_{h}(\vect x_s,\vect x_r)
=\frac{1}{2048}\sum_{j=1}^{2048}\nabla_{\vect x_s}
T(\vect x_r,\vect x_s+\vect\delta_j),\qquad
v_h=\|\overline{\vect g}_h\|^{-1}.
\end{equation}
Only after computing each fixed-receiver speed do we take its median across receivers. Mixing differently directed station gradients before taking the norm is not used. Averaging reciprocal gradient norms is a separate nonlinear operation, retained as a diagnostic in the machine-readable results. An initial 64-versus-128-query check found the unperturbed averages stable but detected sensitivity to the high-frequency intervention. We therefore increased quadrature to 2,048 queries for every bandwidth, without changing weights, distance windows or model selection. Geometry alone defines the refined common support. All initial results are preserved, with intermediate 512/1,024-query estimates and an exact clipped-Gaussian sinusoidal-transfer calculation used only to audit integration error. That analytic calculation never corrects the extracted gradient. Generated queries do not increase the number of independent events.

The imposed oscillation has physical source-coordinate wavenumber $k_j=64\pi d_j/a_j$, where $a_j$ is the field's coordinate scale. Its depth-direction wavelengths are 0.930~km in California and 0.442~km in the known medium. Its maximum gradient perturbations are 0.338 and 0.714~s~km$^{-1}$, respectively, despite a time amplitude of only 0.05~s. For an ideal Gaussian averaging kernel with covariance $\Sigma$, the sinusoidal component is attenuated by $\exp(-\vect k^{\mathsf T}\Sigma\vect k/2)$. Thus horizontal-only averaging would poorly suppress these depth-dominated perturbations. The actual bounded query design is evaluated directly rather than replaced by this ideal formula. Automatic differentiation computes the derivative of the fitted function; agreement between CPU and MPS gradients to approximately $10^{-6}$~s~km$^{-1}$ rules out floating-point differentiation discrepancies as the source of this much larger imposed failure.

The training intervention starts separately from each of the three scalar-only checkpoints. Every arm receives 40 additional epochs on the same full original training cache, identical batches of 8,192, AdamW learning rate $2\times10^{-5}$ with cosine decay to $10^{-6}$, weight decay $10^{-5}$ and gradient clipping at 5. A zero-weight arm controls for the extra optimization. The other arms add
\begin{equation}
\mathcal L_{\rm norm}=\mathbb E\left[
\left(\frac{\|\nabla_sT(\vect x_r,\vect x_s+\vect\delta)\|-
\|\nabla_sT(\vect x_r,\vect x_s)\|}{0.2~\mathrm{s\,km}^{-1}}\right)^2\right]
\end{equation}
at weights 10 or 100. Each batch samples 192 supported training paths in the 10--50-km window, with fixed receivers and offsets drawn uniformly within $\pm(1,1,0.25)$~km. The penalty supplies no target gradient magnitude, velocity value or governing-equation residual. Unlike shrinking the gradient or penalizing all vector curvature, it vanishes for spherical wavefronts in a homogeneous medium. An analytic homogeneous check gives zero norm penalty despite nonzero vector variation, and a parameter finite-difference check agrees with double backpropagation to relative error $1.1\times10^{-9}$.

Each seed and weight selects its checkpoint by the original full validation-time MAE, including the initial checkpoint as an eligible candidate. All 18 fits finish before the new comparisons are scored. We retain every predetermined dose and bandwidth rather than selecting by MUSCAL or true velocity. The control separates added smoothness from additional training, while the finite continuation budget does not establish the optimum of training a regularized model from scratch. This generic regularizer is an added smoothness prior and is distinguished from the original scalar-only emergence evidence.

The refined support retains 743 California and 160 controlled sources before the per-phase receiver-count requirement. In the 10--50-km window, the final comparisons contain 655/708 California P/S events and 137 controlled events for each phase. These denominators differ from the preceding tomography comparison because the query-support and distance conditions differ. Restricting distance alone does not improve the pointwise readout. On 604 common California P events, changing from 20--70 to 10--50~km increases MUSCAL MAE from 0.2976 to 0.3150~km~s$^{-1}$; the paired increase is 0.0174 (event-bootstrap 95\% interval 0.0092--0.0252). S changes from 0.2090 to 0.2094, with an interval spanning zero. Known-medium P/S errors similarly increase from 0.0601/0.0260 to 0.1111/0.0549. The restriction changes the number and directional coverage of usable receivers as well as distance, so it does not isolate a causal effect of distance itself.

Fixed-receiver averaging suppresses the imposed derivative failure. In California, $h=2$~km gives perturbed P/S MAEs of 0.3166/0.2099~km~s$^{-1}$, compared with 2.1085/0.7292 without averaging and 0.3191/0.2112 for the original unperturbed gradient. In the known medium, $h=1$~km gives 0.1122/0.0553, compared with 3.1231/1.2885 for the perturbed direct gradient and 0.1111/0.0549 before perturbation. These are illustrative members of the complete fixed bandwidth scan, not velocity-selected estimators. Unperturbed averages do not reveal new California lateral structure: the $h=2$ P/S depth-detrended correlations remain $-0.102/-0.124$. The operation recovers the pre-perturbation readout but does not correct its systematic reference discrepancy.

The numerical refinement supports the same qualitative rescue while qualifying its precision. At the highlighted California $h=2$ setting, 1,024-to-2,048-query event velocities differ by median 0.0021/0.0007~km~s$^{-1}$ for P/S (95th percentiles 0.0093/0.0029). The exact transfer audit bounds the imposed-gradient integration error at this setting by $2.3\times10^{-4}$~s~km$^{-1}$, and at controlled $h=1$ by $1.9\times10^{-4}$~s~km$^{-1}$. Controlled $h=2$ remains more sensitive, with a corresponding bound of 0.0081~s~km$^{-1}$; it is retained rather than declared converged. Local averaging also has geometric bias: for a 6-km~s$^{-1}$ homogeneous medium, 10-km source depth and 10-km epicentral distance, this $h=2$ query design initially gives approximately 6.09~km~s$^{-1}$ because neighbouring gradient directions differ. A finite-scale derivative need not equal the pointwise physical slowness even in the absence of noise.

Training-time norm smoothness provides a separate positive control in the known medium. Weight 100 reduces P/S MAEs to 0.0688/0.0275~km~s$^{-1}$, versus 0.0919/0.0432 for the equal-budget zero-weight continuation. Paired reductions are 0.0231/0.0156, with event-bootstrap intervals of 0.0107--0.0377/0.0080--0.0244. Depth-detrended correlations increase from 0.679/0.801 to 0.742/0.895. In California, weight 10 gives only unresolved MAE changes of $-0.00064/-0.00029$ relative to the zero-weight continuation, with intervals $-0.00208$--0.00087/$-0.00111$--0.00058. Weight 100 worsens scalar validation behaviour, and every manual seed selects its initial checkpoint; its velocity rows therefore reproduce the original field rather than an accepted strongly regularized solution. Smoothness helps in the controlled setting but does not establish that high-frequency roughness explains the remaining observational mismatch. These intervals resample held-out events conditional on the three fitted seeds and do not propagate catalogue or reference-model uncertainty.

\begin{table}[htbp]
\centering\small
\caption{California external-reference consistency in the 10--50-km distance window under common local-sampling support. $h$ is horizontal sampling width in km, with vertical width $h/4$; 2,048 fixed-receiver source queries are averaged before taking the gradient norm. The scalar and gradient penalties contain no velocity labels or PDE residual. All predetermined bandwidths and training weights are reported. MAE is in km~s$^{-1}$; $\rho_z$ removes a cubic depth trend.}
\begin{adjustbox}{max width=\linewidth}\begin{tabular}{lrrrr}
\toprule
Readout & P MAE & P $\rho_z$ & S MAE & S $\rho_z$ \\
\midrule
Original gradient & 0.3191 & -0.074 & 0.2112 & -0.121 \\
Perturbed gradient & 2.1085 & -0.071 & 0.7292 & -0.060 \\
Local average, $h=0.5$ & 0.3189 & -0.075 & 0.2111 & -0.122 \\
Local average, $h=1$ & 0.3185 & -0.080 & 0.2108 & -0.123 \\
Local average, $h=2$ & 0.3170 & -0.102 & 0.2100 & -0.124 \\
Perturbed average, $h=0.5$ & 1.4854 & -0.082 & 0.4876 & -0.061 \\
Perturbed average, $h=1$ & 0.4776 & -0.104 & 0.2279 & -0.085 \\
Perturbed average, $h=2$ & 0.3166 & -0.104 & 0.2099 & -0.124 \\
40 epochs, no added penalty & 0.3181 & -0.079 & 0.2105 & -0.122 \\
40 epochs, norm penalty 10 & 0.3174 & -0.077 & 0.2102 & -0.116 \\
40 epochs, norm penalty 100 & 0.3191 & -0.074 & 0.2112 & -0.121 \\
\bottomrule
\end{tabular}\end{adjustbox}
\end{table}

\begin{table}[htbp]
\centering\small
\caption{Known-medium recovery in the 10--50-km distance window under common local-sampling support. $h$ is horizontal sampling width in km, with vertical width $h/4$; 2,048 fixed-receiver source queries are averaged before taking the gradient norm. The scalar and gradient penalties contain no velocity labels or PDE residual. All predetermined bandwidths and training weights are reported. MAE is in km~s$^{-1}$; $\rho_z$ removes a cubic depth trend.}
\begin{adjustbox}{max width=\linewidth}\begin{tabular}{lrrrr}
\toprule
Readout & P MAE & P $\rho_z$ & S MAE & S $\rho_z$ \\
\midrule
Original gradient & 0.1111 & 0.627 & 0.0549 & 0.742 \\
Perturbed gradient & 3.1231 & -0.107 & 1.2885 & -0.099 \\
Local average, $h=0.5$ & 0.1114 & 0.627 & 0.0550 & 0.742 \\
Local average, $h=1$ & 0.1122 & 0.626 & 0.0553 & 0.741 \\
Local average, $h=2$ & 0.1159 & 0.625 & 0.0568 & 0.737 \\
Perturbed average, $h=0.5$ & 0.5245 & 0.054 & 0.1646 & 0.216 \\
Perturbed average, $h=1$ & 0.1122 & 0.627 & 0.0553 & 0.741 \\
Perturbed average, $h=2$ & 0.1170 & 0.597 & 0.0574 & 0.723 \\
40 epochs, no added penalty & 0.0919 & 0.679 & 0.0432 & 0.801 \\
40 epochs, norm penalty 10 & 0.0847 & 0.702 & 0.0373 & 0.836 \\
40 epochs, norm penalty 100 & 0.0688 & 0.742 & 0.0275 & 0.894 \\
\bottomrule
\end{tabular}\end{adjustbox}
\end{table}

\clearpage
\section{Quantum Field-Value Readouts}
\label{si:sec:quantum_supplement}

This exploratory synthetic study tests whether phase-sensitive scalar values support unsupervised derivative and trajectory queries under Schr\"odinger evolution. Each task is fitted independently. The test does not use a PINN: physical equations generate reference fields and define post-training readouts, but neither equation residuals nor hidden physical quantities supervise fitting. The initial neural, sampling and scoring protocol was frozen before neural training and derivative scoring. Trajectory sampling was specified during scalar-only training, before derivative or trajectory scoring. A global smooth baseline was added later, with its timing and solver amendment retained.

\subsection{States, information content and known readout}

We use two spatial dimensions and dimensionless units $\hbar=m=1$, with no vector potential and no spin. Writing $\psi=u+iw=R e^{iS}$ gives
\begin{equation}
\rho=u^2+w^2,\qquad
\vect{j}=u\nabla w-w\nabla u,\qquad
\vect{v}_{\mathrm B}=\vect{j}/\rho=\nabla S
\quad (\rho>0).
\label{si:eq:quantum_si_readout}
\end{equation}
The general identity is $\vect{v}_{\mathrm B}=(\hbar/m)\operatorname{Im}(\nabla\psi/\psi)$~\citep{Bohm1952,StruyveValentini2009}. The imaginary-part operator is essential. This Bohmian velocity is a local probability-flow velocity, and $\nabla S$ is its associated momentum field; neither is a general momentum distribution or a group-velocity measurement. The readout is known in advance and applied after training, rather than discovered from data. Complex-field tasks supervise both $u$ and $w$, so they contain phase information. The phase-only task supervises $S$ and supplies only a velocity prediction; hidden reference amplitudes are not used to construct a predicted current.

For a free Gaussian packet, define $\tau=2\sigma^2$, $b=1+it/\tau$, $\vect{d}=\vect{x}-\vect{q}-\vect{k}t$ and
\begin{equation}
G(\vect{x},t;\vect{q},\vect{k},\sigma)=
\frac{1}{\sqrt{2\pi\sigma^2}\,b}
\exp\left[-\frac{\|\vect{d}\|^2}{4\sigma^2 b}
+i\vect{k}\mathbin{\cdot}(\vect{x}-\vect{q})
-\frac{i\|\vect{k}\|^2t}{2}\right].
\end{equation}
The Gaussian-phase task uses $\vect{q}=(-0.7,0.2)$, $\vect{k}=(1,0.35)$ and $\sigma=0.85$. Its unwrapped analytic phase is
\begin{equation}
S=\vect{k}\mathbin{\cdot}(\vect{x}-\vect{q})-\frac{\|\vect{k}\|^2t}{2}
+\frac{t\|\vect{d}\|^2}{4\sigma^2\tau[1+(t/\tau)^2]}-\arctan(t/\tau),
\qquad
\vect{v}_{\mathrm B}=\vect{k}+\frac{t\vect{d}}{\tau^2+t^2}.
\end{equation}
We train on this real phase, not on a wrapped angle or the probability density.

The vortex task uses the normalized harmonic-oscillator state
\begin{equation}
\psi_+(x,y,t)=\frac{x+iy}{\sqrt{\pi}}e^{-(x^2+y^2)/2}e^{-2it},
\qquad V=(x^2+y^2)/2,
\end{equation}
with $\rho=r^2 e^{-r^2}/\pi$ and $\vect{v}_+=(-y,x)/r^2$. The opposite-circulation state $\psi_-=(x-iy)e^{-r^2/2}e^{-2it}/\sqrt{\pi}$ has the same density for every time and $\vect{j}_-=-\vect{j}_+$. This exact paired counterexample establishes the information ambiguity of density-only recovery, consistent with the distinction between density and state information~\citep{Gale1968}. Only $\psi_+$ is fitted as a complex field. The separate density-only network fits its $\rho$; no current recovery is assigned to that fit, and $\psi_-$ is an analytic control rather than a second trained result.

The interference task superposes two free packets:
\begin{equation}
\psi=\frac{G_1+a e^{i\varphi}G_2}{N},\qquad
N^2=1+a^2+2a\operatorname{Re}(e^{i\varphi}\langle G_1|G_2\rangle),
\end{equation}
where $a=0.9$, $\varphi=0.7$, $\sigma=0.75$, $\vect{q}_1=(-1.2,-0.3)$, $\vect{k}_1=(2,0.4)$, $\vect{q}_2=(1.2,0.3)$ and $\vect{k}_2=(-1.7,-0.3)$. The constant overlap is
\begin{equation}
\langle G_1|G_2\rangle=
\exp\left[-\frac{\|\vect{q}_1-\vect{q}_2\|^2}{8\sigma^2}
-\frac{\sigma^2\|\vect{k}_1-\vect{k}_2\|^2}{2}
+\frac{i}{2}(\vect{k}_1+\vect{k}_2)\mathbin{\cdot}(\vect{q}_1-\vect{q}_2)\right].
\end{equation}
Both packet tasks have $V=0$. Analytic derivatives were checked by central differences (maximum absolute discrepancy below $1.1\times10^{-8}$). Independently propagating the interference initial field by a Fourier solver on a $256^2$ grid over $[-12,12)^2$ gives maximum discrepancies below $3.3\times10^{-13}$ at $t=0.45$ and 1.1; the initial normalization differs from unity by less than $3\times10^{-16}$.

\subsection{Sampling, optimization and frozen selection}

All tasks use $0\leq t\leq1.4$. The vortex and density tasks occupy $[-3,3]^2$; the Gaussian and interference tasks occupy $[-4,4]^2$. Uniform rejection sampling supplies 8,192 training and 2,048 validation coordinates per task, excluding the union of the spatial box $0.35<x<1.05$, $-0.7<y<0$ and the temporal slab $0.56<t<0.70$. A separate domain test contains 8,192 coordinates over the full domain, including those excluded regions. Two further tests each contain 2,048 coordinates: the spatial box outside the time slab and the time slab outside the spatial box. All split coordinates are distinct. These are interpolation tests within each state's coordinate domain, not future-time extrapolation or held-out-state tests.

Each network has four fully connected hidden layers of width 128 with SiLU activations and one real output for phase/density or two for a complex field. Inputs are affine-normalized $(x,y,t)$. Output-channel means and one pooled output scale are computed from training values. Physical-coordinate differentiation includes both input and output scales. Three seeds (101, 102, 103) are fitted for each of four tasks, giving 12 models. Adam uses batches of 2,048 for 600 epochs, an initial learning rate of $0.003$ with cosine decay to $10^{-5}$ and a gradient-norm clip of 5. The sole loss is normalized field-value mean squared error. Validation occurs every ten epochs and selects the minimum scalar validation loss. Selected epochs are (600,580,570) for Gaussian phase, (580,600,600) for vortex, (600,600,600) for interference and (600,580,600) for density. Current, derivative and trajectory errors never select a checkpoint.

The primary ensemble is the arithmetic mean of the three predicted fields, differentiated before the nonlinear current/velocity readout. It is not an average of three predicted velocities. Individual-seed metrics use the same coordinates and supports. Table~\ref{si:tab:quantum_primary} gives ensemble errors; Table~\ref{si:tab:quantum_seeds} reports each optimization seed. Variation across these seeds is conditional on fixed data and fixed underlying states.

\begin{table}[htbp]
\centering\small
\caption{Quantum readouts from the mean of three scalar-trained fields.}
\label{si:tab:quantum_primary}
\begin{adjustbox}{max width=\linewidth}\begin{tabular}{lrrrrr}
\toprule
Field & Value (\%) & Current (\%) & Velocity (\%) & Median ($\degree$) & P95 ($\degree$) \\
\midrule
Gaussian phase & 0.292 & -- & 0.441 & 0.134 & 0.825 \\
Vortex & 0.299 & 0.585 & 0.708 & 0.528 & 7.224 \\
Interference & 1.128 & 2.481 & 2.387 & 0.445 & 4.016 \\
\bottomrule
\end{tabular}\end{adjustbox}
\par\smallskip\begin{minipage}{0.97\textwidth}\footnotesize
Value and current errors are relative $L_2$ over all 8,192 test coordinates. Velocity error is probability-weighted relative $L_2$ on true $\rho\geq10^{-4}$; direction additionally requires true speed $\geq0.05$. The phase-only Gaussian has no predicted current.
\end{minipage}
\end{table}

\begin{table}[htbp]
\centering\small
\caption{Individual optimization seeds on the same fixed quantum test set.}
\label{si:tab:quantum_seeds}
\begin{adjustbox}{max width=\linewidth}\begin{tabular}{llrrr}
\toprule
Field & Seed & Value (\%) & Current (\%) & Velocity (\%) \\
\midrule
Gaussian phase & 101 & 0.326 & -- & 0.557 \\
Gaussian phase & 102 & 0.304 & -- & 0.447 \\
Gaussian phase & 103 & 0.351 & -- & 0.561 \\
Vortex & 101 & 0.410 & 0.884 & 1.205 \\
Vortex & 102 & 0.420 & 0.898 & 1.015 \\
Vortex & 103 & 0.445 & 0.982 & 1.165 \\
Interference & 101 & 1.364 & 3.016 & 3.008 \\
Interference & 102 & 1.376 & 2.968 & 3.006 \\
Interference & 103 & 1.339 & 3.115 & 3.050 \\
\bottomrule
\end{tabular}\end{adjustbox}
\par\smallskip\begin{minipage}{0.97\textwidth}\footnotesize
Seeds quantify optimization variation conditional on one training sample and one state per task; they are not independent quantum states. Error definitions match Table~\ref{si:tab:quantum_primary}.
\end{minipage}
\end{table}

\subsection{Metrics, support and classical comparisons}

Value error is $\|\widehat f-f\|_2/\|f\|_2$, where $f$ is phase, density or the two-component complex field as appropriate. Probability-current error is $\|\widehat{\vect{j}}-\vect{j}\|_2/\|\vect{j}\|_2$ over all test coordinates, without a node mask. Velocity error on $M_\epsilon=\{i:\rho_i\geq\epsilon\}$ is
\begin{equation}
E_v(\epsilon)=\left[
\frac{\sum_{i\in M_\epsilon}\rho_i\|\widehat{\vect{v}}_i-\vect{v}_i\|^2}
{\sum_{i\in M_\epsilon}\rho_i\|\vect{v}_i\|^2}\right]^{1/2}.
\end{equation}
The primary threshold is $\epsilon=10^{-4}$; $10^{-5}$ and $10^{-3}$ are also reported. Direction errors use the same mask and additionally require reference speed $\geq0.05$. Their median and 95th percentile are unweighted over retained coordinates. Any predicted density below $10^{-10}$ on a valid true-density point counts as a failure; a $10^{-12}$ denominator floor is numerical protection, not a means to remove such points. No such failures occur for the primary ensemble or its reported threshold scan. The primary masks retain more than 99.95\% of the estimated probability mass within the sampled test domains. Table~\ref{si:tab:quantum_nodes} reports retained counts and the complete ensemble threshold scan; this finite-domain mass estimate is not the integral over all space.

The initial classical comparator is a thin-plate radial-basis interpolator using 96 nearest training coordinates, degree-one polynomials and smoothing selected from $\{0,10^{-5},10^{-3}\}$ by scalar validation. Derivatives use physical-coordinate central differences with step $10^{-3}$. Its changing neighbour sets can switch local interpolants. On the first 1,024 test points, halving the step to $5\times10^{-4}$ changes Gaussian, vortex and interference velocity errors from 1.906/8.925/14.254\% to 2.708/10.794/15.303\%. Those sensitivity results and the original full metrics remain archived; they do not establish a neural advantage over smooth classical representations.

After this diagnostic, we added a global $C^2$ cubic tensor-product spline with analytic derivatives, using all the same training values. Candidate basis counts are $(14,14,9)$ and $(20,20,12)$ in $(x,y,t)$, crossed with coefficient-ridge strengths $10^{-8}$, $10^{-5}$ and $10^{-3}$. Only scalar validation selects a candidate. An initial iterative solver did not converge adequately for two weakly regularized candidates; the same objectives and candidates were solved directly using sparse ridge normal equations before any spline test scoring. All normalized normal-equation residuals are below $10^{-10}$, and derivative finite-difference checks are below $10^{-6}$. Gaussian selects $(14,14,9)$ with $10^{-8}$; the other tasks select $(20,20,12)$ with $10^{-5}$. Neural checkpoints were not changed. The amendment, all six candidates and solver diagnostics are archived.

Table~\ref{si:tab:quantum_baseline} reports the smooth comparison and dedicated holdouts. Splines outperform the neural fields for the Gaussian and vortex. Interference-current errors are similar; neural velocity error is lower on the domain test but higher in the time holdout (6.595\% versus 5.383\%). Thus this comparison supports recovery through multiple representations, not uniform neural superiority. Both use value-only validation, but neither final value accuracy nor computational budget was matched.

\begin{table}[htbp]
\centering\small
\caption{Global smooth comparison and dedicated interpolation holdouts.}
\label{si:tab:quantum_baseline}
\begin{adjustbox}{max width=\linewidth}\begin{tabular}{llrrrrr}
\toprule
Field & Representation & Value (\%) & Current (\%) & Domain & Space & Time \\
\midrule
Gaussian phase & Neural mean & 0.2920 & -- & 0.4406 & 0.3480 & 0.5226 \\
Gaussian phase & $C^2$ spline & 0.0079 & -- & 0.0027 & 0.0021 & 0.0024 \\
Vortex & Neural mean & 0.2988 & 0.585 & 0.7081 & 0.3803 & 0.6346 \\
Vortex & $C^2$ spline & 0.1003 & 0.354 & 0.3805 & 0.4620 & 0.2898 \\
Interference & Neural mean & 1.1283 & 2.481 & 2.3870 & 3.4044 & 6.5951 \\
Interference & $C^2$ spline & 1.0909 & 2.488 & 3.8062 & 4.6508 & 5.3829 \\
\bottomrule
\end{tabular}\end{adjustbox}
\par\smallskip\begin{minipage}{0.97\textwidth}\footnotesize
The final three columns are velocity errors (\%). Domain: 8,192 uniform coordinates, including the excluded regions. Space/time: 2,048 dedicated coordinates each, excluding their intersection. Value/current columns use the domain test. The spline was added after initial neural/RBF scoring; its grid and ridge strength were selected only by scalar validation.
\end{minipage}
\end{table}

\begin{table}[htbp]
\centering\small
\caption{Node-threshold sensitivity of the neural mean-field readout.}
\label{si:tab:quantum_nodes}
\begin{adjustbox}{max width=\linewidth}\begin{tabular}{lrrrrr}
\toprule
Field & $\rho$ cutoff & Retained $n$ & Mass (\%) & Velocity (\%) & P95 ($\degree$) \\
\midrule
Gaussian phase & $10^{-5}$ & 6,752 & 99.9974 & 0.441 & 0.914 \\
Gaussian phase & $10^{-4}$ & 5,452 & 99.9550 & 0.441 & 0.825 \\
Gaussian phase & $10^{-3}$ & 3,857 & 99.4547 & 0.439 & 0.696 \\
Vortex & $10^{-5}$ & 7,760 & 99.9993 & 0.728 & 15.427 \\
Vortex & $10^{-4}$ & 7,069 & 99.9860 & 0.708 & 7.224 \\
Vortex & $10^{-3}$ & 5,510 & 99.6788 & 0.621 & 2.767 \\
Interference & $10^{-5}$ & 6,650 & 99.9976 & 2.485 & 7.966 \\
Interference & $10^{-4}$ & 5,574 & 99.9609 & 2.387 & 4.016 \\
Interference & $10^{-3}$ & 3,942 & 99.4430 & 2.158 & 2.581 \\
\bottomrule
\end{tabular}\end{adjustbox}
\par\smallskip\begin{minipage}{0.97\textwidth}\footnotesize
Mass is $\sum_{\rho_i\geq\epsilon}\rho_i/\sum_i\rho_i$ on sampled test coordinates; it is not a full-space normalization integral. No predicted-density failures ($\widehat\rho<10^{-10}$) occur on these true-density supports. Angular P95 also applies the fixed speed threshold.
\end{minipage}
\end{table}

\subsection{Independent trajectories and circulation}

Each task uses 32 initial points sampled by density-weighted rejection at $t=0$ with seed 2026091981, true density $\geq10^{-3}$ and a 0.5-unit margin inside the spatial domain. Vortex radii additionally satisfy $0.4\leq r\leq1.8$. These restrictions precede trajectory scoring; all 32 points are retained in reported counts. The Gaussian reference is
\begin{equation}
\vect{x}(t)=\vect{q}+\vect{k}t+(\vect{x}_0-\vect{q})\sqrt{1+(t/\tau)^2}.
\end{equation}
The vortex reference preserves $r_0$ and advances the polar angle by $t/r_0^2$. Interference reference paths integrate the analytic field with DOP853 at relative tolerances $10^{-10}$ and $10^{-11}$, corresponding absolute tolerances 100 times smaller and maximum step 0.01. Neural paths use an independent batched RK4 integrator with 280 and 560 steps, float64 positions and CPU float32 field evaluation. Predicted density below $10^{-10}$, nonfinite values or exit from the trained spatial domain would stop and count a path as failed. All ensemble and individual-seed paths complete (32/32 in each task), with no reference-domain exits. No reference path passes below $\rho=10^{-4}$, so this sample does not test traversal arbitrarily close to a node.

Table~\ref{si:tab:quantum_paths} gives dimensionless Euclidean endpoint errors and whole-path RMS. Refining RK4 from 280 to 560 steps changes positions by at most $2.18\times10^{-8}$, $5.63\times10^{-8}$ and $2.29\times10^{-4}$ for Gaussian, vortex and interference, respectively. For vortex circulation, 12 prescribed loops use radii 0.4, 0.8, 1.2 and 1.8 at times 0, 0.6 and 1.2, with 1,024 points per loop. The ensemble winding $\oint\vect{v}_{\mathrm B}\mathbin{\cdot}d\vect{l}/(2\pi)$ differs from the reference $+1$ by at most $1.65\times10^{-8}$. This is a consistency check of recovered topology; it is not a discovery of a quantization rule. Figure~7 in the main text displays the first eight interference initial conditions without selection by fit quality.

\begin{table}[htbp]
\centering\small
\caption{Independent trajectory comparison over $0\leq t\leq1.4$.}
\label{si:tab:quantum_paths}
\begin{adjustbox}{max width=\linewidth}\begin{tabular}{lrrrr}
\toprule
Field & Completed & Median endpoint & P95 endpoint & Path RMS \\
\midrule
Gaussian phase & 32/32 & 0.00238 & 0.00437 & 0.00136 \\
Vortex & 32/32 & 0.00307 & 0.01472 & 0.00380 \\
Interference & 32/32 & 0.01888 & 0.03971 & 0.01869 \\
\bottomrule
\end{tabular}\end{adjustbox}
\par\smallskip\begin{minipage}{0.97\textwidth}\footnotesize
Position errors are dimensionless Euclidean distances. Each of the three individual seeds also completes all 32 identical initial conditions per task. No paths are dropped from the reported sample. The Gaussian and vortex references are closed form; interference uses independent DOP853 integration.
\end{minipage}
\end{table}

\subsection{Phase sensitivity and density-only ambiguity}

The fixed intervention $\phi=0.01\sin(40x+29y)$ is added to the frozen Gaussian phase or multiplies a frozen complex field as $\widehat\psi\mapsto\widehat\psi e^{i\phi}$. In the latter case the predicted density is exactly unchanged and $\widehat{\vect{v}}\mapsto\widehat{\vect{v}}+\nabla\phi$ wherever the unfloored readout is defined. The oscillatory derivative therefore amplifies a small value perturbation. The perturbed field is not claimed to remain a solution for the original Hamiltonian. Table~\ref{si:tab:quantum_perturbation} reports all three interventions. This experiment establishes derivative sensitivity; it does not test a quantum training-time gradient regularizer.

The density-only mean field has 0.349\% relative value error. Its unconstrained real output is negative at 3.10\% of low-density test coordinates; the summed negative magnitude is 0.0093\% of the summed true test density. It is not treated as a valid quantum state or used to infer current. The decisive non-identifiability result comes from the exact $\psi_+/\psi_-$ pair, not from that fitter's imperfections: even exact density at all sampled times cannot determine the circulation sign. Phase-sensitive data, an appropriate representation and derivative regularity are distinct requirements.

\begin{table}[htbp]
\centering\small
\caption{Fixed phase perturbation of each frozen neural mean field.}
\label{si:tab:quantum_perturbation}
\begin{adjustbox}{max width=\linewidth}\begin{tabular}{llrrr}
\toprule
Field & Readout & Value (\%) & Current (\%) & Velocity (\%) \\
\midrule
Gaussian phase & Original & 0.292 & -- & 0.441 \\
Gaussian phase & Phase perturbation & 0.396 & -- & 31.002 \\
Vortex & Original & 0.299 & 0.585 & 0.708 \\
Vortex & Phase perturbation & 0.762 & 34.951 & 34.908 \\
Interference & Original & 1.128 & 2.481 & 2.387 \\
Interference & Phase perturbation & 1.344 & 33.598 & 21.745 \\
\bottomrule
\end{tabular}\end{adjustbox}
\par\smallskip\begin{minipage}{0.97\textwidth}\footnotesize
The phase perturbation is $\phi=0.01\sin(40x+29y)$. Multiplication by $e^{i\phi}$ leaves the predicted density unchanged for the complex fields. It is a sensitivity intervention, not a claim that the altered field solves the original Hamiltonian dynamics.
\end{minipage}
\end{table}

\subsection{Provenance, reproduction and scope}

The companion \path{quantum_value_pilot_20260920.zip} archive contains all 12 selected weights, scalar caches and splits, individual and ensemble scores, trajectory arrays, the original local-RBF results, global-spline candidates, locked protocols and source snapshots. An independent extracted copy verified 88 manifest entries on CPU. The training/generator source matches the locked version, all checkpoint hashes pass, all split coordinates are unique and neither training nor validation contains an excluded-block point. Reconstructed CPU field values and derivatives agree with archived MPS inference within $1.5\times10^{-6}$, and 45 ensemble scalar/current/velocity metrics were recomputed independently. This verifies frozen-weight inference and scoring, not a complete retraining run. The recorded environment uses Python 3.14.0, NumPy 2.3.4, SciPy 1.16.2, PyTorch 2.9.1 and Matplotlib 3.10.7 on macOS arm64.

Compact protocols, scores, figure arrays and audit records accompany the manuscript under \path{supplementary/revision/quantum_value_pilot/}; source snapshots and the asset builder are under \path{code/revision/quantum_value_pilot/}. The results concern synthetic observations of three fixed phase-bearing fields and one density control. They demonstrate neither experimental quantum-state reconstruction, transfer to unseen states or Hamiltonians, a cross-system reliability theorem, nor a quantum benefit from gradient regularization. Their role is to test the same conditional value-to-geometry mechanism under a different evolution equation while exposing information and regularity limits explicitly.

\FloatBarrier

\end{document}